\documentclass[epj,nopacs]{svjour}
\pdfoutput=1

\usepackage[utf8]{inputenc} 
\usepackage[T1]{fontenc}    
\usepackage{hyperref}       
\usepackage{url}            
\usepackage{booktabs}       
\usepackage{amsfonts}       
\usepackage{nicefrac}       
\usepackage{microtype}      
\usepackage{graphicx}
\usepackage{capt-of}        
\usepackage{amsmath}
\usepackage{authblk}
\usepackage{multirow}
\usepackage{stfloats}
\usepackage{listings}

\newcommand{\pizero}{\ensuremath{\pi^0}}
\newcommand{\chpi}{\ensuremath{\pi^\pm}}

\begin{document}

\title{Calorimetry with Deep Learning: Particle Simulation and Reconstruction for Collider Physics}
\titlerunning{Deep Learning for Calorimetry}

\author{Dawit Belayneh\inst{1} \and
Federico Carminati\inst{2} \and
Amir Farbin\inst{3} \and
Benjamin Hooberman\inst{4} \and
Gulrukh Khattak\inst{2}\inst{5} \and
Miaoyuan Liu\inst{6} \and
Junze Liu\inst{4} \and
Dominick Olivito\inst{7} \and
Vitória Barin Pacela\inst{8} \and
Maurizio Pierini\inst{2} \and
Alexander Schwing\inst{4} \and
Maria Spiropulu\inst{9} \and
Sofia Vallecorsa\inst{2} \and
Jean-Roch Vlimant\inst{9} \and
Wei Wei\inst{4} \and
Matt Zhang\thanks{corresponding author, mzhang60@illinois.edu}\inst{4}}
\authorrunning{Zhang et al.}

\institute{Univ. of Chicago \and
European Organization for Nuclear Research (CERN) \and
Univ. of Texas Arlington \and
Univ. of Illinois at Urbana-Champaign \and
UET Peshawar \and
Fermi National Accelerator Laboratory \and
Univ. of California, San Diego \and
Univ. of Helsinki \and
California Institute of Technology}

\abstract{
Using detailed simulations of calorimeter showers as training data, we investigate
the use of deep learning algorithms for the simulation and reconstruction of particles produced in
high-energy physics collisions. We train neural networks on shower data at the
calorimeter-cell level, and show significant improvements for simulation and reconstruction when using these networks compared to methods which rely on currently-used
state-of-the-art algorithms. We define two models: an end-to-end reconstruction network which performs simultaneous particle identification and energy regression of particles when given calorimeter shower data, and a generative network which can provide reasonable modeling of calorimeter showers for different particle types at specified angles and energies. We investigate the optimization of our models with hyperparameter scans. Furthermore, we demonstrate the applicability of the reconstruction model to shower inputs from other detector geometries, specifically ATLAS-like and CMS-like geometries.  These networks can serve as fast and computationally light methods for particle shower simulation and reconstruction for current and future experiments at particle colliders.}

\maketitle

\section{Overview}

In high energy physics (HEP) experiments, detectors act as imaging devices, allowing physicists to take snapshots of decay products from particle collision "events". Calorimeters are key components of such detectors. When a high-energy primary particle travels through dense calorimeter material, it deposits its energy and produces a shower of secondary particles. Detector "cells" within the calorimeter then capture these energy depositions, forming a set of voxelized images which are characteristic of the type and energy of the primary particle. 

The starting point of any physics analysis is the identification of the types of particles produced in each collision and the measurement of the momentum carried by each of these particles. These tasks have traditionally used manually-designed algorithms, producing measurements of physical features such as shower width and rate of energy loss for particles traversing calorimeter layers.
In the last few years, researchers have started realizing that machine learning (ML) techniques are well suited for such tasks, 
e.g. using boosted decision trees (BDTs) on calculated features for doing particle classification. Indeed, ML has long been applied to various other tasks in HEP~\cite{Denby:1987rk,Peterson:1988gs,Abreu:1992jp}, including the 2012 discovery of the Higgs boson~\cite{HiggsATLAS,HiggsCMS} at the ATLAS~\cite{Aad:2008zzm} and CMS~\cite{Chatrchyan:2008aa} experiments at the Large Hadron Collider (LHC). 

In the next decade, the planned High Luminosity Large Hadron Collider (HL-LHC) upgrade~\cite{Apollinari:2284929} will enhance
the experimental sensitivity to rare phenomena by increasing the number of collected proton-proton collisions by a factor of ten. In addition, many next-generation detector components, such as the sampling calorimeters proposed for the ILC~\cite{ILC}, CLIC~\cite{CLIC}, and CMS~\cite{CMSCollaboration:2015zni} detectors, will improve physicists' ability to identify and measure particles by using much finer 3D arrays of voxels. These and future accelerator upgrades will lead to higher data volumes and pose a variety of technological and computational challenges in tasks, such as real-time particle reconstruction.

In addition to actual collision data, physics analyses typically require extremely detailed and precise simulations of detector data, generated using software packages such as GEANT4\cite{GEANT4}. These simulations are used to develop and test analysis techniques. They rely on calculations of the micro-physics governing the interaction of particles with matter, and are generally very CPU intensive. In some cases, such as the ATLAS experiment, simulation currently requires roughly half of the experiment's computing resources\cite{GEANT_usage}. This fraction is expected to increase significantly for the HL-LHC. These challenges require novel computational and algorithmic techniques, which has prompted recent efforts in HEP to apply modern ML to calorimetry~\cite{ML1,ML2,ML3,ML4}.

With this work, we aim to demonstrate the applicability of neural-network based approaches to reconstruction and simulation tasks, looking at a real use case. To do this, we use fully simulated calorimeter data for a typical collider detector to train two models: (i) a network for end-to-end particle reconstruction, receiving as input a particle shower and acting both as a particle identification algorithm and as a regression algorithm for the particle's energy; (ii) a generative adversarial network (GAN)~\cite{Goodfellow} for simulating particle showers, designed to return calorimeter-cell voxelized images like those generated by GEANT4.
Both models aim to preserve the accuracy of more traditional approaches while drastically reducing the required computing resources and time, thanks partly to a built-in portability to heterogeneous CPU+GPU computing environments.


This paper is a legacy document summarizing two years of work. It builds upon initial simulation, classification, and regression results which we presented at the 2017 Workshop on Deep Learning for Physical Sciences at the NeurIPS conference. Those results were derived using simplified problem formulations~\cite{Mau2017}.  For instance, we only used particles of a single fixed energy for classification, and had only considered showers produced by particles traveling perpendicularly to the calorimeter surface. The results presented in this paper deal with a more realistic use case and supersede the results in Ref.~\cite{Mau2017}. 

For the studies presented in this paper, we used two computing clusters: at the University of Texas at Arlington (UTA), and at the Blue Waters supercomputing network, located at the University of Illinois at Urbana Champaign (UIUC). The UTA cluster has 10 NVIDIA GTX Titan GPUs with 6 GB of memory each. Blue Waters uses NVDIA Kepler GPUs, also with 6 GB of memory each. 

GAN models were implemented and trained using Keras~\cite{keras} and  Tensorflow~\cite{tensorflow2015-whitepaper}. Reconstruction models were implemented and trained using PyTorch~\cite{PyTorch}. The sample generation and training frameworks were both written in Python, with the sample generation codebase at \url{https://github.com/UTA-HEP-Computing/CaloSampleGeneration} and the TriForce reconstruction training framework at \url{https://github.com/BucketOfFish/Triforce_CaloML}.

This document is structured as follows: In Section~\ref{sec:data}, we describe how we created and prepared the data used in these studies. Section~\ref{sec:problems} introduces the two physics problems, particle simulation and reconstruction. Sections~\ref{sec:GAN}~and~\ref{sec:reco} describe the corresponding models, how they were trained, and the performances they reached. In particular, Section~\ref{sec:reco} compares our results to those of more traditional approaches, and also extends those comparisons to simulated performances on detector geometries similar to those of the ATLAS and CMS calorimeters. Conclusions are given in Section~\ref{sec:conclusion}.

\section{Dataset}\label{sec:data}

This study is based on simulated data produced with GEANT4~\cite{GEANT4}, using the geometric layout of the proposed Linear Collider Detector (LCD) for the CLIC accelerator~\cite{Lebrun:2012hj}. We limit the study to the central region (barrel) of the LCD detector, where the electromagnetic calorimeter (ECAL) consists of a cylinder with inner radius of 1.5 m, structured as a set of 25 silicon sensor planes, segmented in $5.1~\times~5.1$ mm$^2$ square cells, alternated with tungsten absorber planes. In the barrel region, the hadronic calorimeter (HCAL) sits behind the ECAL, at an inner radius of 1.7 m. The HCAL 
consists of 60 layers of polystyrene scintillators, segmented in cells with  $3~\times~3$ cm$^2$ area and alternated with layers of steel absorbers. 

The event simulation considers the full detector layout, including the material in front of the calorimeter and the effect of the solenoidal magnetic field. From the full data we take slices centered around the barycenter of each ECAL energy deposit and we represent the ECAL and HCAL slices as 3D arrays of energy deposits in the cells. 

We consider four kinds of particles (electrons $e$, photons $\gamma$, charged pions $\pi$, and neutral pions $\pi^0$) with energies uniformly distributed between 10 and 510 GeV, and with incident angles uniformly distributed between a polar angle $\theta$ between 1.047 and 2.094 radians with respect to the beam direction.

We get the barycenter of a shower by taking the 2D projection of its energy deposit on the ECAL inner surface. Then, knowing the point of origin of the incoming particle, we use the barycenter to estimate the particle's polar and azimuthal angles $\theta$ and $\phi$. The estimated pseudorapidity $\eta$ is then computed as $\eta=-\log[\tan\frac{\theta}{2}]$. Each single-shower event is prepared by taking a slice of the ECAL in a window around the shower barycenter, as well as the corresponding HCAL slice behind. Depending on the task (generation or reconstruction), we take:
\begin{itemize}
  \item {\bf GEN dataset}: A $51 \times 51 \times 25$ cell window in the
    ECAL, for electrons in the energy range $100-200$ GeV. Used in the shower generation task.
  \item {\bf REC dataset}: A $25 \times
    25 \times 25$ cell slice of the electromagnetic calorimeter (ECAL)
    and a corresponding $11 \times 11 \times 60$ cell slice of the
    hadronic calorimeter (HCAL), for $e,~\gamma,~\pi,$~or~$\pi^0$ in the energy range $10-510$ GeV and with $\eta$ from $-0.524-0.524$. Used in the particle reconstruction task.
\end{itemize}

\begin{figure*}[htbp]
\centering
\includegraphics[width=0.45\textwidth]{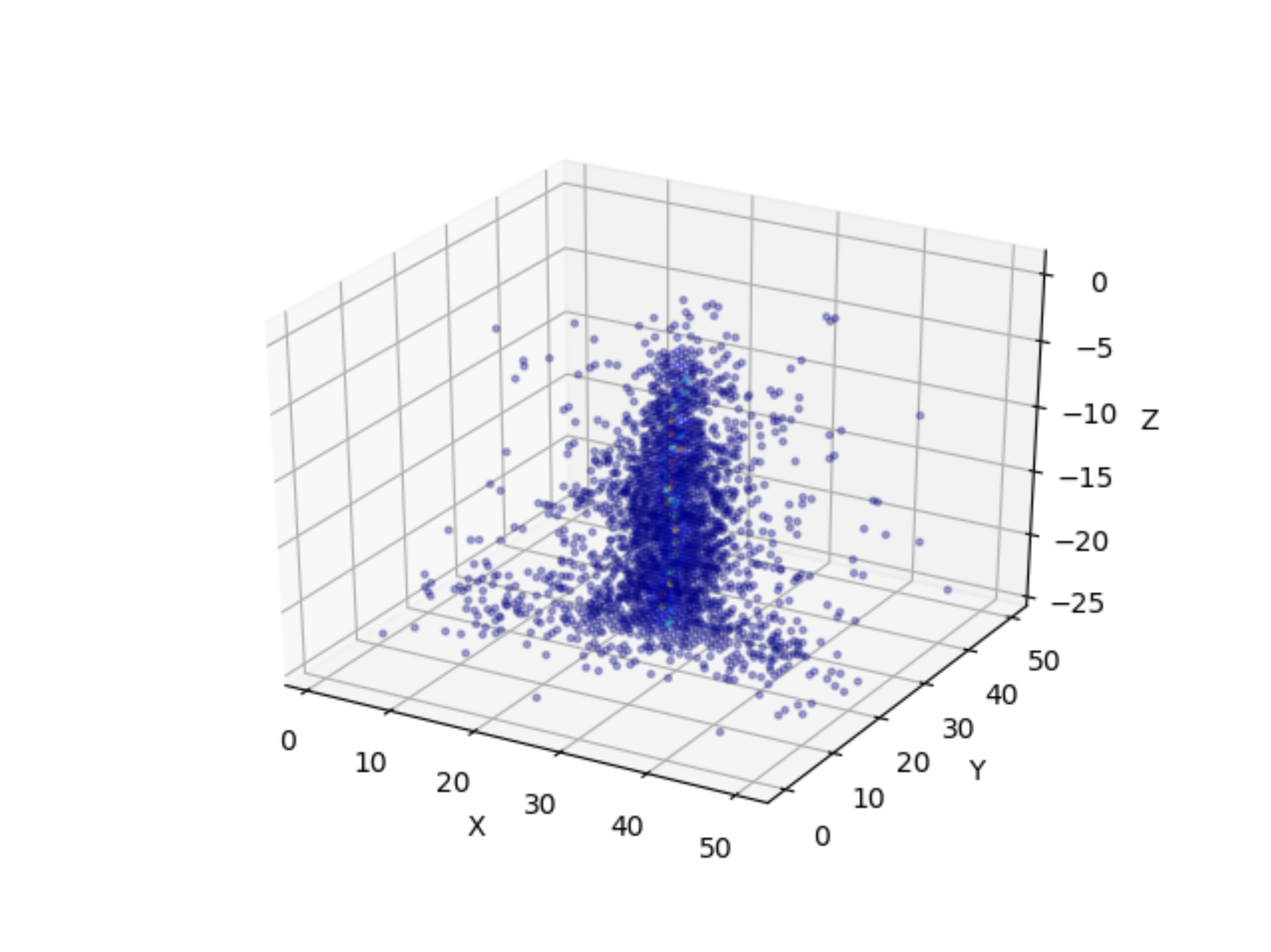}
\includegraphics[width=0.45\textwidth]{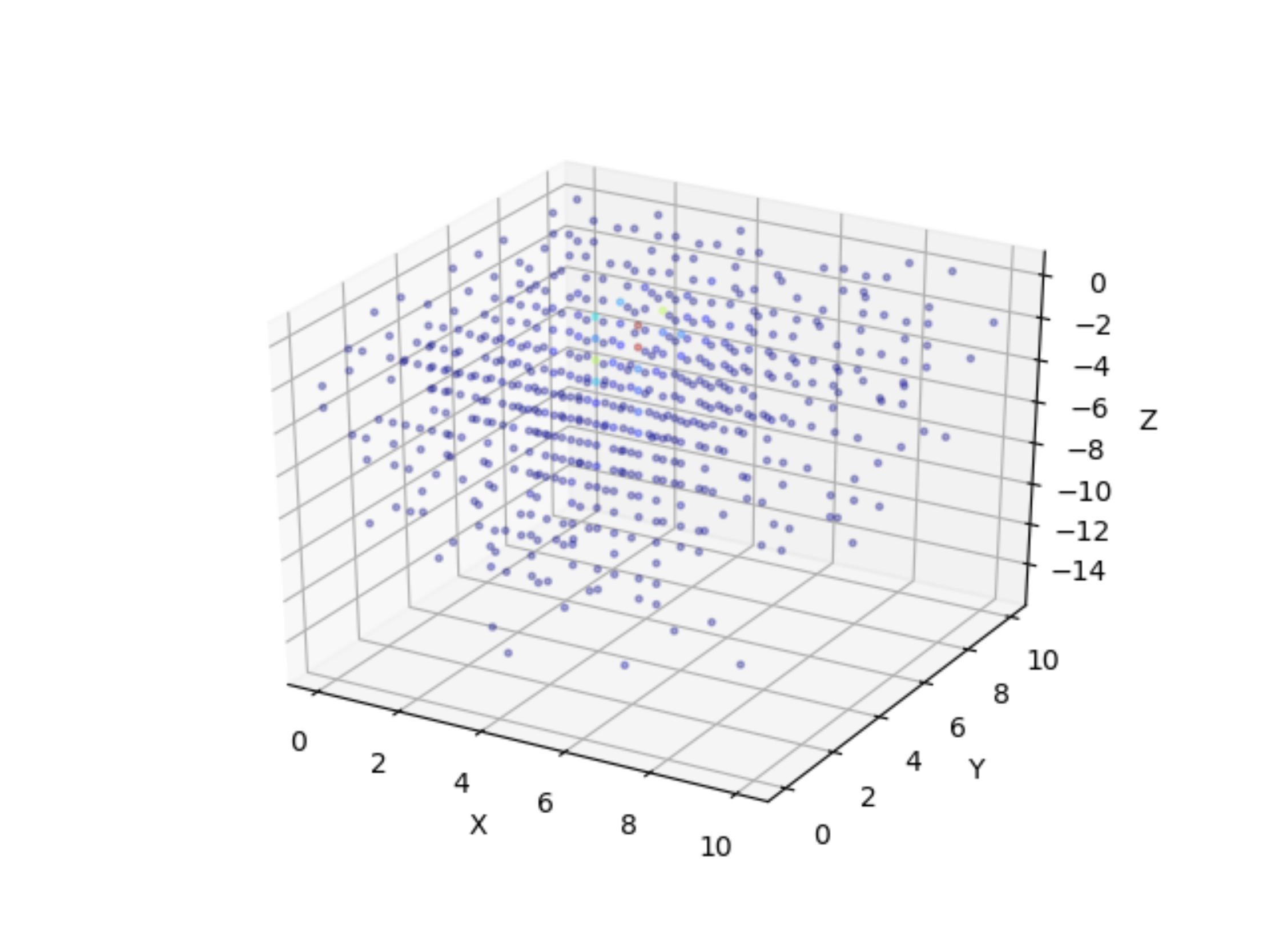} \\
\includegraphics[width=0.45\textwidth]{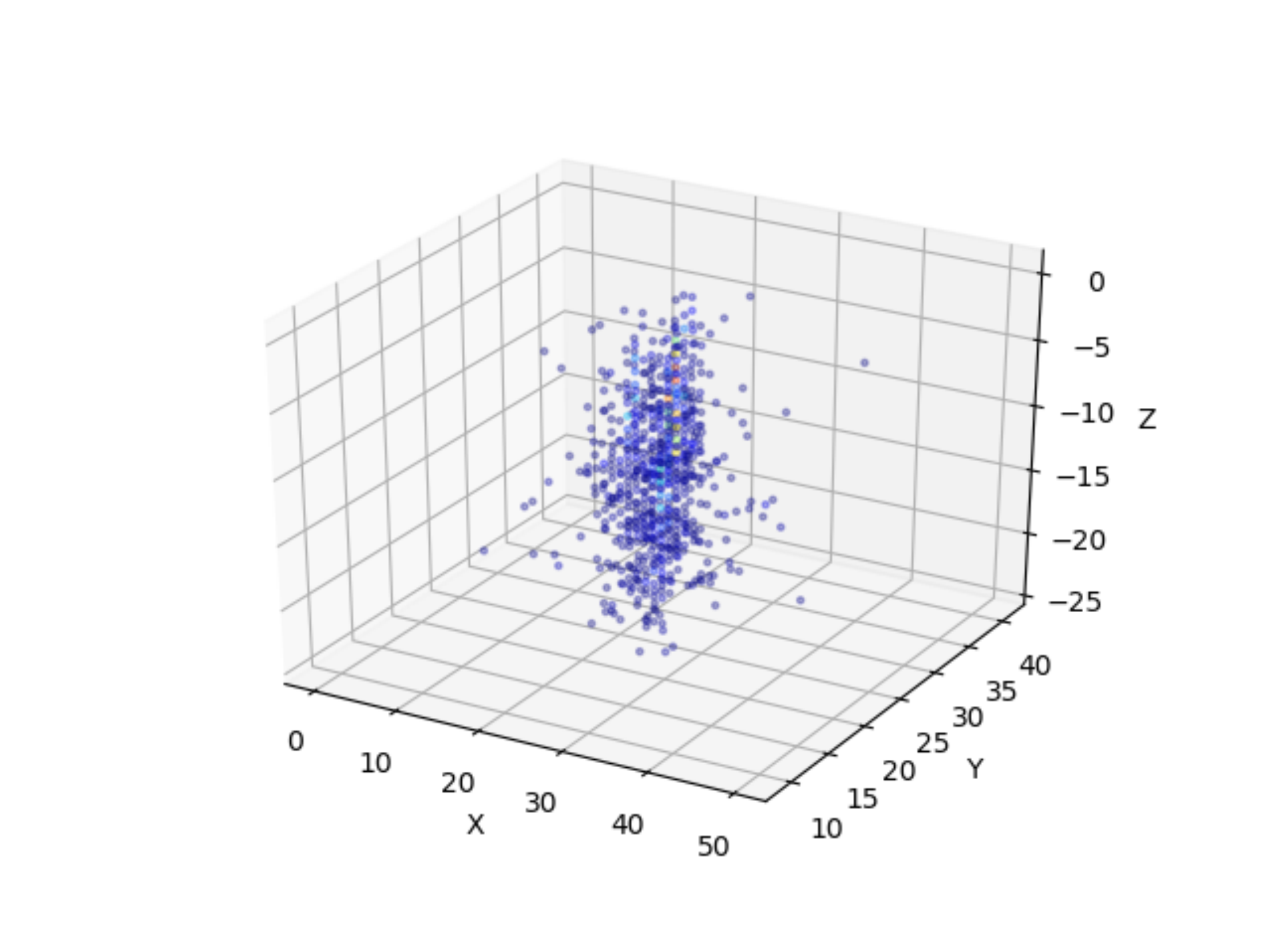}
\includegraphics[width=0.45\textwidth]{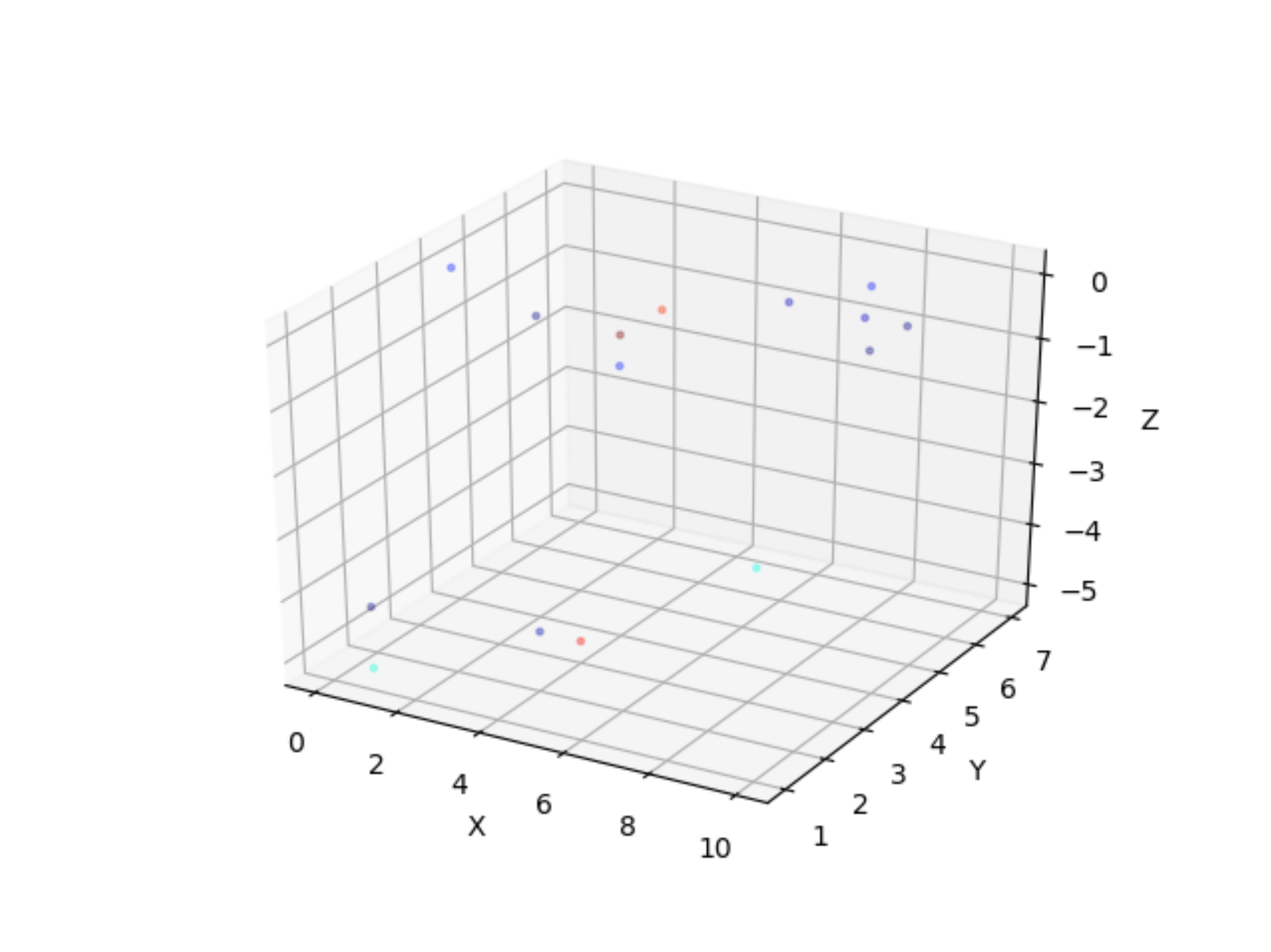}
\caption{3D image of a photon (top) and neutral pion (bottom) shower in ECAL (left) and HCAL (right).}
\label{fig:sample}
\end{figure*}

Examples of an electron shower and a charged-pion shower can be seen in Figure~\ref{fig:sample}. The incoming particles enter from the top ($z=0$), at the center of the $(x,y)$ transverse plane ($x=y=25)$. The electron event has left more hits in both the ECAL and HCAL. We can also see the presence of two subtracks in the neutral pion event. 

The window size for the GEN dataset has been defined in order to contain as much of the shower information as practically possible.  
Motivated by the need of reducing the memory footprint for some of the models, we used a smaller window size for the REC dataset. When training classification models on these data, a negligible accuracy increase was observed when moving to larger windows, as described in Appendix~\ref{app:window_size}.

\begin{figure}[htbp]
\centering
\includegraphics[width=0.3\textwidth]{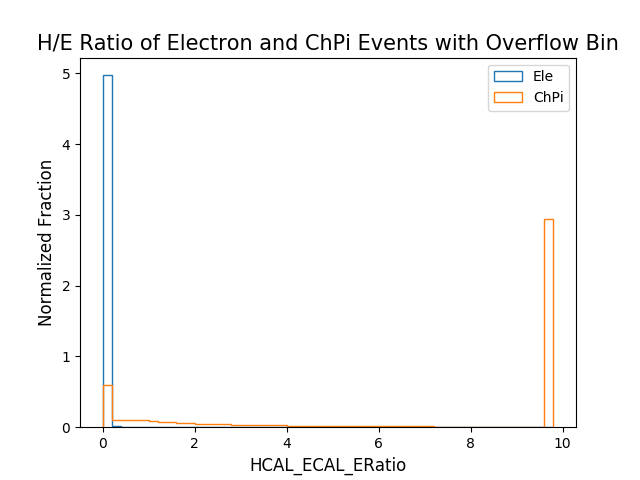}
\includegraphics[width=0.3\textwidth]{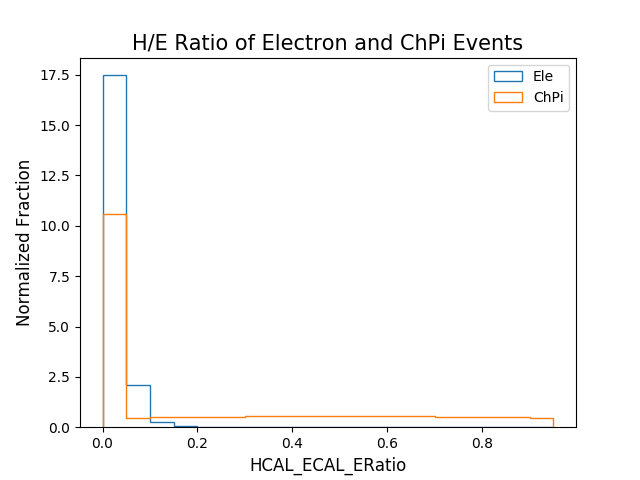}
\caption{HCAL/ECAL energy ratios for electrons and charged pions. The bottom plot is a zoomed-in version of the top plot.}
\label{fig:HE_ratio}
\end{figure}

\begin{figure}[htbp]
\centering
\includegraphics[width=0.3\textwidth]{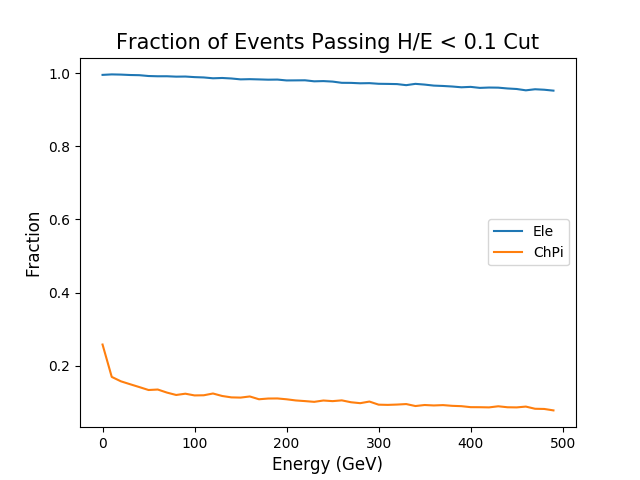}
\includegraphics[width=0.3\textwidth]{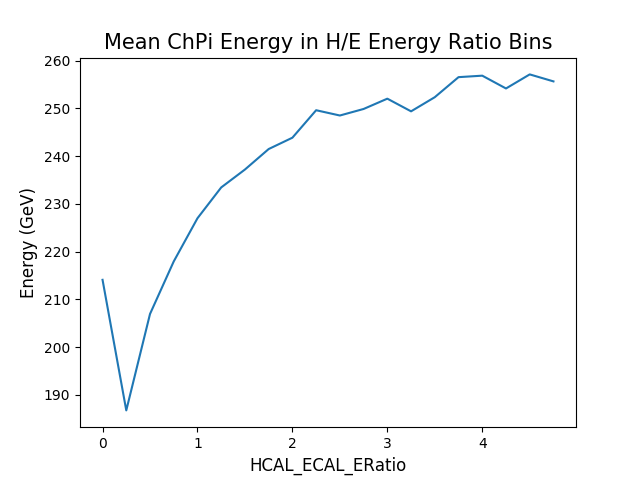}
\caption{Fractions of electrons and charged pions passing a HCAL/ECAL energy selection at various particle energies (top). The mean charged pion energy at each energy ratio (bottom).}
\label{fig:HE_ratio_energy}
\end{figure}

\begin{figure}[htbp]
\centering
\includegraphics[width=0.3\textwidth]{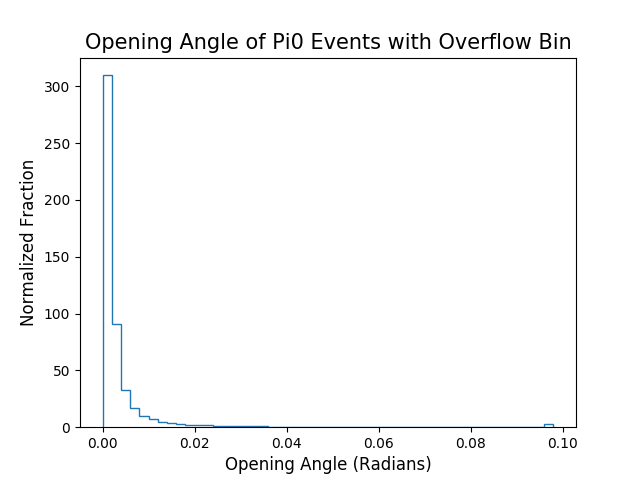}
\caption{Opening angle distribution for neutral pions decaying into two photons.}
\label{fig:opening_angle}
\end{figure}

\begin{figure}[htbp]
\centering
\includegraphics[width=0.3\textwidth]{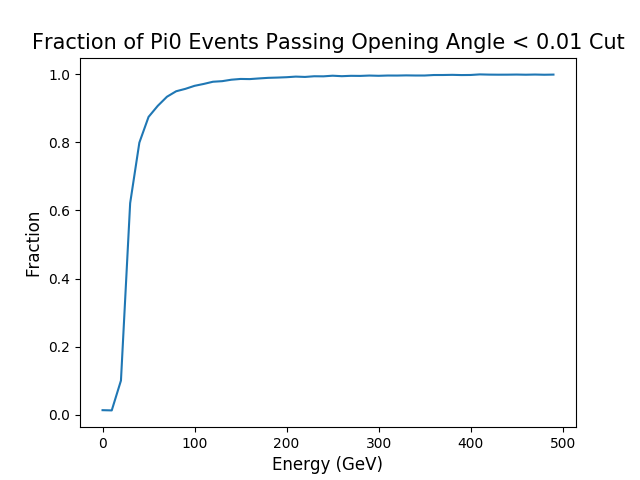}
\includegraphics[width=0.3\textwidth]{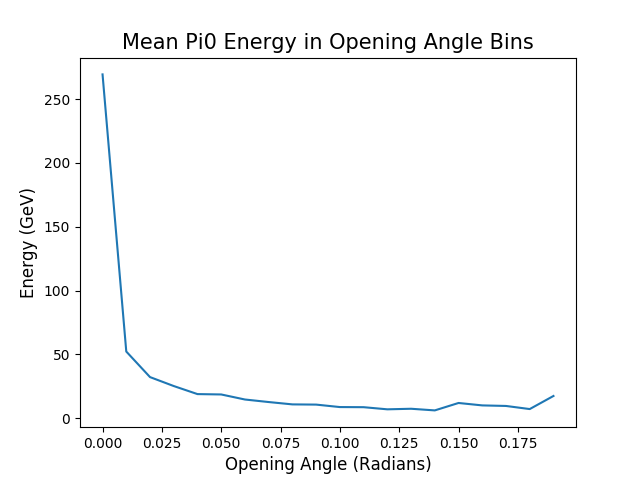}
\caption{The fraction of photons and neutral pions passing an opening angle < 0.01 radian selection at various particle energies (top). The mean neutral pion energy at each opening angle (bottom).}
\label{fig:opening_angle_energy}
\end{figure}

We apply a task-dependent filtering of the REC dataset, in order to select the subset of examples for which the task at hand is not trivial. For instance, distinguishing a generic pion from a generic electron is an easy task, and can be accomplished with high accuracy by looking at the HCAL/ECAL energy ratio. On the other hand, it is difficult to distinguish a generic electron from a pion that has a small HCAL/ECAL energy ratio. Thus, we ignore charged-pion showers with a large HCAL/ECAL energy ratio. To be more specific, we see in Figure~\ref{fig:HE_ratio} that the ratio of total ECAL energy to total HCAL energy is very different for electrons and charged pions, with the heavier charged pions tending to leave little energy in the ECAL. In order to make the particle-identification task more challenging, we only consider showers with HCAL/ECAL < 0.1 cut. The results of this selection are shown in Figure~\ref{fig:HE_ratio_energy}. We can see that this selection favors mostly low-energy charged pions, which tend to leave more of their energy in the ECAL rather than punching through to the HCAL. Discriminating accurately between electrons and charged pions in this range is thus crucial for compressed-mass physics analyses, where we search for decay products with low energy.

Photons and neutral pions are more similar to each other. The easily distinguishable events are mostly due to the fact that neutral pions decay into two photons which are separated by a small angle. If the pion has a low energy, the opening angle between the two photons is larger and the shower is easily identified as originating from a neutral pion. High-energy neutral pions produce more collimated photon pairs, which are more easily mistaken as a single high-energy photon. The opening angle distribution for neutral pions is shown in Figure~\ref{fig:opening_angle}. In order to limit the study to the most challenging case, we filter the neutral-pion dataset by requiring the opening angle between the two photons to be smaller than 0.01 radian.  The effect of  this requirement on the otherwise uniform energy distribution is shown in Figure~\ref{fig:opening_angle_energy}. As expected, the selection mostly removes low-energy neutral pions. 


The ECAL and HCAL 3D arrays are passed directly to our neural networks. We also compute a set of expert features, as described in Ref.~\cite{NIPS}. These features are used to train alternative benchmark algorithms (see Appendices~\ref{app:BDT}~and~\ref{app:regression_baseline}), representing currently-used ML algorithms in HEP.

\section{Benchmark Tasks}
\label{sec:problems}

In this section, we introduce the two benchmark tasks that we aim to solve with ML algorithms: 
\begin{itemize}
    \item Particle reconstruction: starting from raw detector hits, determine the nature of a particle and its momentum.
    \item Particle simulation: starting from a generator-level information of an incoming particle, generate the detector response (raw detector hits) using random numbers to model the stochastic nature of the process.
\end{itemize}

Both tasks represent heavy loads for central computing systems of large-scale high-energy physics experiments. A sizable acceleration of these processes in terms of resource reduction and execution time would generate a resource saving that could be invested in new opportunities. 

This paper extends upon previous ML investigations in ATLAS. Some prior classification studies on ATLAS data can be found at~\cite{Nachman_DNN}, and work involving the generation of electron showers at ATLAS can be found at~\cite{Nachman_GAN} and~\cite{Nachman_GAN2}. Since the ATLAS data was much less granular than the CLIC datasets we examine here, we were able to examine more complex neural architectures. Furthermore, we demonstrate the use of a single tool which performs multiple aspects of particle reconstruction simultaneously, simply starting from a calorimeter image.

\subsection{Simulation}
\label{sec:sim}

It is common in HEP to generate large amounts of detailed synthetic data from Monte Carlo simulations. This simulated data allows physicists to determine the expected outcome of a given experiment based on known physics. Having this prior expectation, one can reveal the presence of new phenomena by observing an otherwise inexplicable difference between real and simulated data. An accurate simulation of a detector response is a computationally heavy task, currently taking a significant fraction of the overall computing resources in a typical HEP analysis. Thus we also investigate the use of ML algorithms to speed up the event simulation process. In particular, we build a generative model to simulate detector showers, similar to those on which we train the end-to-end reconstruction algorithm. Such a generator could drastically reduce Monte Carlo simulation time, and turn event generation into an on-demand task.

In order to create realistic calorimetric shower data, we train a generative adversarial network (GAN) on the GEN dataset defined in Section~\ref{sec:data}. We restrict the study to ECAL showers for incoming electrons with energy between 100 and 200 GeV. The task is to create a model that can take an electron's energy and flight direction as inputs and generate a full ECAL shower, represented as a $51 \times 51 \times 25$ array of energy deposits along the trajectory of the incoming electron. 
The advantage of using a GAN is that it's much faster and less computationally intense than traditional Monte Carlo simulation, and the results may more accurately reproduce physical behavior if the GAN is trained on real data.

\subsection{Reconstruction}

At particle-collider experiments, data consist of sparse sets of hits recorded by various detector components at beam collision points. A typical analysis begins with a complex reconstruction algorithm that processes these raw data to produce a set of physics objects (jets, electrons, muons, etc.), which are then used further down the line. Traditionally, the reconstruction software consists of a set of rule-based algorithms that are designed based on physics knowledge of the specific problem at hand (e.g., the bending of particles in a solenoidal magnetic field, due to the Lorentz force). Over the past decade or so, machine-learning algorithms have been integrated into certain aspects of particle reconstruction (e.g., particle identification). One example is the identification of electrons and photons via a BDT, taking as input for each event a set of high-level features quantifying the shape of the energy cluster deposited in a calorimeter shower~\cite{BDT_thesis}.

Event reconstruction is one of the most CPU-intensive tasks at the LHC. In order to reduce the resource needs, one could imagine using ML techniques to extract the required information directly from the raw data, without first computing high-level features. Following this idea, we investigate here an end-to-end ML model based on computer vision techniques, treating the calorimeter input as a 3D image. Using a combined architecture, the model is designed to simultaneously perform particle identification and energy measurement.

When dealing with particle reconstruction, one is interested in identifying a particle's type (electron, photon, etc.) and its momentum. An end-to-end application aiming to provide a full reconstruction of a given particle should thus be able to simultaneously solve a multi-class classification problem and a regression problem. In our study, we filter the REC dataset to make the classification task non-trivial, as described in Section~\ref{sec:data}. Since differentiating charged and uncharged particles is trivial, we judged the classification of our model on its ability to distinguish electrons from charged pions, and photons from neutral pions.

Our reconstruction networks were thus given the following three tasks:
\begin{itemize}
\item {\bf Identify electrons over a background of charged pions}: Charged pions are the most abundant particles produced in LHC collisions. They are typically arranged into jets, which are collimated sprays resulting from the showering and hadronization processes of quarks and gluons. On the other hand, electrons are rarely produced, and their presence is typically an indication of an interesting event occurring in the collision. A good electron identification should aim at misidentifying at most 1 in 10,000 pions as an electron. In our problem, we consider the background as originating from single pions, which is a case more typical of electron-positron colliders. Initial studies of dense neural network (DNN) based classification of unfiltered events containing the four particles types in our simulated samples yielded extremely good results, with area under curve (AUC) of receiver-operator curves (ROC) near 1, when using equally sized and unbiased samples of each particle class. In order to test ML capabilities with a more challenging problem and to approach the kind of task that one faces at the LHC, we perform this binary classification on the REC dataset of Section~\ref{sec:data}, after applying an HCAL/ECAL energy ratio cut.
\item {\bf Identify photons over a background of neutral pions}: At particle colliders, the main background to photon identification comes from neutral pions decaying to a photon pair. The two photons from the $\pi^0$ decays are produced with an angular distance that tends to zero in the limit of high $\pi^0$ momentum. In general, a generic $\gamma/\pi^0$ classification task is relatively easy, since the presence of two nearby clusters is a clear signature of $\pi^0$. On the other hand, at high $\pi^0$ momentum the energy deposits from the two photons merge into a single cluster, and $\gamma$ and $\pi^0$ become very similar. In this paper, we focus on this case, running the binary classification task on the REC dataset of Section~\ref{sec:data}, considering only $\pi^0$s with an opening angle < 0.01 radian.
\item {\bf Energy measurement}: Once the particle is identified, it is very important to accurately determine its energy (and by extension, its momentum), since this allows physicists to calculate all the relevant high-level features, such as the mass of new particles that generated the detected particles when decaying. In this study, we address this problem on the same dataset used for the classification tasks, restricting the focus to range of energies from 10 to 510 GeV, and at various incident angles ($\eta$). Regression results using various neural network architectures were compared with results from linear regression, comparing both resolution and bias. The models we consider are designed to return the full particle momentum (energy, $\eta$, and $\phi$) of the incoming particle momentum. At this stage, this functionality is not fully exploited and only the energy determination is considered. An extension of our work to include the determination of $\eta$ and $\phi$ could be the matter of future studies.
\end{itemize}
\section{Generative Model}\label{sec:GAN}

Generative Adversarial Networks are composed of two networks, a discriminator and a generator. Our model, 3DGAN, implements an architecture inspired by the auxiliary classifier GAN~\cite{acgan}. The generator takes as input a specific particle type, flight direction, and energy, and generates the 3D image of an energy deposit using an auxiliary input vector of random quantities (latent vector). 
The output has the same format as the 3D array of ECAL hits in the GEN sample (see Section~\ref{sec:data}). The discriminator network receives as input an ECAL 3D array and classifies it as {\it real} (coming from the GEANT4-generated GEN dataset) or {\it fake} (produced by the generator).

 Our initial 3DGAN prototype~\cite{NIPS} successfully simulated detector outputs for electrons which were orthogonally incident to the calorimeter surface. In addition, the discriminator performed an auxiliary regression task on the input particle energy. This task was used to cross check the quality of the generation process. 
 
 In this study, we consider a more complex dataset, e.g., due to the variable incident angle of the incoming electron on the inner ECAL surface. To monitor this additional complexity, we add additional components to the loss function, related to the regression of the particle direction and the pixel intensity distribution (energy deposition in cells). This will be described in more detail below.

Before training our GAN, we pre-processed the GEN dataset by replacing each cell energy content $E$ with $E^\alpha$, where $\alpha<1$ is a fixed hyperparameter. This pre-processing compensates for the large energy range (about 7 orders of magnitude) covered by individual cell energies, and mitigates some performance degradation we previously observed at low energies. After testing for different values of $\alpha$, we observed optimal performance for $\alpha=0.85$.

\begin{figure*}[htbp]
\centering
    \includegraphics[scale=0.65, trim={0cm 6cm 3.5cm 1.8cm}]{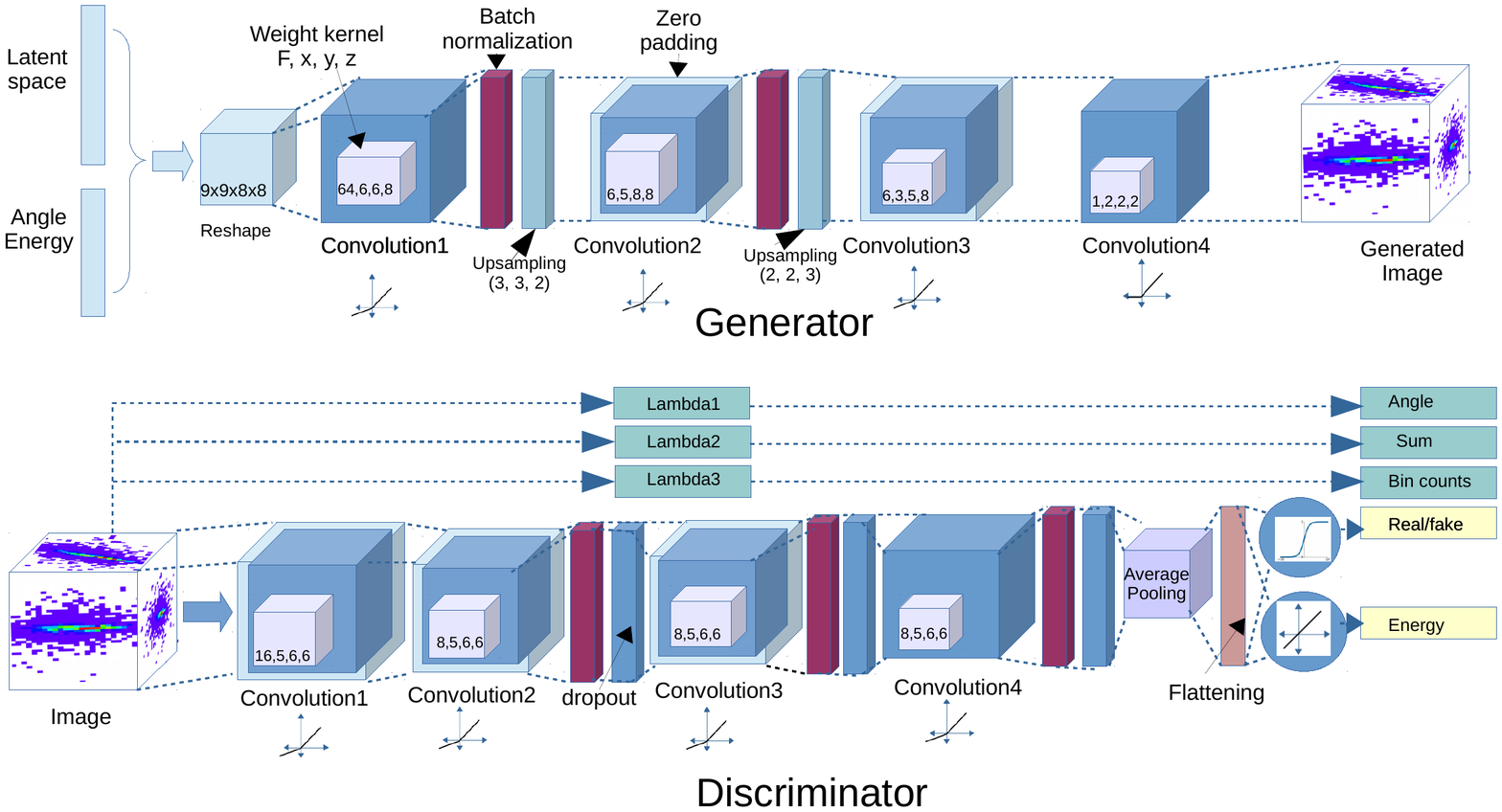}
    \caption{3DGAN generator and discriminator network architectures}
    \label{fig:GAN_arch}
\end{figure*}

\subsection{GAN Architecture}
\label{sec:GANarch}

The 3DGAN architecture is based on 3-dimensional convolutional layers, as shown in Figure~\ref{fig:GAN_arch}. The generator takes as input a vector with a desired particle energy and angle, and concatenates a latent vector of 254 normally distributed random numbers. This goes through a set of alternating upsampling and convolutional layers. The first convolution layer has $64$ filters with $6 \times 6 \times 8$ kernels. The next two convolutional layers have $6$ filters of $5 \times 8 \times 8$ and $3 \times 5 \times 8$ kernels, respectively. The last convolutional layer has a single filter with a $2 \times 2 \times 2$ kernel. The first three layers are activated by leaky ReLU functions, while ReLU functions are used for the last layer. Batch normalization and upscaling layers were added after the first and second convolutional layers.

The discriminator takes as input a $51  \times 51  \times 25$ array and consists of four 3D convolutional layers. The first layer has $16$ filters with $5 \times 6 \times 6$ kernels. The second, third, and fourth convolutional layers each have $8$ filters with $5 \times 6 \times 6$ kernels. There are leaky ReLU activation functions in each convolutional layer. Batch normalization and dropout layers are added after the second, third, and fourth convolutional layers. The output of the final convolution layer is flattened and connected to two output nodes: a classification node, activated by a sigmoid and returning the probability of a given input to be true or fake; and a regression node, activated by a linear function and returning the input particle energy.
The 3DGAN model is implemented in KERAS~\cite{keras} and Tensorflow~\cite{tensorflow2015-whitepaper}. 

\subsection{Training and Results}
The 3DGAN loss function
\begin{equation}
   L_{Tot}  = W_{G}L_{G} + W_{P}L_{P} + W_{A}L_{A} + W_{E}L_{E} + W_{B}L_{B} 
\label{eq:loss}
\end{equation}
is built as a weighted sum of several terms: a binary cross entropy ($L_{G}$) function of the real/fake probability returned by the discriminator, mean absolute percentage error terms (MAPE) related to the regression of the primary-particle energy ($L_{P}$) , the total deposited energy ($L_{E}$) and the binned pixel intensity distribution ($L_{B}$), and a mean absolute error (MAE) for the incident angles measurement ($L_{A}$). The binary cross entropy term, percentage errors and absolute error are weighted by $3.0$, $0.1$ and $25$ respectively. The weights $W$ are tuned to balance the relative importance of each contribution. The predicted energy and incident angle provide a feedback on the conditioning of the image. The binned pixel intensity distribution loss compares the counts in different bins of pixel intensities. 

The model training is done using the RMSprop \cite{rmsProp} optimiser. We alternately train the discriminator on a batch of real images and a batch of generated images, applying label switching. We then train the generator while freezing the discriminator weights.

\begin{figure}[htbp]
    \includegraphics[width=0.48\textwidth]{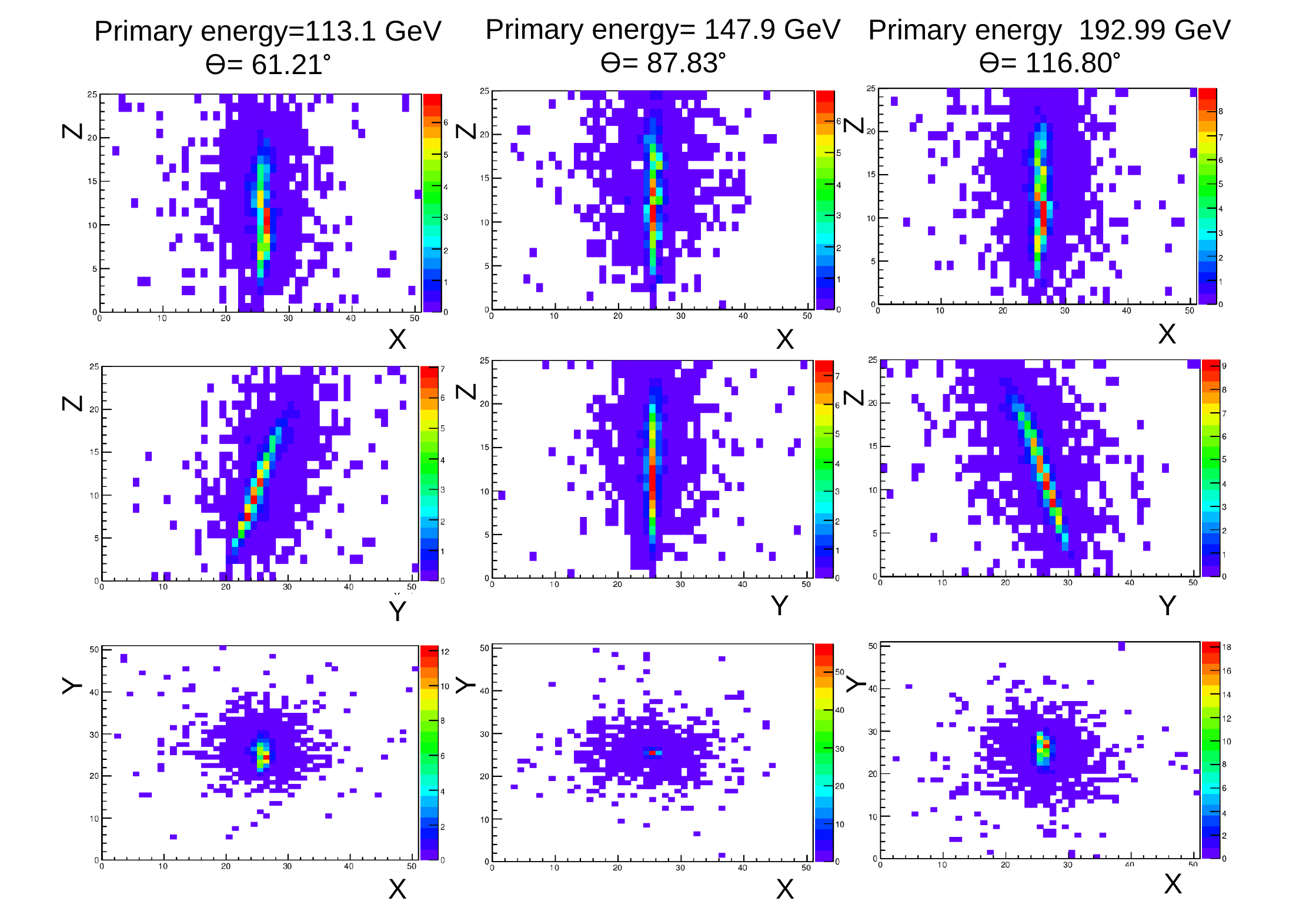}
    \caption{GEN sample: electrons with different primary particle energies and angles.}
    \label{fig:GEANT4_events}
\end{figure}

\begin{figure}[htbp]
    \includegraphics[width=0.48\textwidth]{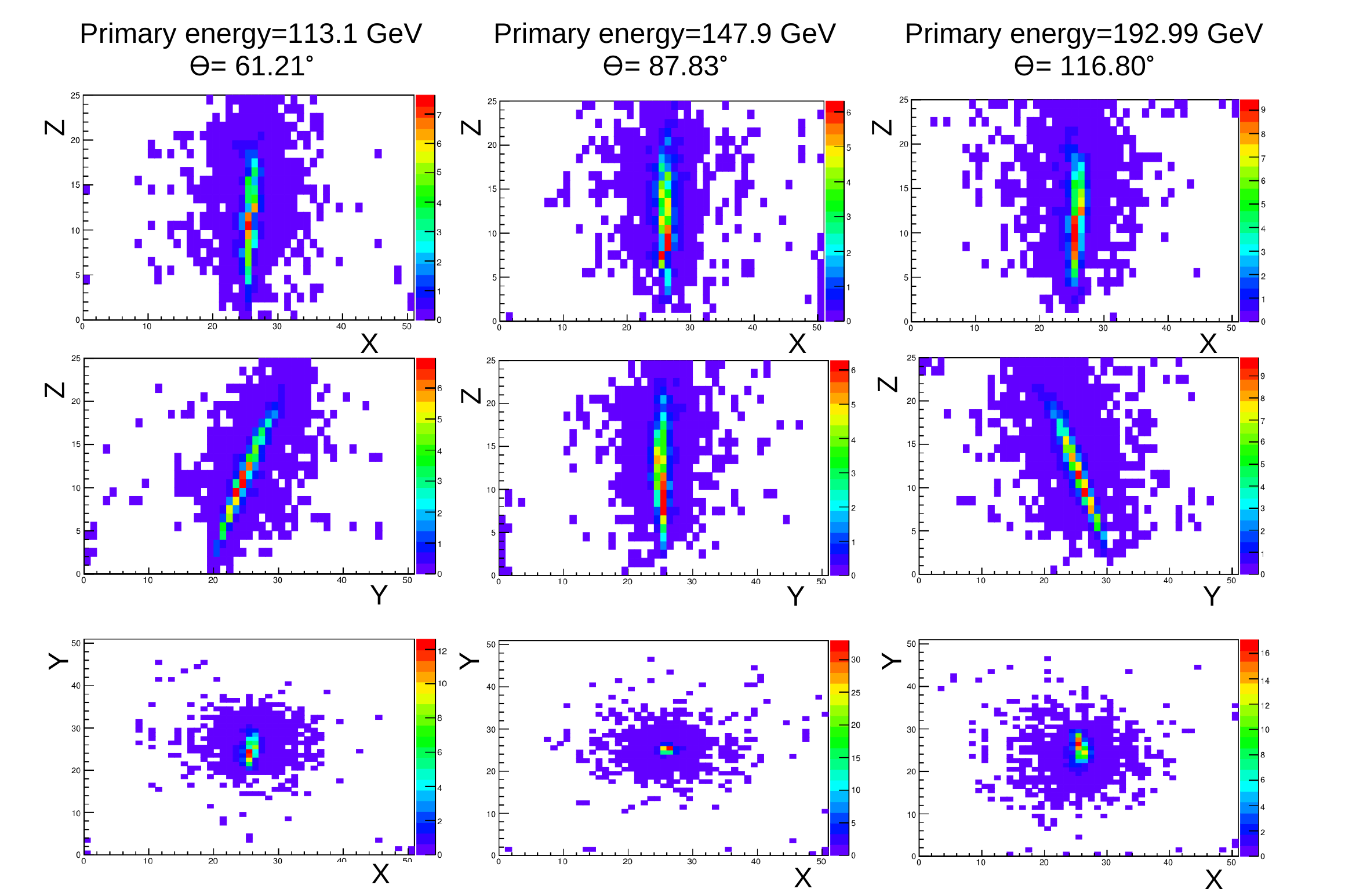}
    \caption{GAN generated electrons with primary energies and angles corresponding to the electrons showed in Figure~\ref{fig:GEANT4_events}.}
    \label{fig:GAN_events}
\end{figure}

Figure~\ref{fig:GEANT4_events} shows a few events from the GEN data set. The events were selected to cover both ends of the primary-particle energy and angle spectrum. Figure~\ref{fig:GAN_events} presents the corresponding generated events with the same primary particle energy and angle as the GEN events in Figure~\ref{fig:GEANT4_events}. Initial visual inspection shows no obvious difference between the original and GAN generated images. A detailed validation based on several energy-shape related features confirms these results. We discuss a few examples below.

\begin{figure}
    \centering
    \includegraphics[width=0.48\textwidth]{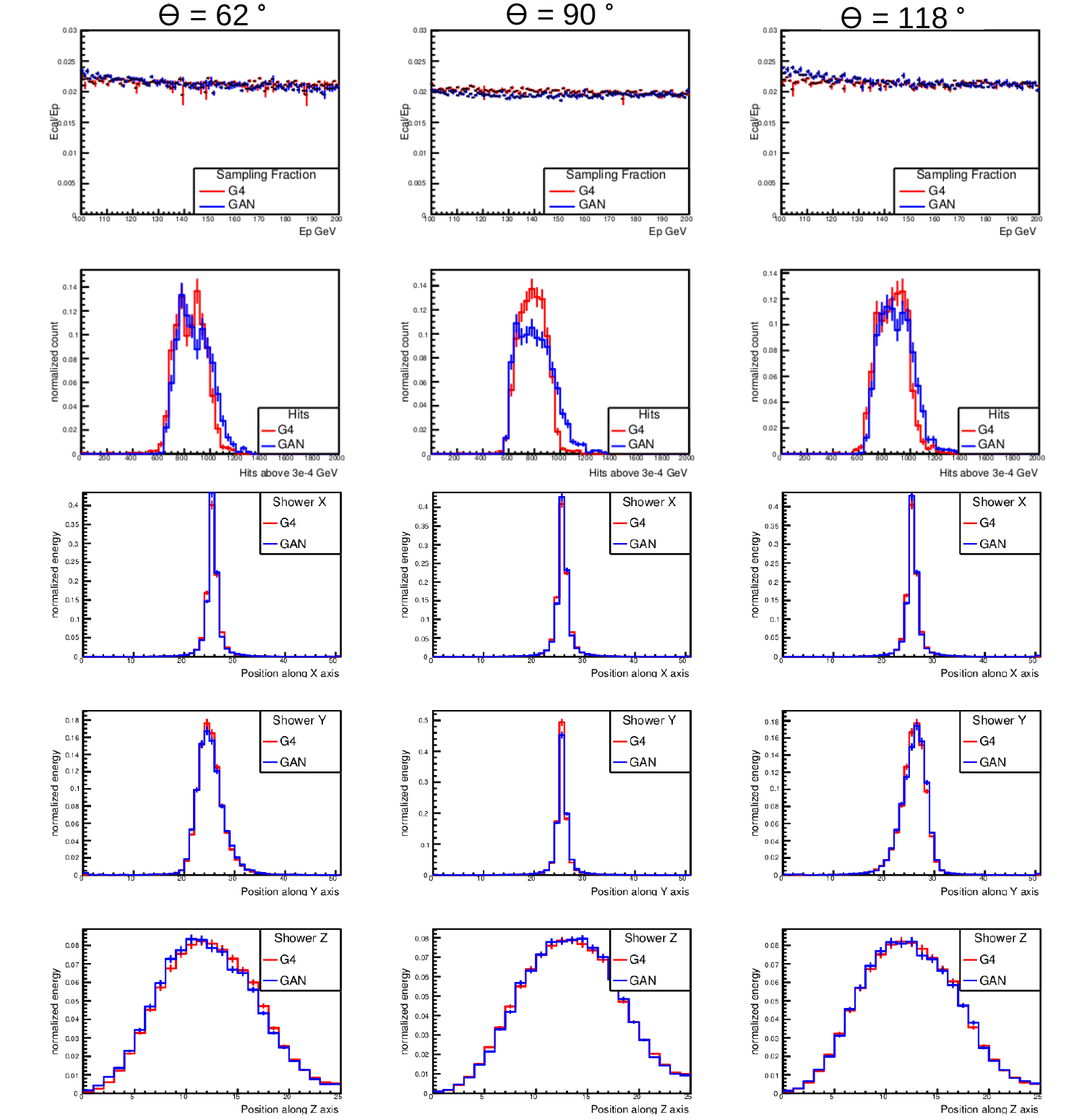}
    \caption{GEANT4 vs. GAN comparison for sampling fraction, number
      of hits and shower shapes along x,y,z axis for different angle
      bins with 100-200 GeV primary particle energies.
      \label{fig:GAN_features1}}
\end{figure}

\begin{figure}
    \centering
    \includegraphics[width=0.48\textwidth]{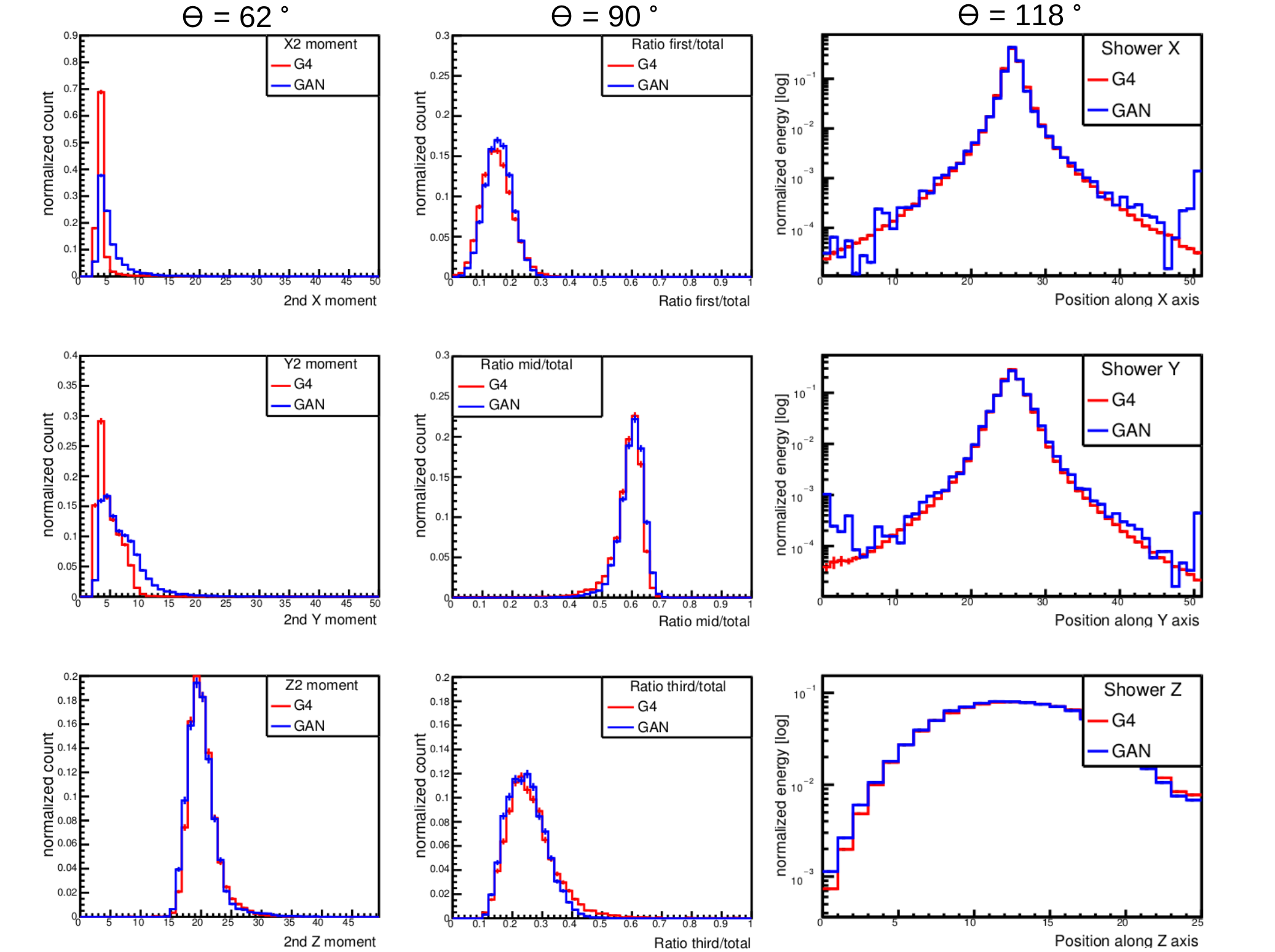}
    \caption{GEANT4 vs. GAN comparison for shower width (second
      moment) in x,y,z, ratio of energy deposited in parts along
      direction of particle traversal to total energy and shower
      shapes along x,y,z axis in log scale for 100-200 GeV primary
      particle energies and 60-120 degrees $\theta$.
      \label{fig:GAN_features2}}
\end{figure}

The top row in Figure~\ref{fig:GAN_features1} shows the ratio between the total energy deposited in the calorimeter and the primary particle energy as a function of the primary particle energy (we refer to it as "sampling fraction") for different angle values. 3DGAN can nicely reproduce the expected behaviour over the whole energy spectrum. The second row in Figure~\ref{fig:GAN_features1} shows the number of hits above a $3 \times 10^{-4}$ MeV threshold. Figure~\ref{fig:GAN_features1} also shows the x, y, z "energy shapes", i.e. the amount of energy deposited along different axes (x and y on the transverse plane and z along the calorimeter depth). The agreement is very good, and in particular 3DGAN is able to mimic the way the energy distributions changes with incident angle. 
Figure~\ref{fig:GAN_features2} shows some additional features aimed at defining the shape of the deposited energy distribution. In particular the second moments along the x,y and z axes are shown on the first column, measuring the width of the deposited energy distribution along those axes. The second column shows the way the energy is deposited along the depth of the calorimeter, by splitting the calorimeter in three parts along the longitudinal direction and measuring the ratios between the energy deposited in each third  and the total deposited energy. Finally, the third column in Figure~\ref{fig:GAN_features2} highlights the tails of the "energy shapes". It can be seen that, while the core of the distribution is perfectly described by 3DGAN, the network tends to overestimate the amount of energy deposited at the edges of the volume. It should be noted however that energy depositions in those cells are very sparse. 

The 3DGAN training runs in around $1.5$ hours per epoch on a single NVIDIA GeForce GTX 1080 card for $60$ epochs. The simulation  time  on a Intel  Xeon 8180  is about $13$ ms/particle  and it goes down to about $4$ ms/particle on a NVIDIA  GeForce  GTX  1080. For  comparison  GEANT4  simulation takes  about $17$ seconds  per  particle on  a  Intel  Xeon  8180 (currently  it  is  not  possible  to  run a full  GEANT4-based  simulation  on  GPUs). Thus our GAN represents a potential simulation speedup of over 4,000 times for this specific aspect of the event simulation.

When given as input to a particle regression and reconstruction model (see section~\ref{sec:reco}), this dataset produces the same output as the original GEANT4 sample, as described in Appendix~\ref{appendix:RECO_on_GAN}.

\section{End-to-End Particle Reconstruction}
\label{sec:reco}

This section describes the use of a deep neural network to accomplish an end-to-end particle reconstruction task. The model consists of a neural architecture which simultaneously performs both particle classification and energy regression. This combined network is trained using the ECAL and HCAL cell arrays as well as the total ECAL energy and total HCAL energy as inputs. The training loss function is written as the sum of a binary cross entropy for particle identification and a mean-square error loss for energy regression. Through experimentation, we found that multiplying the energy loss by a factor of 200 gave the best results, as it was easier to quickly achieve low loss values for energy regression.

We compare three different architectures for our reconstruction model, each trained using calorimeter cell-level information as inputs:
\begin{itemize}
\item A dense (i.e, fully connected) neural network (DNN).
\item A 3D convolutional network (CNN).
\item A network based on GoogLeNet (GN)~\cite{GoogLeNet}, using layers of inception modules.
\end{itemize}

In order to compare the model performance to a typical state-of-the-art particle reconstruction algorithm, we also consider the following alternatives:
\begin{itemize}
    \item A feature-based BDT (see Appendix~\ref{app:BDT}) for the classification task.
    \item A linear regression for the regression task.
    \item A BDT for the regression task (for more info on regression baselines see Appendix~\ref{app:regression_baseline}).
\end{itemize}

In a previous study~\cite{NIPS}, we compared the classification accuracy obtained with a neural model taking as input the energy cells, a feature-based neural models, and a feature-based BDTs. In that context, we demonstrated that feature-based BDTs and neural networks perform equally well, and are both equally capable of correctly classify particles from a small set of calculated features. 
We do not compare feature-based neural networks in this paper, and use feature-based BDTs to represent the current state-of-the-art classification algorithms.

\subsection{Deep Network Models}

The three ML models take as input the ECAL and HCAL 3D energy arrays of the REC dataset (see Section~\ref{sec:data}), together with the total energies recorded in ECAL and in HCAL (i.e., the sum of the values stored in the 3D arrays), as well as the estimated $\phi$ and $\eta$ angles of the incoming particle, calculated using the collision origin and the barycenter of the event. The architecture of each model is defined with a number of floating parameters (e.g. number of hidden layers), which are refined through a hyperparameter optimization, as described in Section~\ref{sec:hpscan}. Each model returns three numbers. After applying a softmax activation, two of these elements are interpreted as the classification probabilities of the current two-class problem. The third output is interpreted as the energy of the particle.

Here we describe in detail the three model architectures:

\begin{itemize}
    \item In the DNN model we first flatten our ECAL and HCAL inputs into 1D arrays. We then concatenate these array along with the total ECAL energy, total HCAL energy, estimated $\phi$, and estimated $\eta$, for an array of total size $25 \times 25 \times 25 + 11 \times 11 \times 60 + 4 = 22889$ inputs. This array is fed as input to the first layer of the DNN, followed by a number of hidden layers each followed by a ReLU activation function and a dropout layer. The number of neurons per hidden layer and the dropout probability are identical for each relevant layer. The number of hidden layers, number of hidden neurons per layer, and dropout rate are hyperparameters, tuned as described in the next session.  Finally, we take the output from the last dropout layer, append the total energies and estimated angles again, and feed the concatenated array into a final hidden layer, which results in a three-element output. 
    \item The CNN architecture consists of one 3D convolutional layer for each of the ECAL and HCAL inputs, each followed by a ReLU activation function and a max pooling layer of kernel size $2 \times 2 \times 2$. The number of filters and the kernel size in the ECAL convolutional layer are treated as optimized hyperparameter (see next session). The HCAL layer is fixed at 3 filters with a kernel size of $2 \times 2 \times 6$. The two outputs are then flattened and concatenated along with the total ECAL and HCAL energies, as well as the estimated $\phi$ and $\eta$ coordinates of the incoming particle. The resulting 1D array is passed to a sequence of dense layers each followed by a ReLU activation function and dropout layer, as in the DNN model. The number of hidden layers and the number of neurons on each layer are considered as hyperparameters to be optimized. The output layer consists of three numbers, as for the DNN model. We found that adding additional convolutional layers to this model beyond the first had little impact on performance. This may be because a single layer is already able to capture important information about localized shower structure, and reduces the dimensionality of the event enough where a densely connected net is able to do the rest.
    \item The third model uses elements of the GoogLeNet~\cite{GoogLeNet} architecture. This network processes the ECAL input array with a 3D convolutional layer with 192 filters, a kernel size of 3 in all directions, and a stride size of 1. The result is batch-normalized and sent through a ReLU activation function. This is followed by a series of inception and MaxPool layers of various sizes, with the full architecture described in Appendix~\ref{app:GoogLeNet}. The output of this sequence is concatenated to the total ECAL energy, the total HCAL energy, the estimated $\phi$ and $\eta$ coordinates, and passed to a series of dense layers like in the DNN architecture, to return the final three outputs. The number of neurons in the final dense hidden layer is the only architecture-related hyperparameter for the GN model. Due to practical limitations imposed by memory constraints, this model does not take the HCAL 3D array as input. This limitation has a small impact on the model performance, since the ECAL array carries the majority of the relevant information for the problems at hand (see Appendix~\ref{app:classification_HCAL}).
\end{itemize}

On all models, the regression task is facilitated by using skip connections to directly append the input total ECAL and HCAL energies to the last layer. The impact of this architecture choice on regression performance is described in Appendix~\ref{app:skip_connections}. In addition to using total energies, we also tested the possibility of using 2D projections of the input energy arrays, summing along the $z$ dimension (detector depth). This choice resulted in worse performance (see Appendix~\ref{app:z_sum_regression}) and was discarded.

\subsection{Hyperparameter Scans}
\label{sec:hpscan}

In order to determine the best architectures for the end-to-end reconstruction models, we scanned over a hyperparameter space for each architecture. Learning rate and decay rate were additional hyperparameters for each architecture. For simplicity, we used classification accuracy for the $\gamma$ vs. $\pi^0$ problem as a metric to determine the overall best hyperparameter set for each architecture. This is because a model optimized for this task was found to generate good results for the other three tasks as well, and because $\gamma$ vs. $\pi^0$ classification was found to be the most difficult problem.

Each hyperparameter point was scanned ten times, in order to obtain an estimate of the uncertainty associated with each quoted performance value. For each scan point, the DNN and CNN architectures trained on 400,000 events, using another sample of 400,000 events for testing. DNN and CNN scan points trained for three epochs each, taking about seven hours each. GN trained on 100,000 events and tested on another 100,000. Due to a higher training time, each GN scan point only trained for a single epoch, taking about twenty hours.

For CNN and DNN training, we used batches of 1,000 events when training. However, due to GPU memory limitations, we could not do the same with GN. Instead, we split each batch into 100 minibatches of ten events each. A single minibatch was loaded on the GPU at a time, and gradients were added up after back-propagation. Only after the entire batch was calculated did we update network weights using the combined gradients.

The best settings were found to be as follows:
\begin{itemize}
    \item For DNN, 4 hidden layers, 512 neurons per hidden layer, a learning rate of 0.0002, decay rate of 0, and a dropout probability of 0.04.
    \item For CNN, 4 hidden layers and 512 neurons per hidden layer, a learning rate of 0.0004, decay rate of 0, a dropout probability of 0.12, 6 ECAL filters with a kernel size of $6 \times 6 \times 6$.
    \item For GN, 1024 neurons in the hidden layer, 0.0001 learning rate, and 0.01 decay rate. 
\end{itemize}

\begin{figure*}[htbp]
\centering
\includegraphics[width=0.45\textwidth]{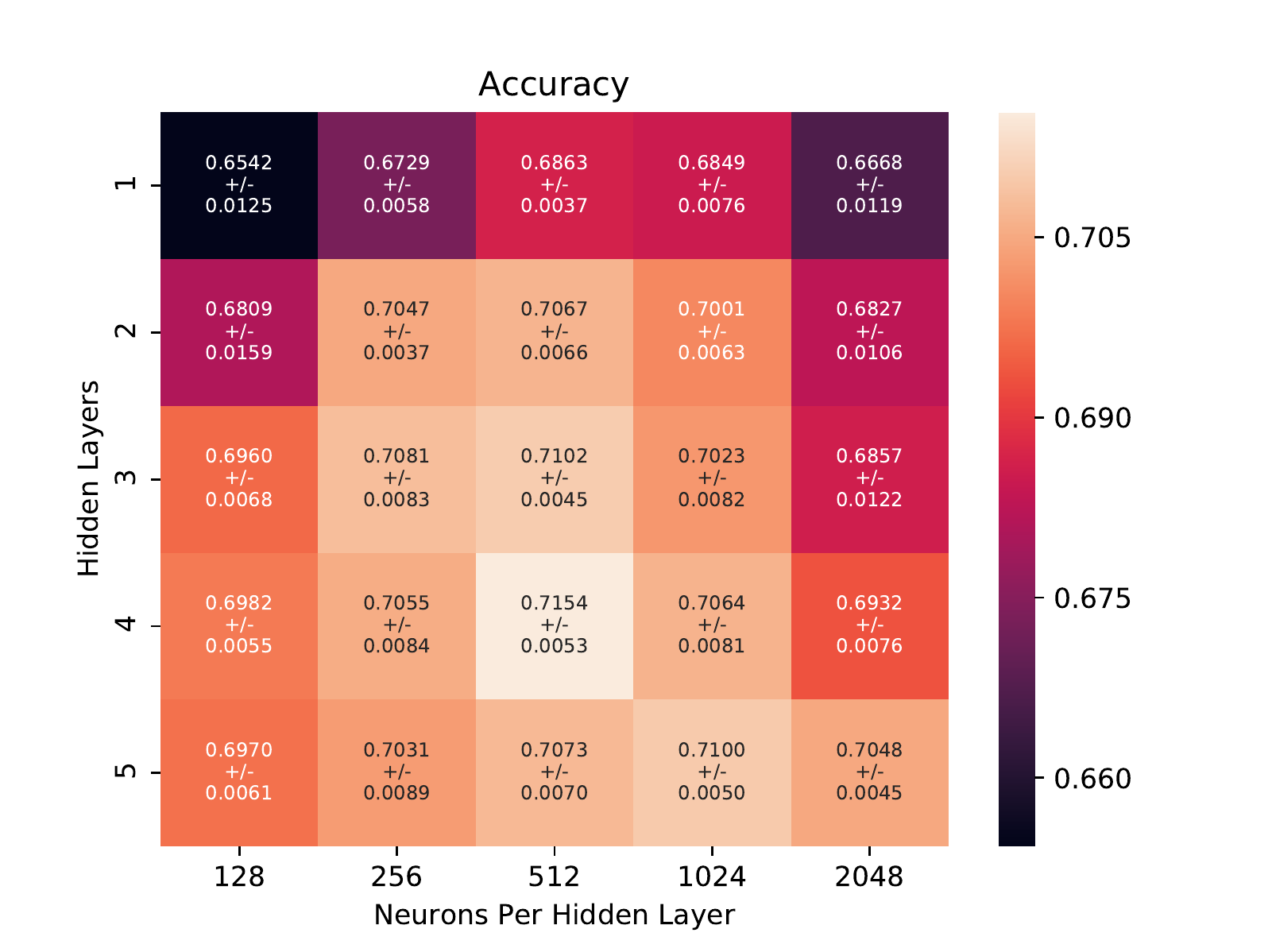}
\includegraphics[width=0.45\textwidth]{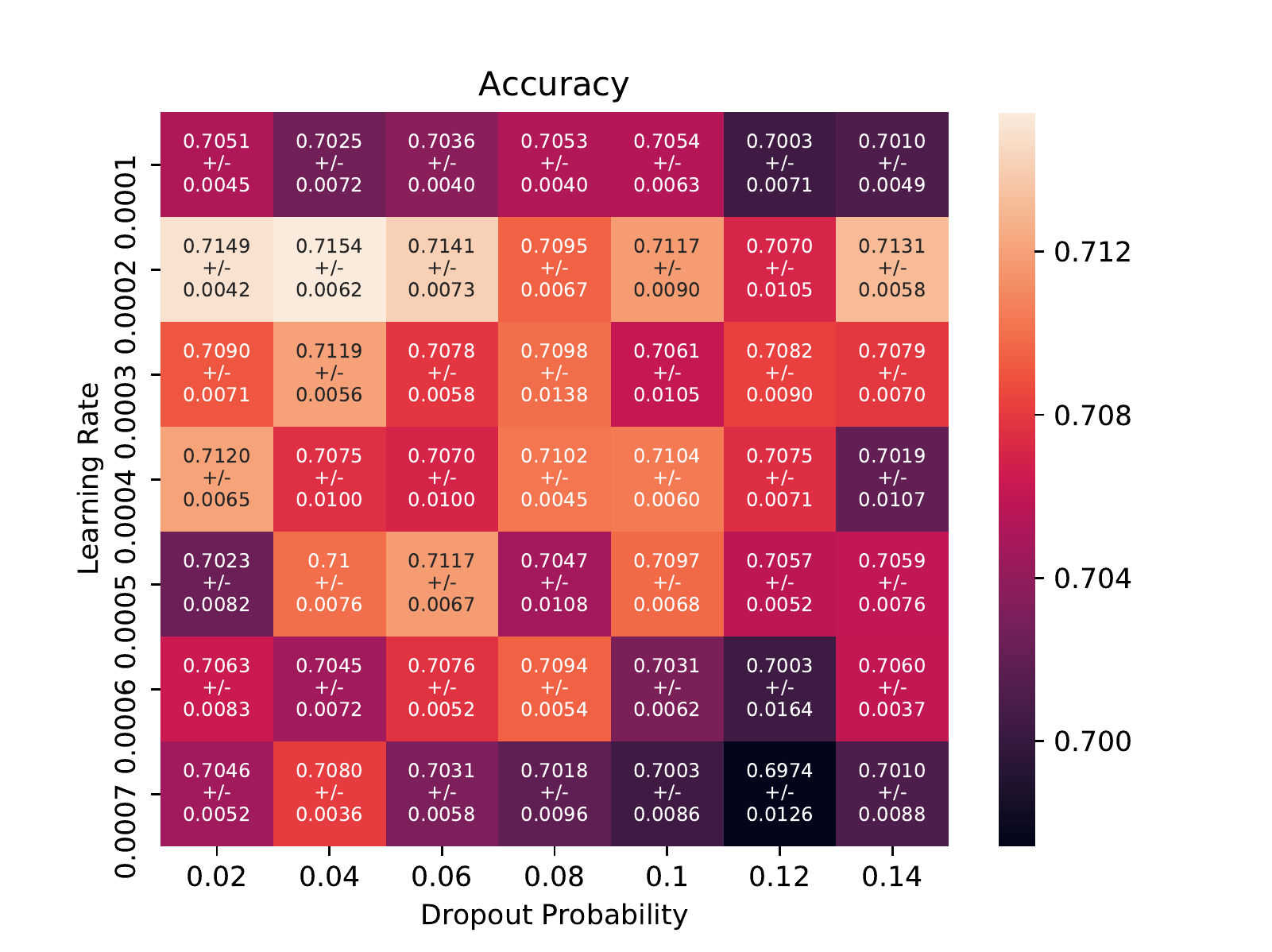} \\
\includegraphics[width=0.45\textwidth]{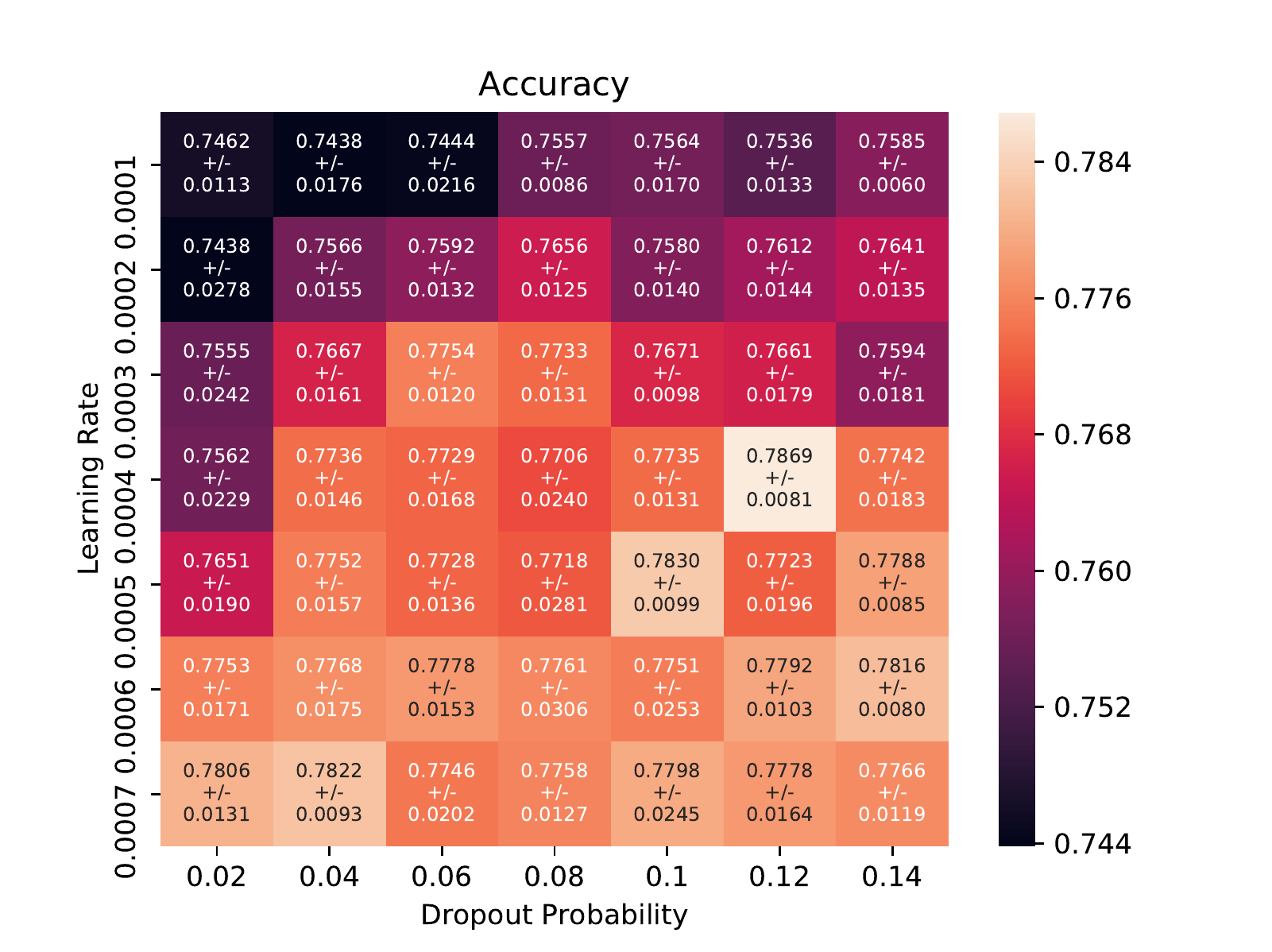}
\includegraphics[width=0.45\textwidth]{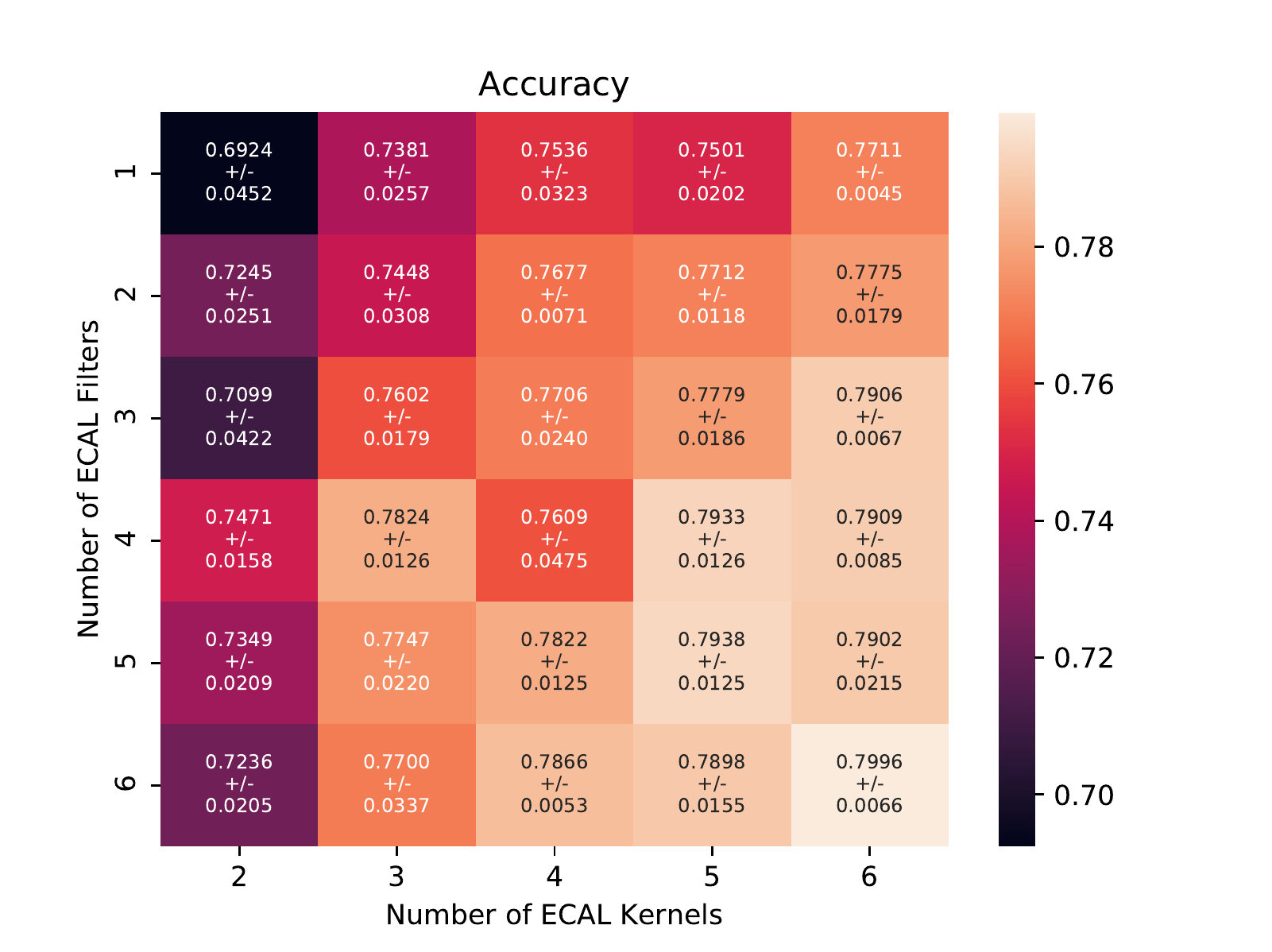} \\
\includegraphics[width=0.45\textwidth]{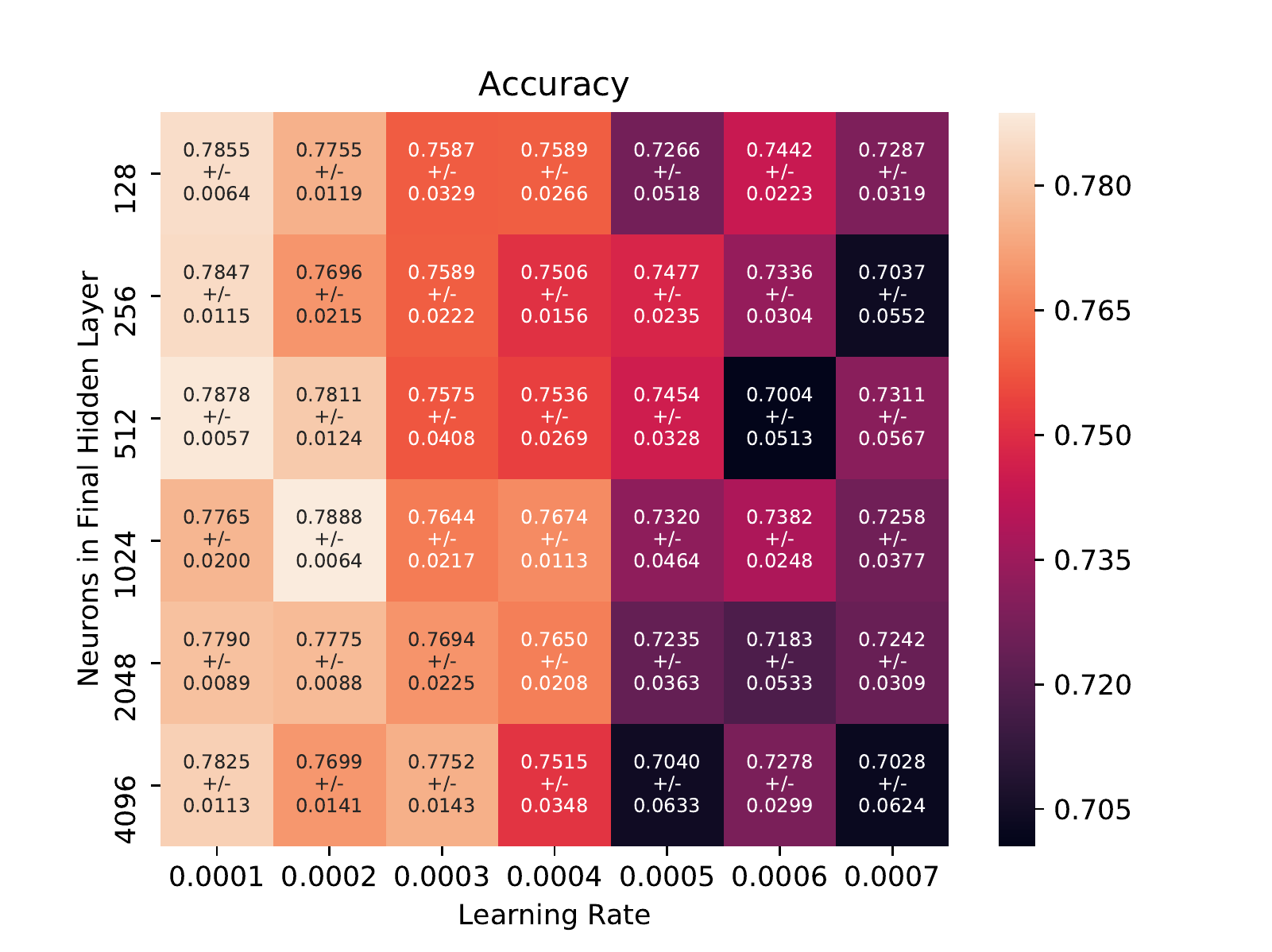}
\includegraphics[width=0.45\textwidth]{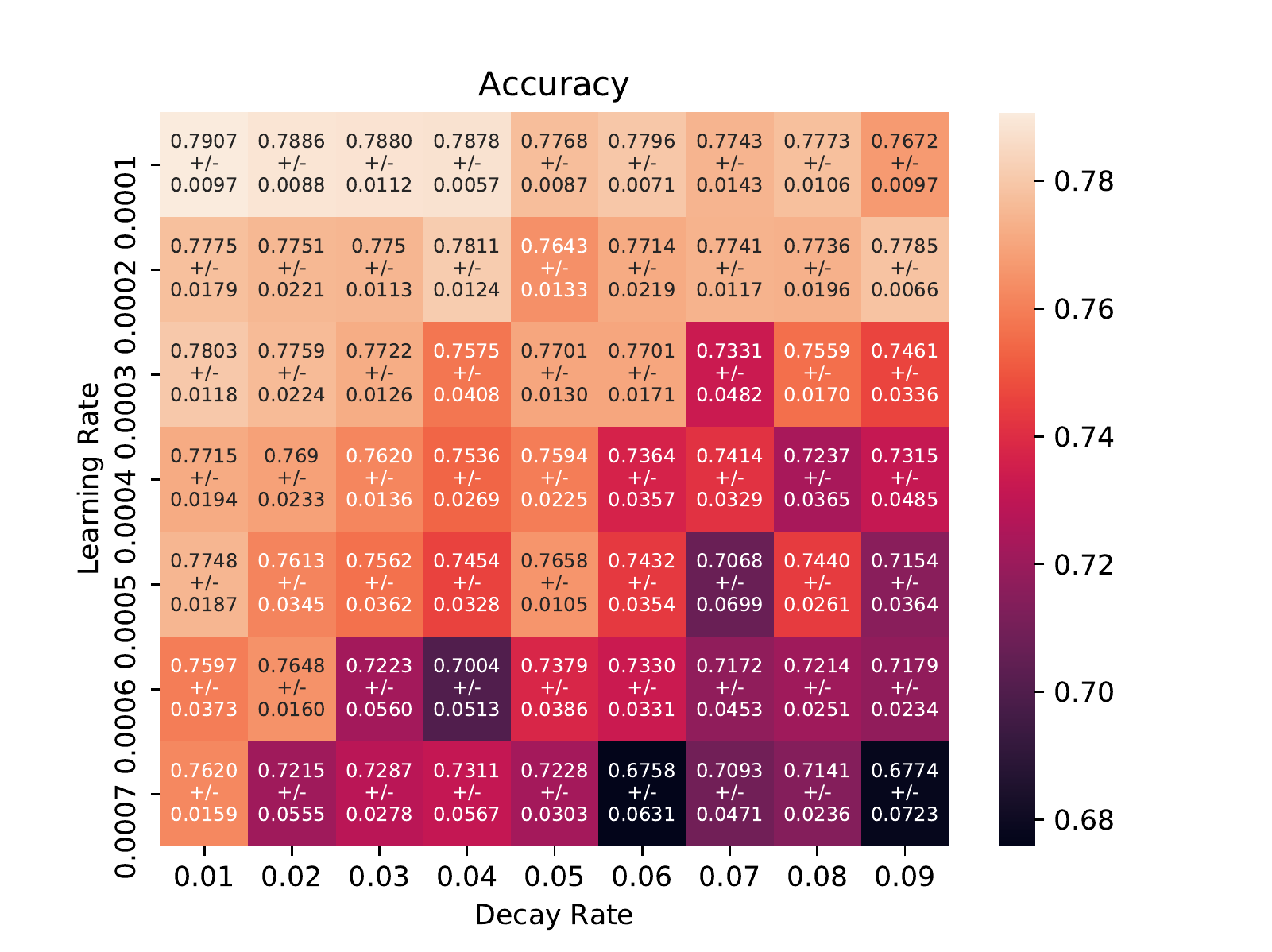}
\caption{Selected hyperparameter scan results for DNN (top), CNN (center), and GN (bottom). In each figure, the classification accuracy is displayed as a function of the hyperparameters reported on the two axes.}
\label{fig:scan_hyperparameter}
\end{figure*}

Selected hyperparameter scan slices are shown in Figure~\ref{fig:scan_hyperparameter}. 
These 2D scans were obtained setting all values besides the two under consideration (i.e., those on the axes) to be fixed at default values: a dropout rate of 0.08, a learning rate of 0.0004, a decay rate of 0.04, three dense layers for CNN and DNN, and 512 neurons per hidden layer. For GN, the default number of ECAL filters was 3, with a kernel size of 4.

\begin{figure}[htbp]
\centering
\includegraphics[width=0.42\textwidth]{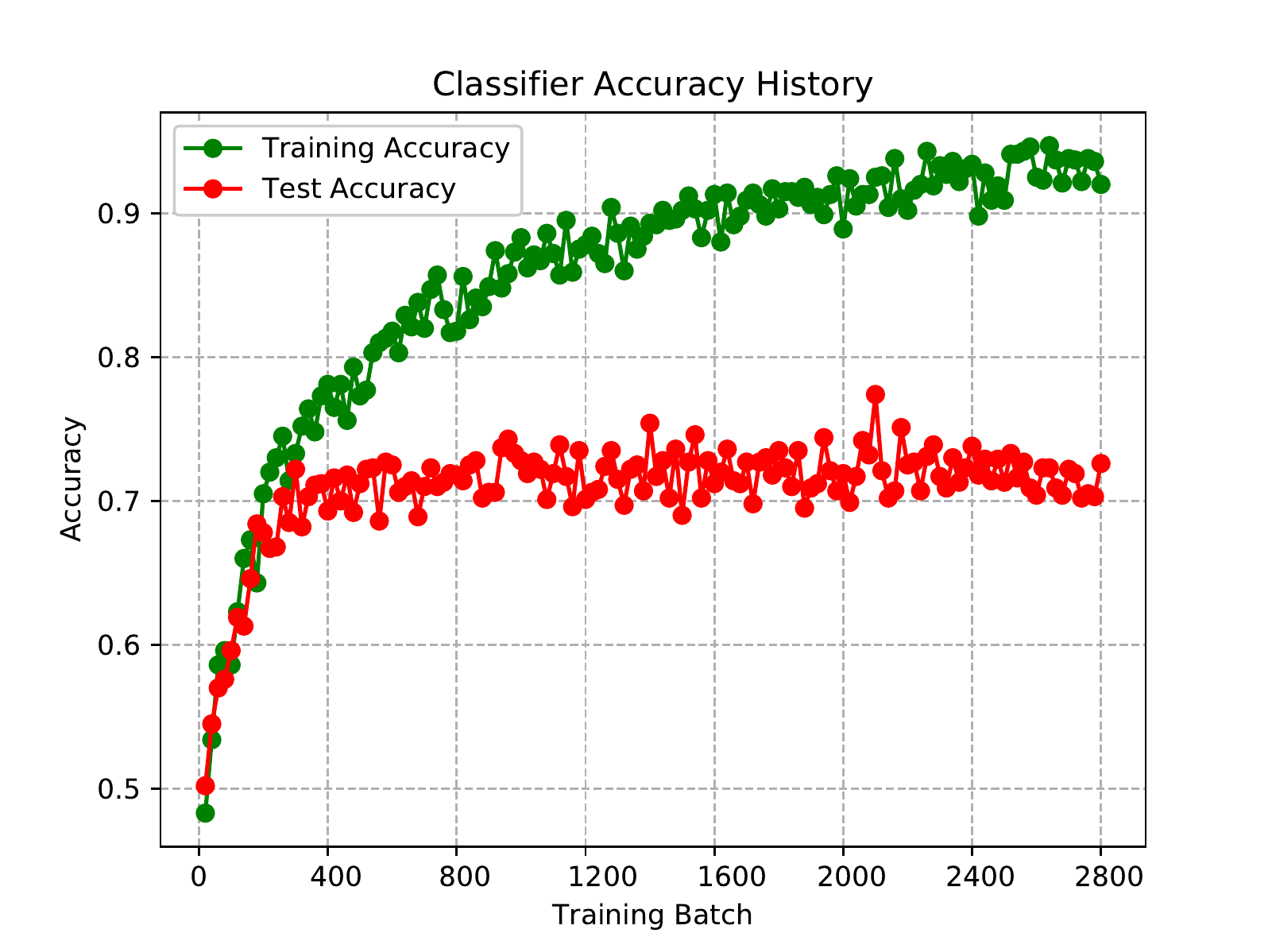}
\includegraphics[width=0.42\textwidth]{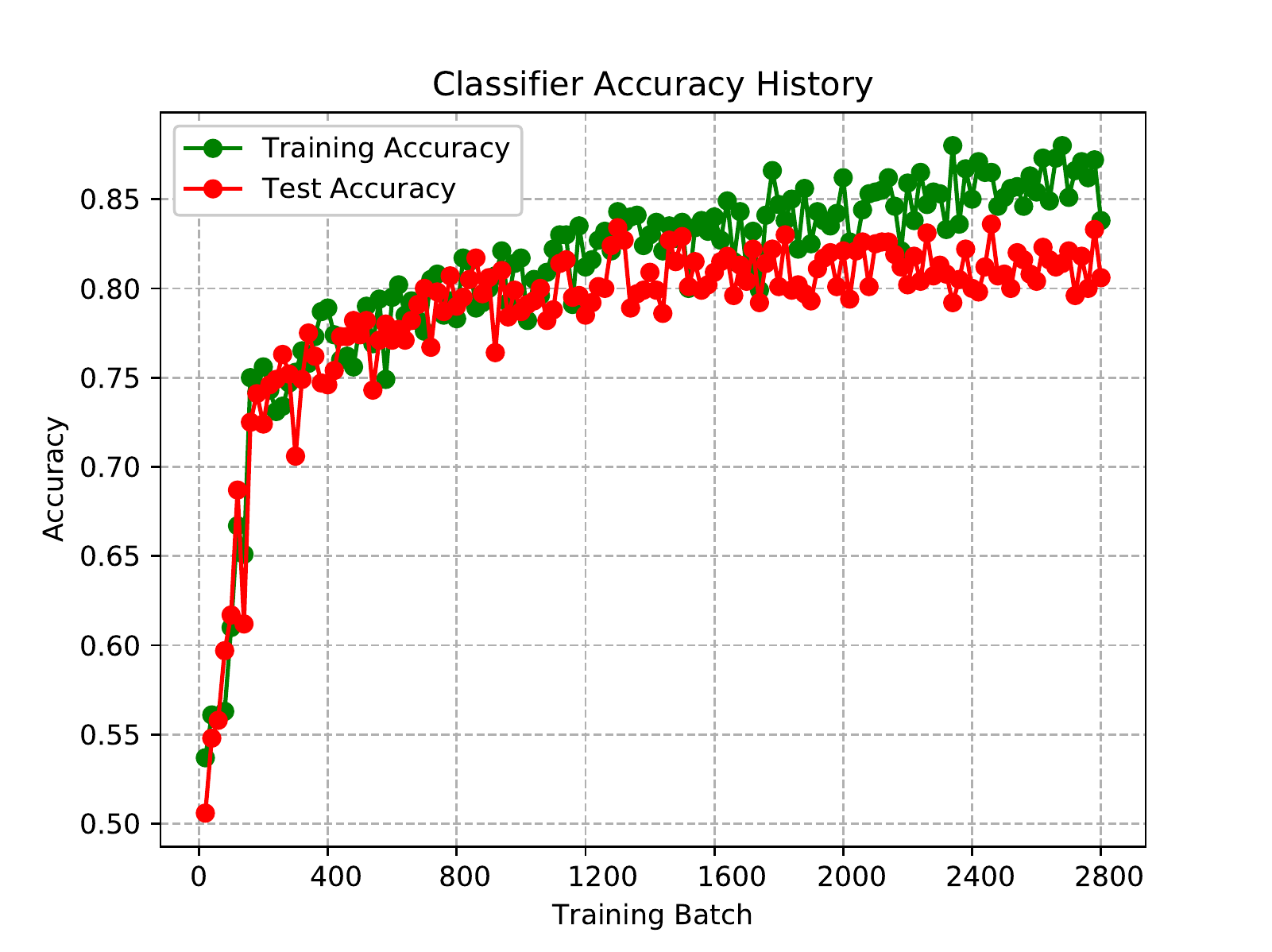}
\includegraphics[width=0.42\textwidth]{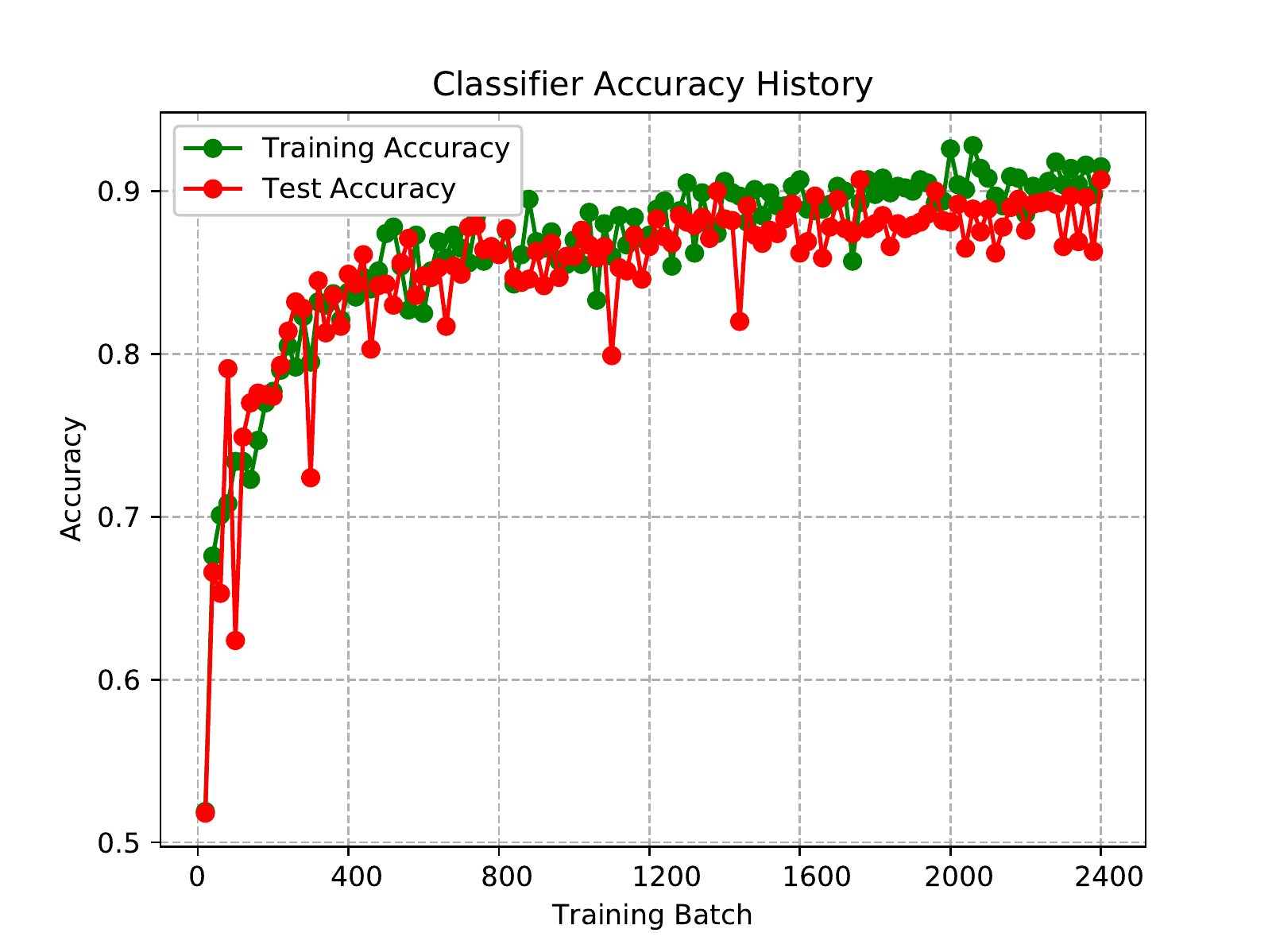}
\caption{Training curves for best DNN (top), CNN (middle), and GN (bottom) hyperparameters, trained on variable-angle $\gamma$/$\pi^0$ samples. We see that the DNN over-trains quickly and saturates at a relatively low accuracy, while the CNN takes longer to over-train and reaches a higher accuracy, and GN performs best of all. Each 400 batches corresponds to a single epoch.}
\label{fig:training_curves_comparison_gamma_pi0}
\end{figure}

After performing the hyperparameter scan, we trained each architecture using its optimal hyperparameters for a greater number of epochs. The evolution of the training and validation accuracy as a function of the batch number for these extended trainings is shown in Figure~\ref{fig:training_curves_comparison_gamma_pi0}.

\subsection{Results}

We apply the best architectures described in the previous section separately to our electron vs. charged pion and photon vs. neutral pion reconstruction problems.

\subsubsection{Classification Performance}
\label{sec:classification}

\begin{figure}[htbp]
\centering
\includegraphics[width=0.45\textwidth]{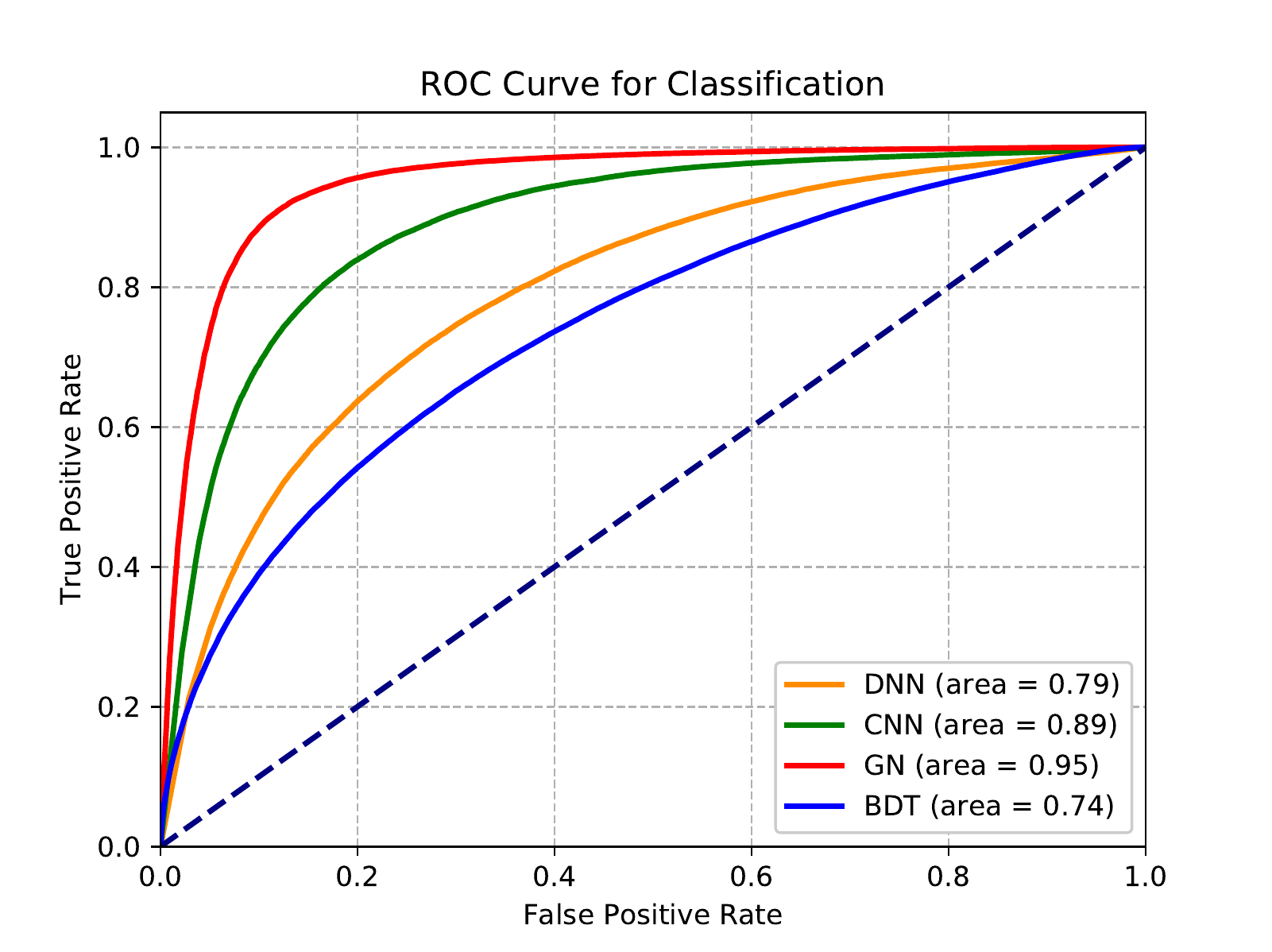}
\includegraphics[width=0.45\textwidth]{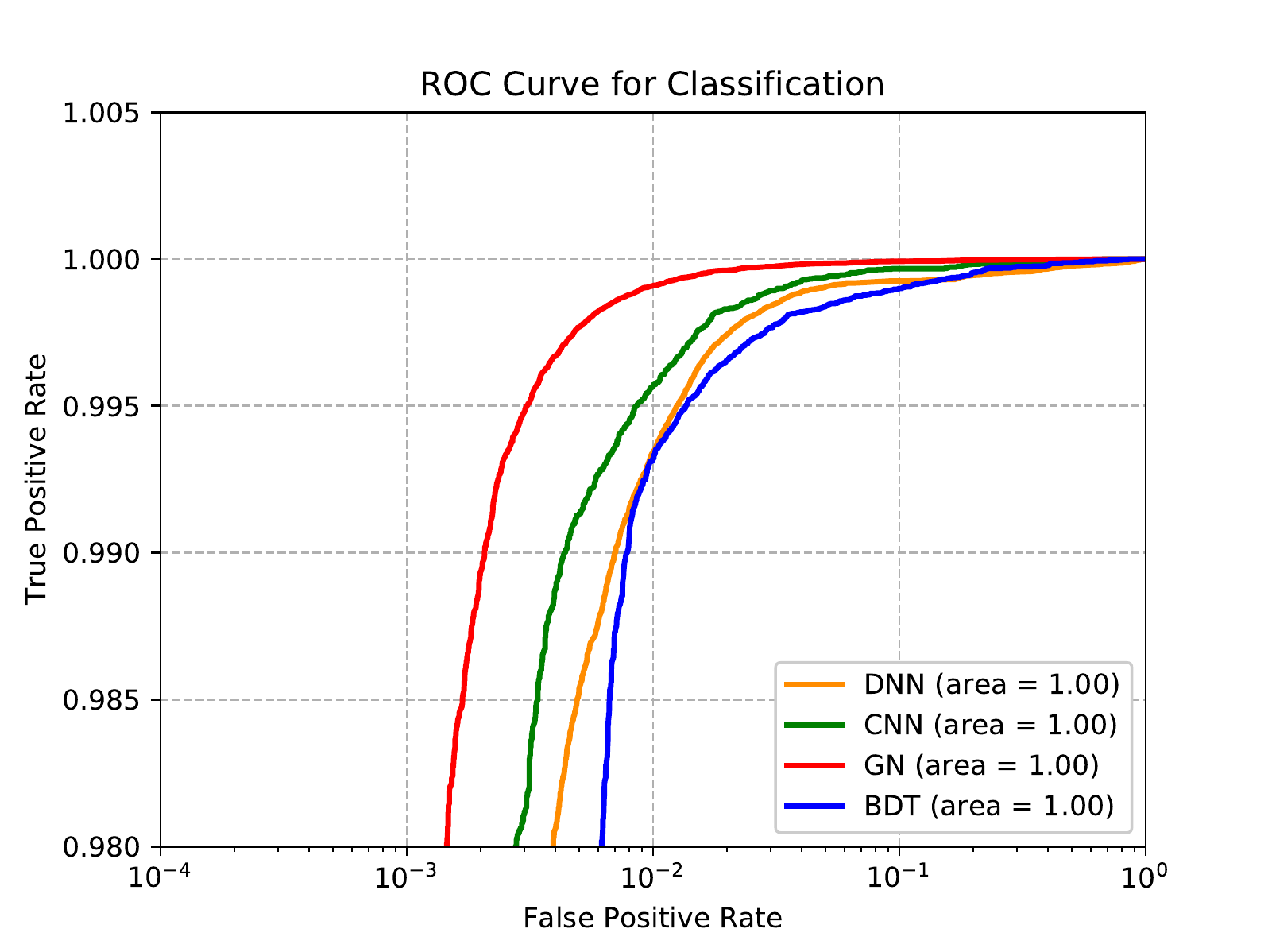}
\caption{ROC curve comparisons for $\gamma$ vs. $\pi^0$ (top) and $e$ vs. $\pi^\pm$ (bottom) classification using different neural network architectures. Samples include particle energies from 10 to 510 GeV, and an inclusive $\eta$ range.}
\label{fig:architectures_ROC_comparisons}
\end{figure}

Figure~\ref{fig:architectures_ROC_comparisons} shows ROC curve comparisons for the two classification tasks. As expected, the electron vs. charged pion classification problem was found to be a simple task, resulting in an area under the curve (AUC) close to $100\%$. For a baseline comparison, the curve obtained for a BDT (see Appendix~\ref{app:BDT}) is also shown. This BDT was optimized using the {\it scikit-optimize} package~\cite{skopt}, and was trained using high-level features computed from the raw 3D arrays. It represents the performance of current ML approaches on these problems.

\begin{figure*}[htbp]
\centering
\includegraphics[width=0.45\textwidth]{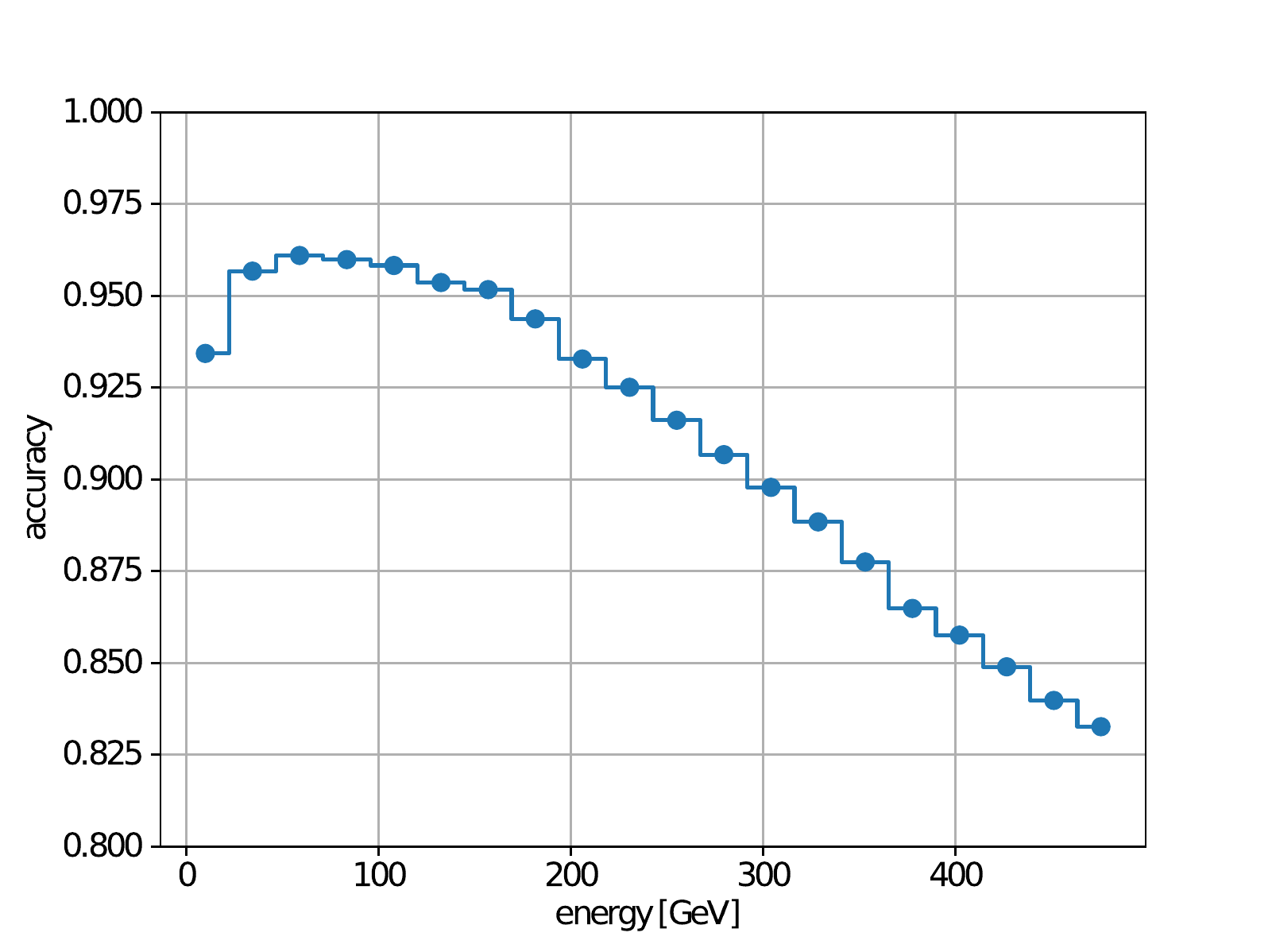}
\includegraphics[width=0.45\textwidth]{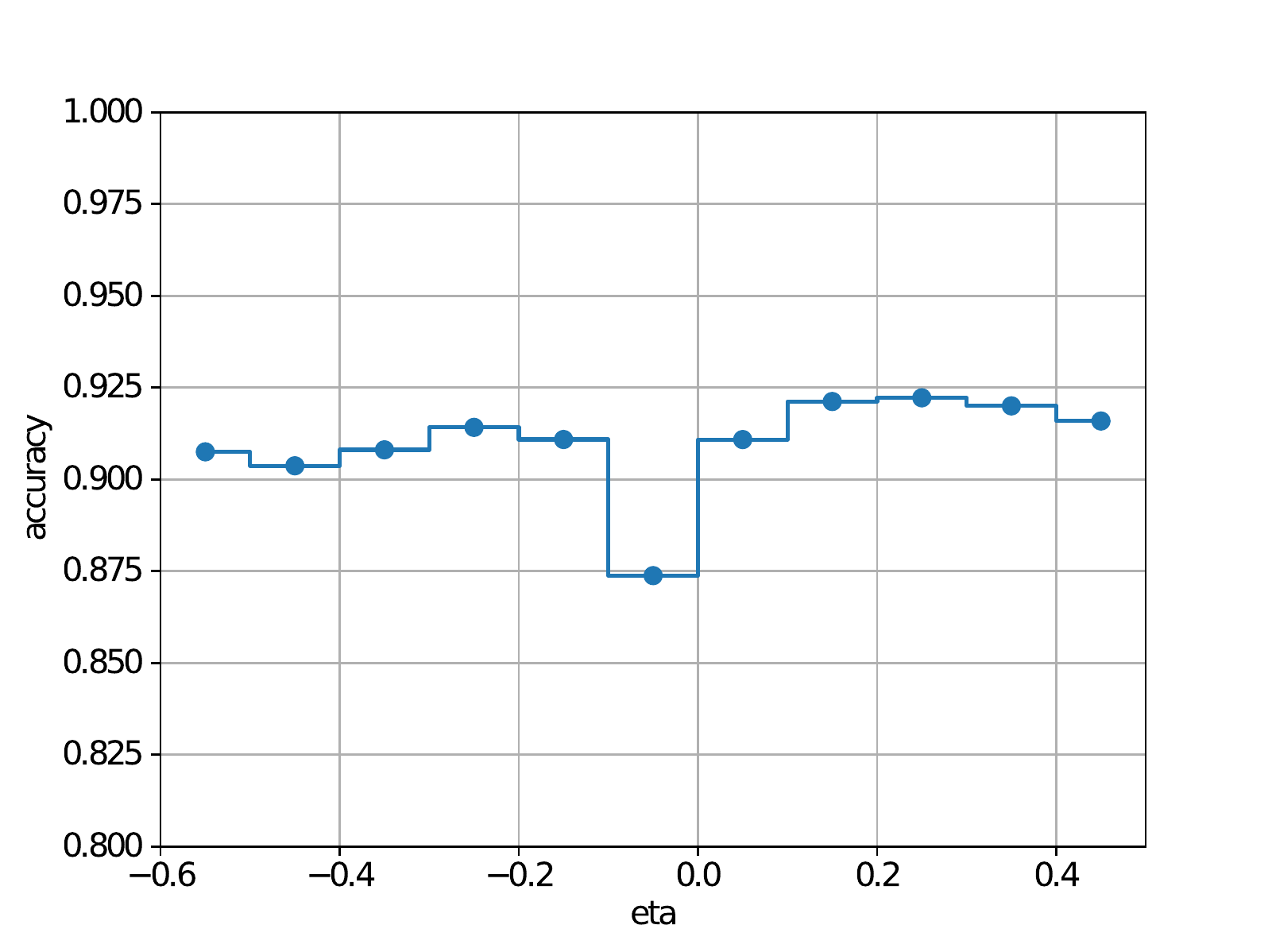} \\
\includegraphics[width=0.45\textwidth]{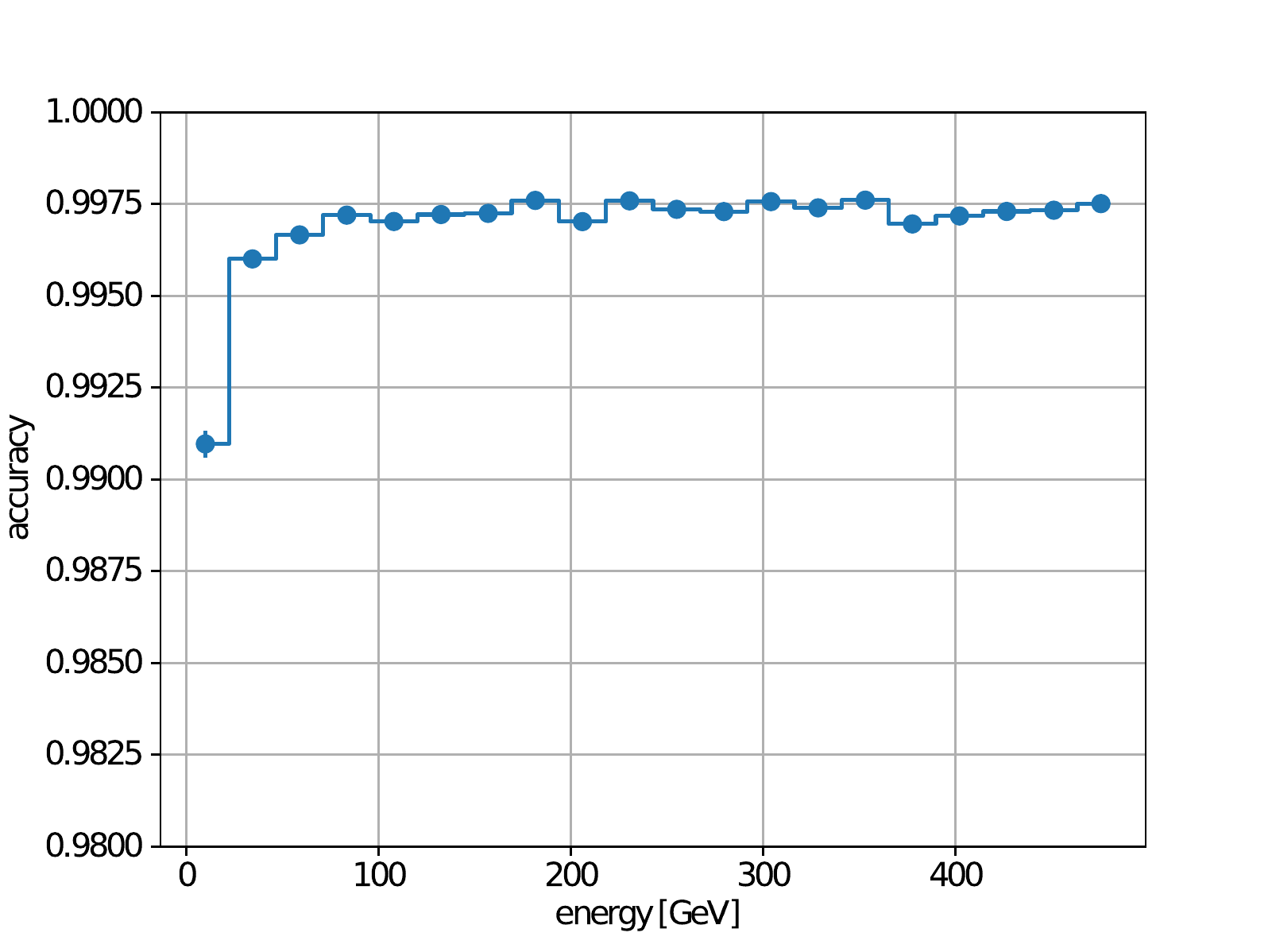}
\includegraphics[width=0.45\textwidth]{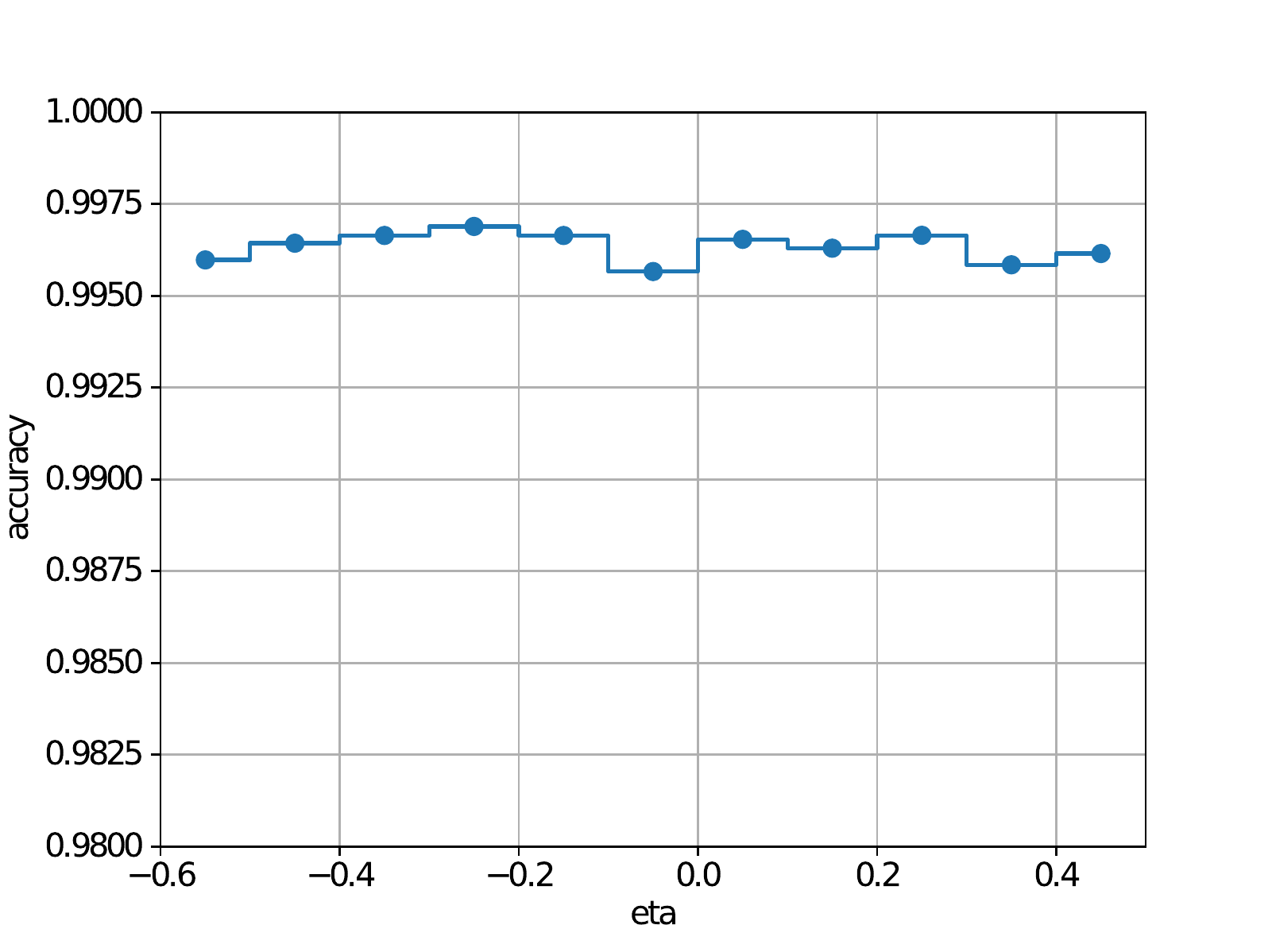}
\caption{Classification accuracy of best performing network for $\gamma$ vs. $\pi^0$ (top) and $e$ vs. $\pi^\pm$ (bottom), in bins of energy (left) and $\eta$ (right).}
\label{fig:accuracy_bins}
\end{figure*}

The ML models outperform the BDT, with the GN reaching the best classification performance on both problems. Figure~\ref{fig:accuracy_bins} shows the best-model performance as a function of the energy and $\eta$ of the incoming particle, for the photon vs. neutral pion and the electron vs. charged pion problems. These figures show that classification accuracy is maintained over a wide range of particle energies and angles. The models appear to perform a bit worse at higher energies for the photon vs. neutral pion case, due to the fact that the pion to two photon decay becomes increasingly collimated at higher energies. Similarly, the performance is slightly worse when particles impact the detector perpendicularly than when they enter at a wide angle, because the shower cross section on the calorimeter inner surface is reduced at $90^{\mathrm o}$, making it harder to distinguish shower features.

\subsubsection{Regression Performance}
\label{sec:regression}

\begin{figure*}[htbp]
\centering
\includegraphics[width=0.38\textwidth]{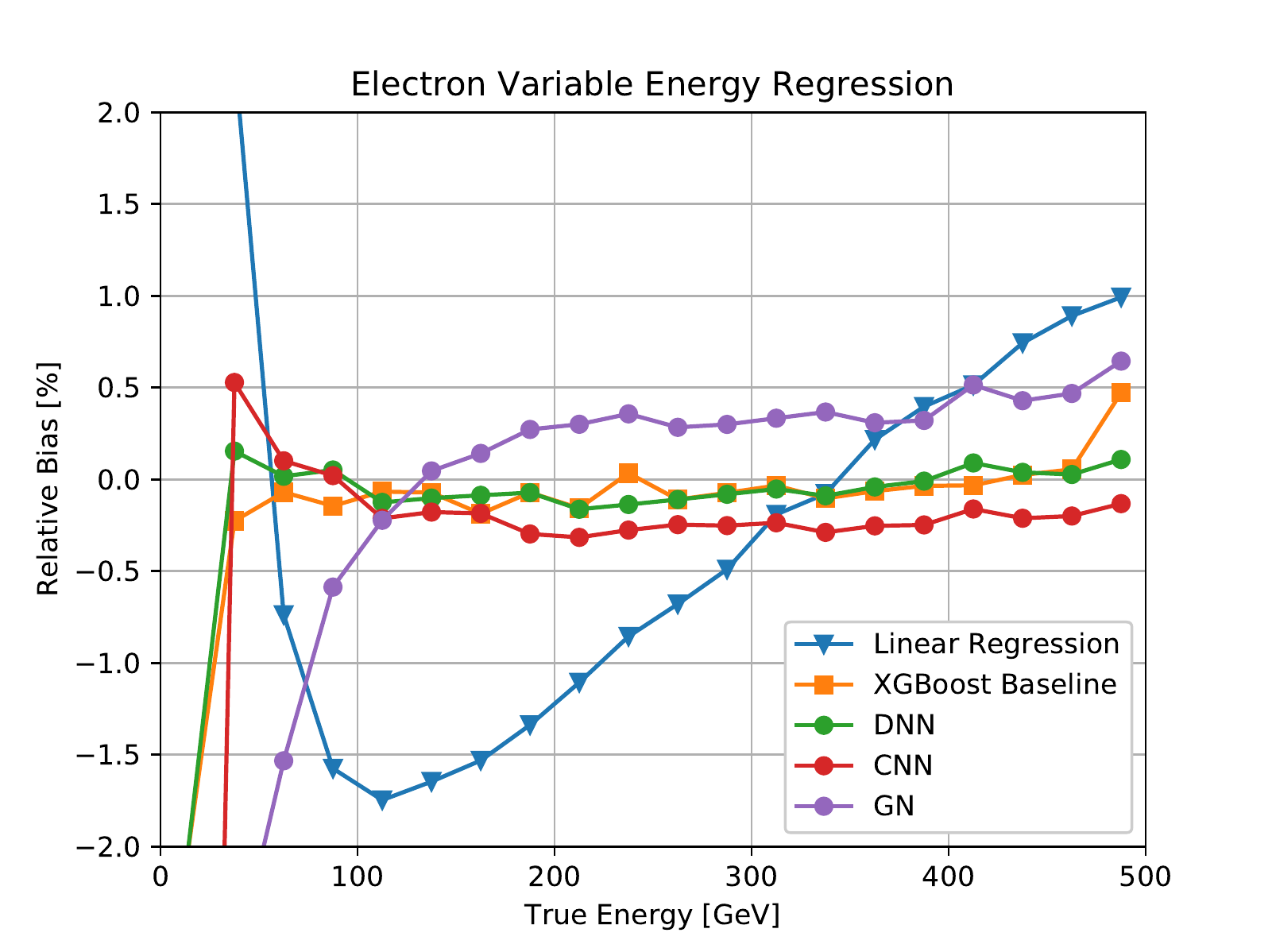}
\includegraphics[width=0.38\textwidth]{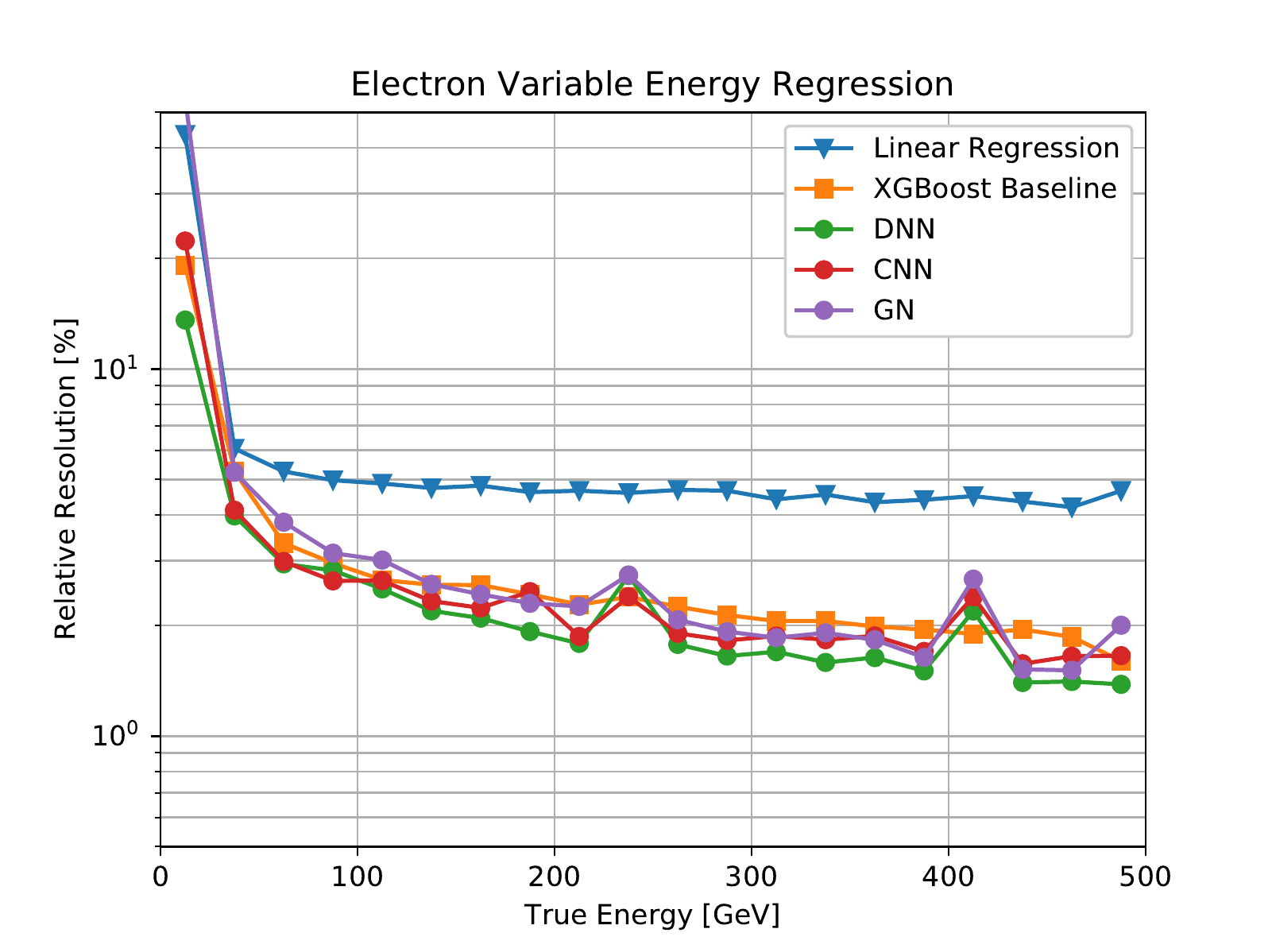} \\
\includegraphics[width=0.38\textwidth]{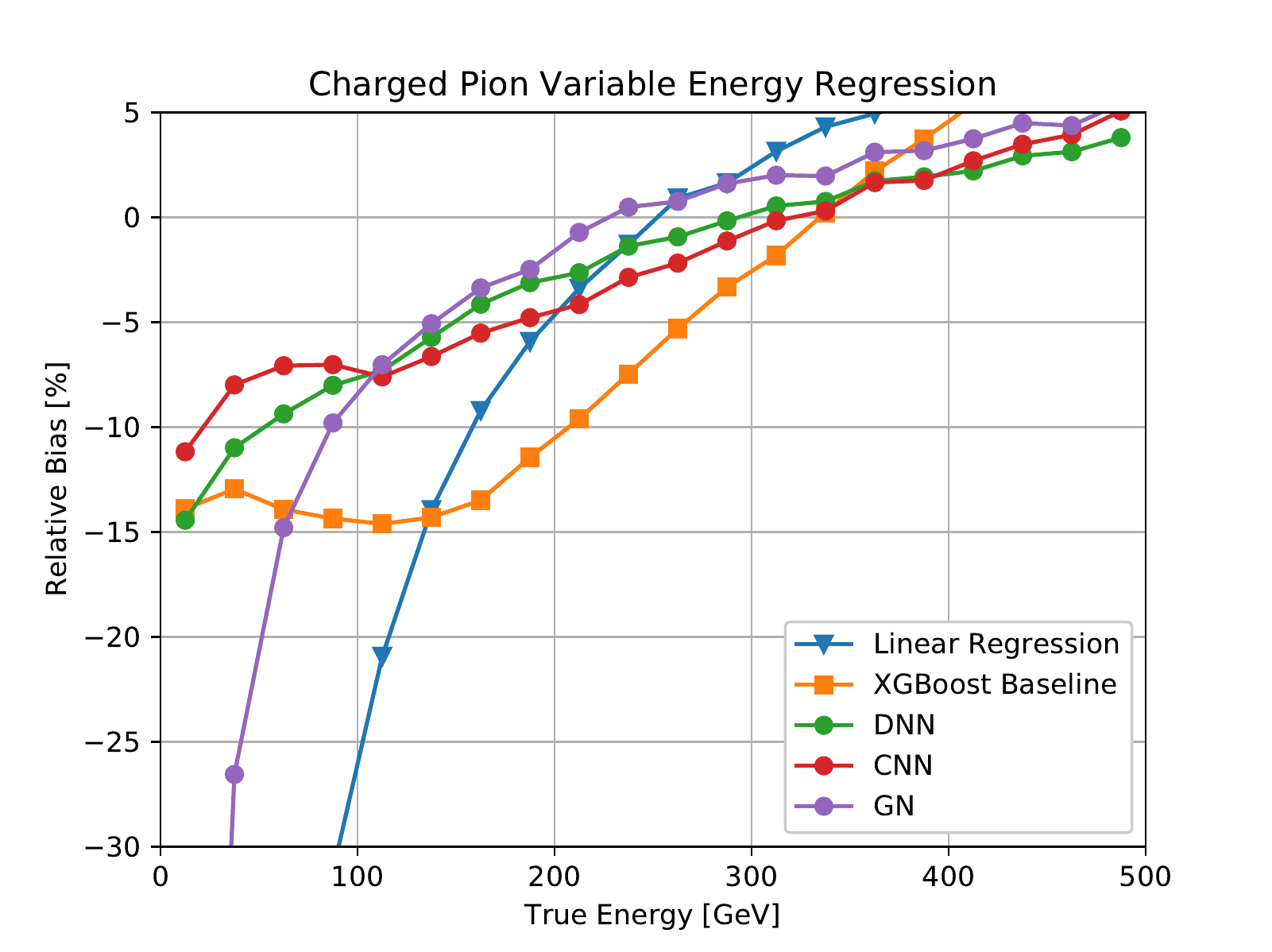}
\includegraphics[width=0.38\textwidth]{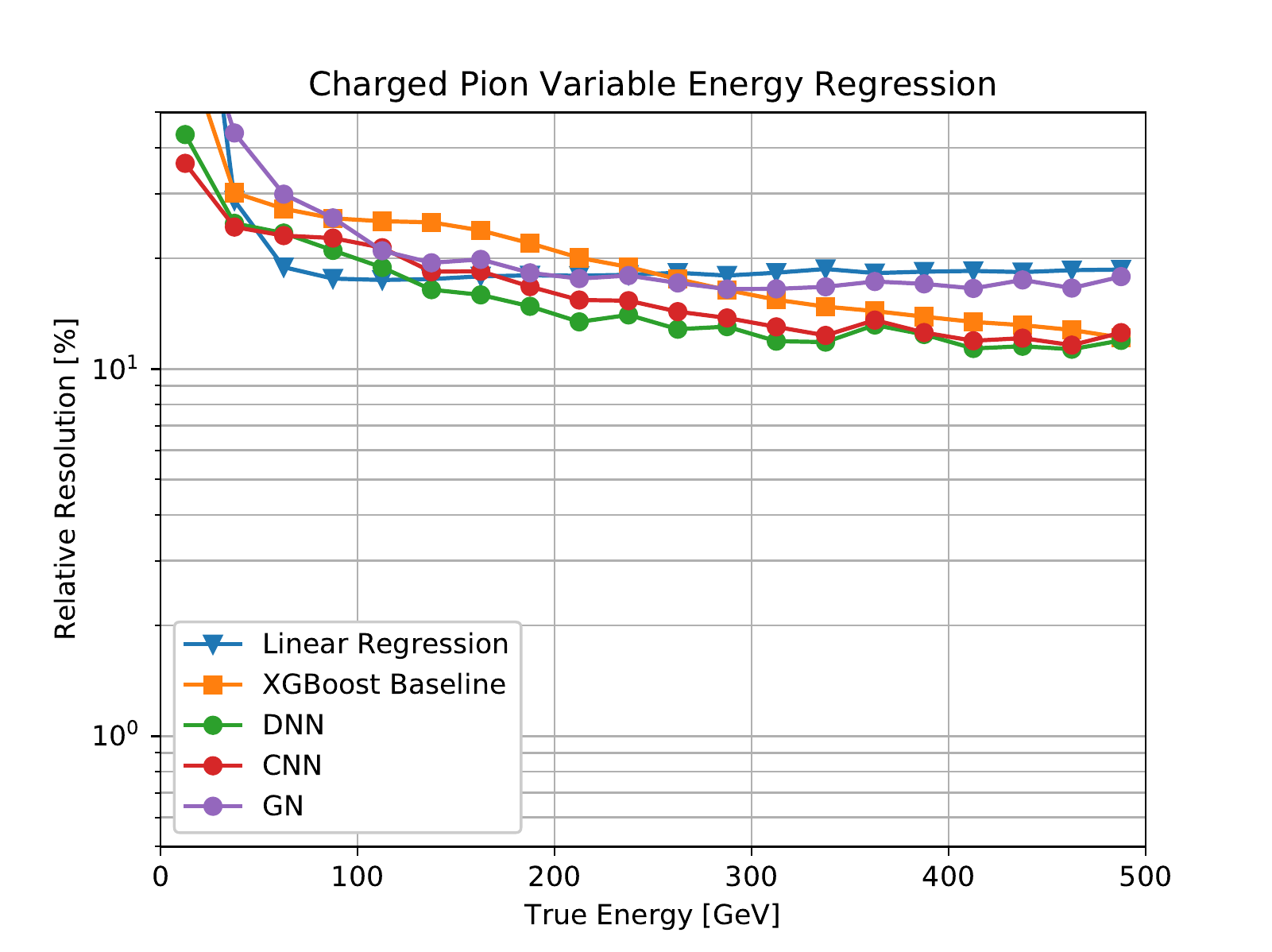}\\
\includegraphics[width=0.38\textwidth]{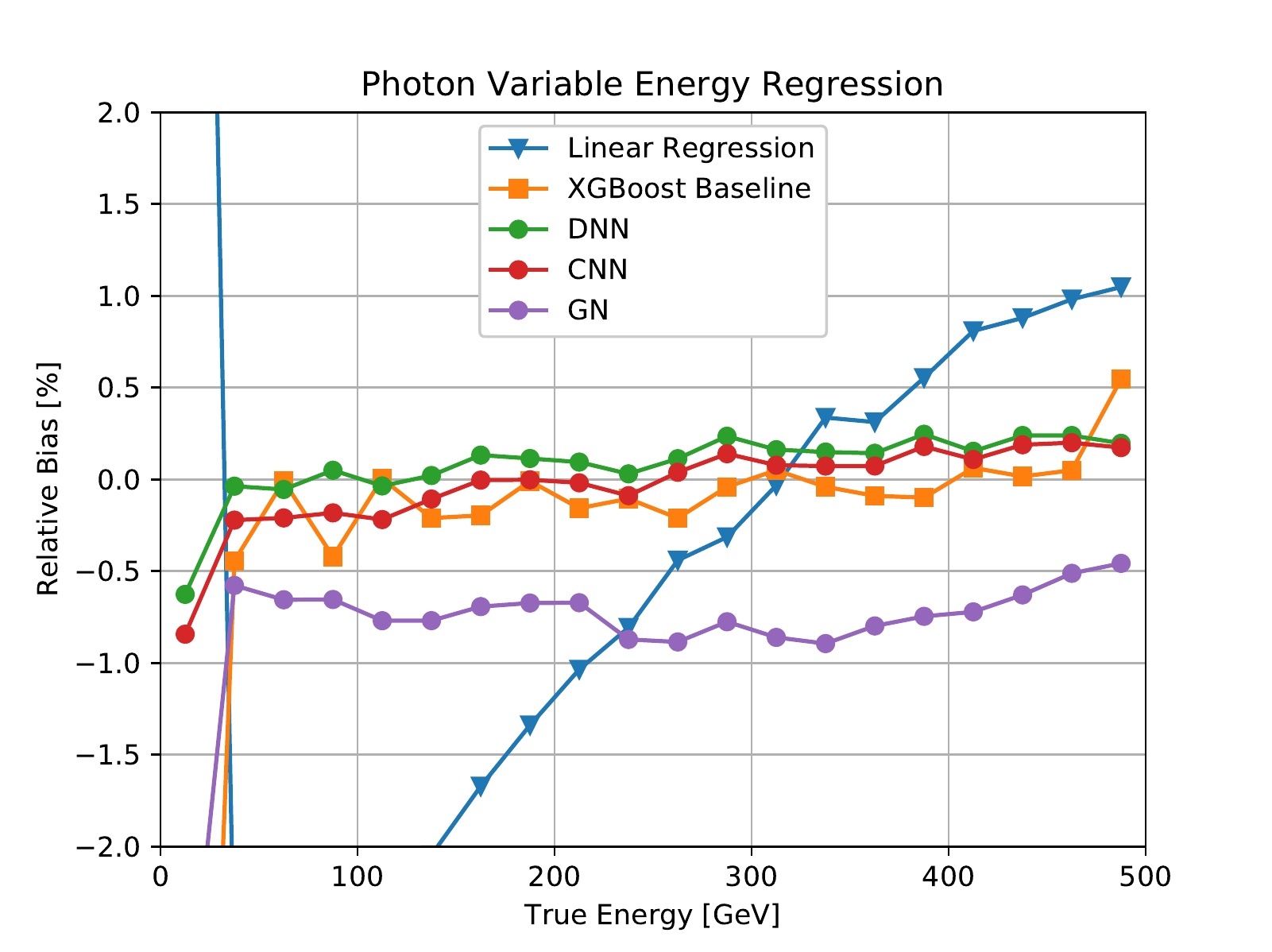}
\includegraphics[width=0.38\textwidth]{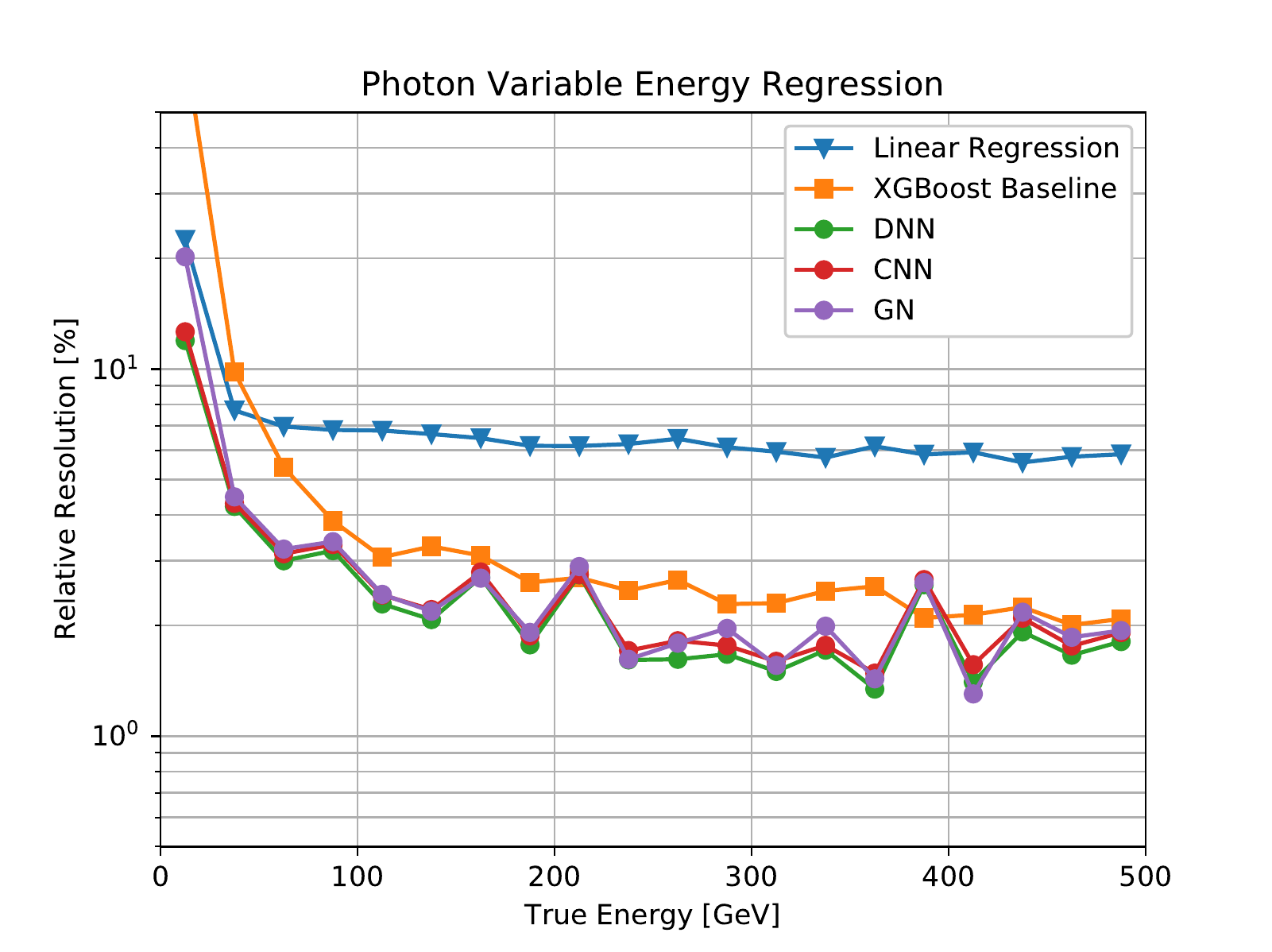}\\
\includegraphics[width=0.38\textwidth]{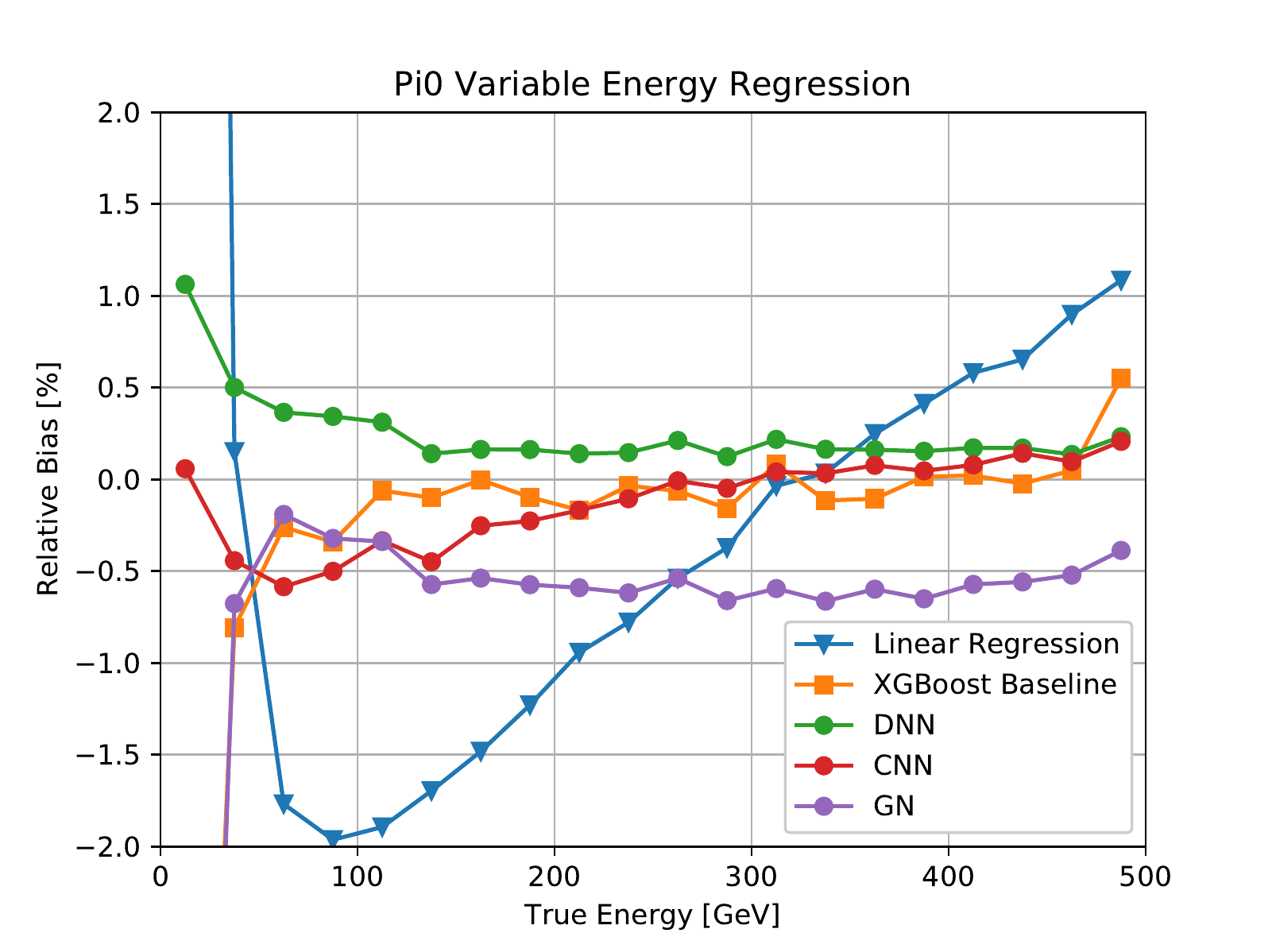}
\includegraphics[width=0.38\textwidth]{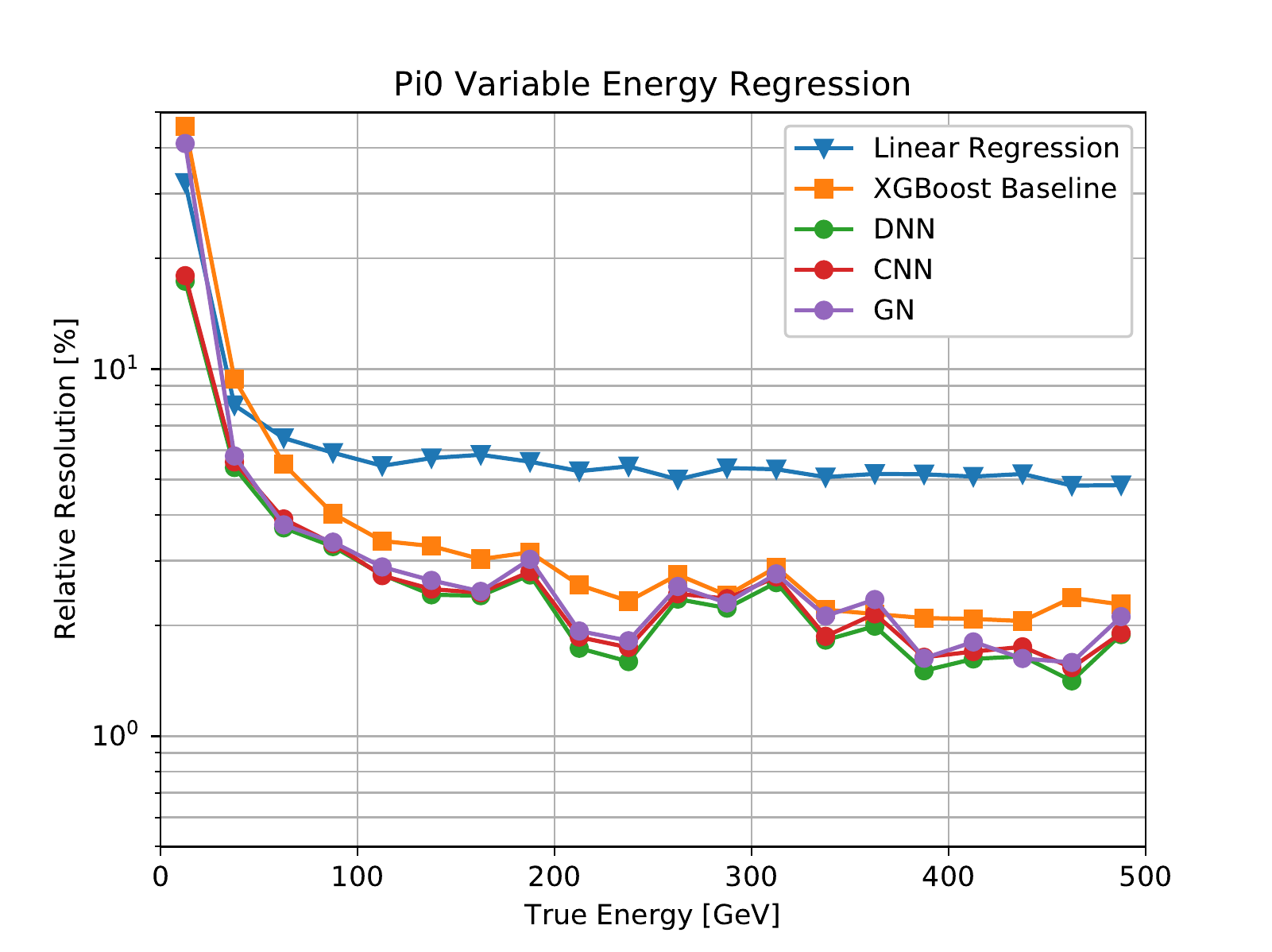}
\caption{Regression bias (top) and resolution (bottom) as a function
  of true energy for energy predictions on the REC dataset with
  variable-angle incident angle. From top to bottom: electrons,
  charged pions, photons, and neutral
  pions.\label{fig:reg_dnn_vs_cnn_variable}}
\end{figure*}

Figure~\ref{fig:reg_dnn_vs_cnn_variable} shows the energy regression performance for each particle type, obtained from the end-to-end reconstruction architectures. In this case, we compare against a linear regression algorithm and a BDT (labelled as "XGBoost") representing the current state-of-the-art, as described in Appendix~\ref{app:regression_baseline}. 

Since the energy regression problem is not as complex as the classification problem, the three architectures (DNN, CNN, GN) perform fairly similarly, with the exception of the GN performance on \chpi, which is a bit worse.
The performance is overall worse for \chpi, both with the networks and with the benchmark baselines (linear regression and XGBoost).

A closer look at the performance boost given by each network can be obtained examining the case of particles entering the calorimeter inner surface at $90^{\mathrm o}$, i.e. with $\eta=0$~\footnote{For these additional fixed-angle regression plots, we did not train GoogLeNet architectures.}. In this case, the problem is more constrained and both the networks and the baseline algorithms are able to perform accurately. The results for fixed angle samples are shown in Appendix~\ref{app:regression_fixed_angle}.

We have also tested the result of training on one class of particle and performing regression on another. These results can be seen in Appendix~\ref{app:xtrain_regression}. In addition, we have looked at the effect on energy regression of increasing the ECAL and HCAL window sizes. This can be seen in Appendix~\ref{app:large_window_regression}.

\subsection{Resampling to ATLAS and CMS Geometries}\label{sec:resampling}

In addition to the results presented so far, we show in this section how the end-to-end reconstruction would perform on calorimeters with granularity and geometry similar to those of the ATLAS and CMS calorimeters. Since the REC dataset (see Section~\ref{sec:data}) is generated using the geometry of the proposed LCD detector, it has a much higher granularity than the current-generation ATLAS and CMS detectors. To visualize how our calorimeter data would look with a coarser detector, we linearly extrapolate the contents of each event to a different calorimeter geometry, using a process we have termed "resampling". To keep the resampling procedure simple, we discard the HCAL information and consider only the ECAL 3D array.

\begin{table*}[tbp]
\centering
\caption{Detailed description of the three detector geometries used in this study: the baseline CLIC ECAL detector and the ATLAS and CMS calorimeters.\label{tab:resampling_geometry}}
\begin{tabular}{c|c|ccc|c}
\hline
\multirow{2}{*}{Parameter} & \multirow{2}{*}{\textbf{CLIC}} & \multicolumn{3}{c|}{\textbf{ATLAS}} & \multirow{2}{*}{\textbf{CMS}} \\
            &               & 1st layer & 2nd layer & 3rd layer & \\
\hline
$\Delta \eta$         & 0.003  & 0.025 /8 & 0.025 & 0.5   & 0.0175 \\
$\Delta \phi$         & 0.003  & 0.1      & 0.025 & 0.025 & 0.0175 \\
Radiation Length [cm] & 0.3504 & 14       & 14    & 14    & 0.8903 \\
Moliere radios [cm]   & 0.9327 & 9.043    & 9.043 & 9.043 & 1.959  \\
\hline 
\end{tabular}
\end{table*}

\begin{figure}[htbp]
    \centering
    \includegraphics[scale=0.3, clip]{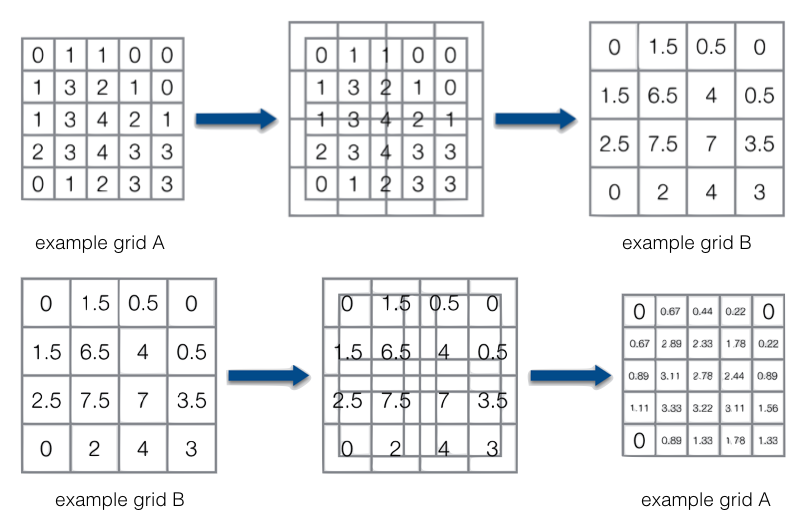}
    \caption{Example of the resampling procedure used to emulate CLIC data on a different detector geometry (the example shown here is simply a larger grid). First, we extrapolate hit information from one geometry to another (top). Next, we extrapolate back to the original geometry (bottom). This allows us to emulate the rougher granularity of the second geometry, while keeping data array sizes constant and enabling us to use the models we have already developed for the CLIC dataset. Note that some information is lost at the edges.}
    \label{fig:resampling}
\end{figure}

A not-to-scale example of the full procedure is shown in Figure~\ref{fig:resampling}. In this example, we resample the input to a regular square grid with lower granularity than the input data. The operation is simplified in the figure, in order to make the explanation easy to visualize. The actual ATLAS and CMS calorimeter geometries are more complex than a regular array, as described in Table~\ref{tab:resampling_geometry}.

In the resampling process, we first extrapolate each energy value from the grid of CLIC cells to a different geometry. To do so, we scale the content of each CLIC cell to the fraction of overlap area between the CLIC cell and the cell of the target geometry. When computing the overlap fraction, we take into account the fact that different materials have different properties (Moliere radius, interaction length, and radiation length). For instance, CLIC is more fine-grained than CMS or ATLAS detectors, but the Moliere radius of the CLIC ECAL is much smaller than in either of those detectors. This difference determines an offset in the fine binning. Thus, when applying our resampling procedure we normalize the cell size by the detector properties. The Moliere radius is used for $x$ and $y$ re-binning, and radiation length is used for the $z$ direction. At this point we have a good approximation for how the event would look in a calorimeter with the target geometry.

To complete the resampling process, we invert the procedure to go back to our original high-granularity geometry. This last step allows us to keep using the model architectures that we have already optimized. It adds no additional information that would not be present in the low-granularity geometry. This up-sampling also allows us to deal with the irregular geometry of the ATLAS calorimeter by turning it into a neat grid. With no up-sampling, it would not be possible to apply the CNN and GN models.

The resampling procedure comes with a substantial simplification of the underlying physics process. First of all, the information at the edge of the grid is imperfectly translated during the resampling process, leading to worse performance than what could theoretically be achieved in the actual CMS and ATLAS detectors. Also, this simple geometrical rescaling doesn't capture many other detector characteristics. For example, the CMS ECAL detector has no depth information, but being homogeneous it provides a very precise energy measurement. Our resampling method only captures geometric effects, and would not be able to model the improvement in energy resolution. Furthermore, we are unable to include second-order effects such as gaps in the detector geometries. Despite these limitations, one can still extract useful information from the resampled datasets, comparing the classification and regression performances of the end-to-end models defined in Sections~\ref{sec:classification} and~\ref{sec:regression} on different detector geometries.

\begin{figure}[htbp]
    \centering
    \includegraphics[scale=0.5, clip]{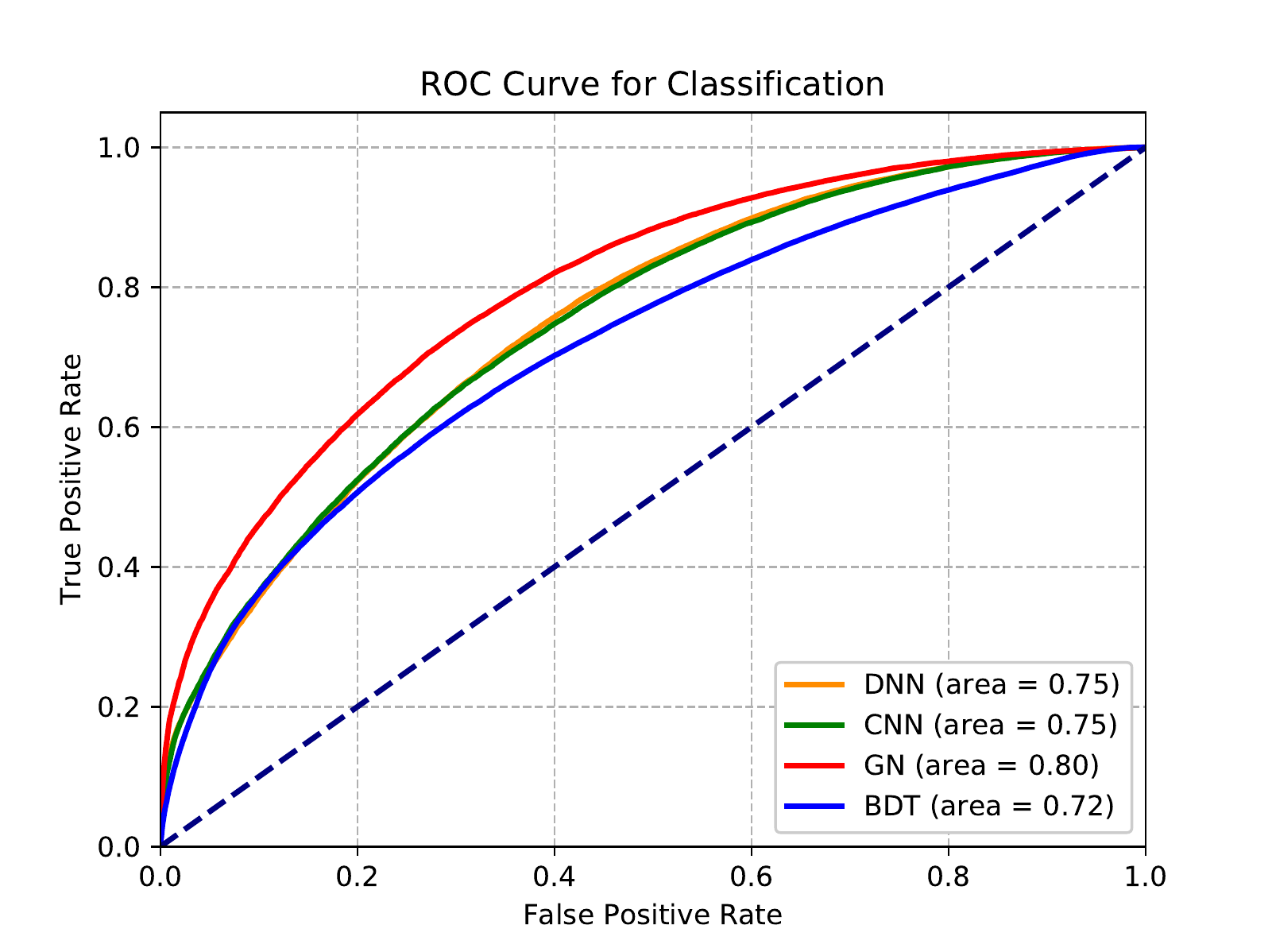}
    \includegraphics[scale=0.5, clip]{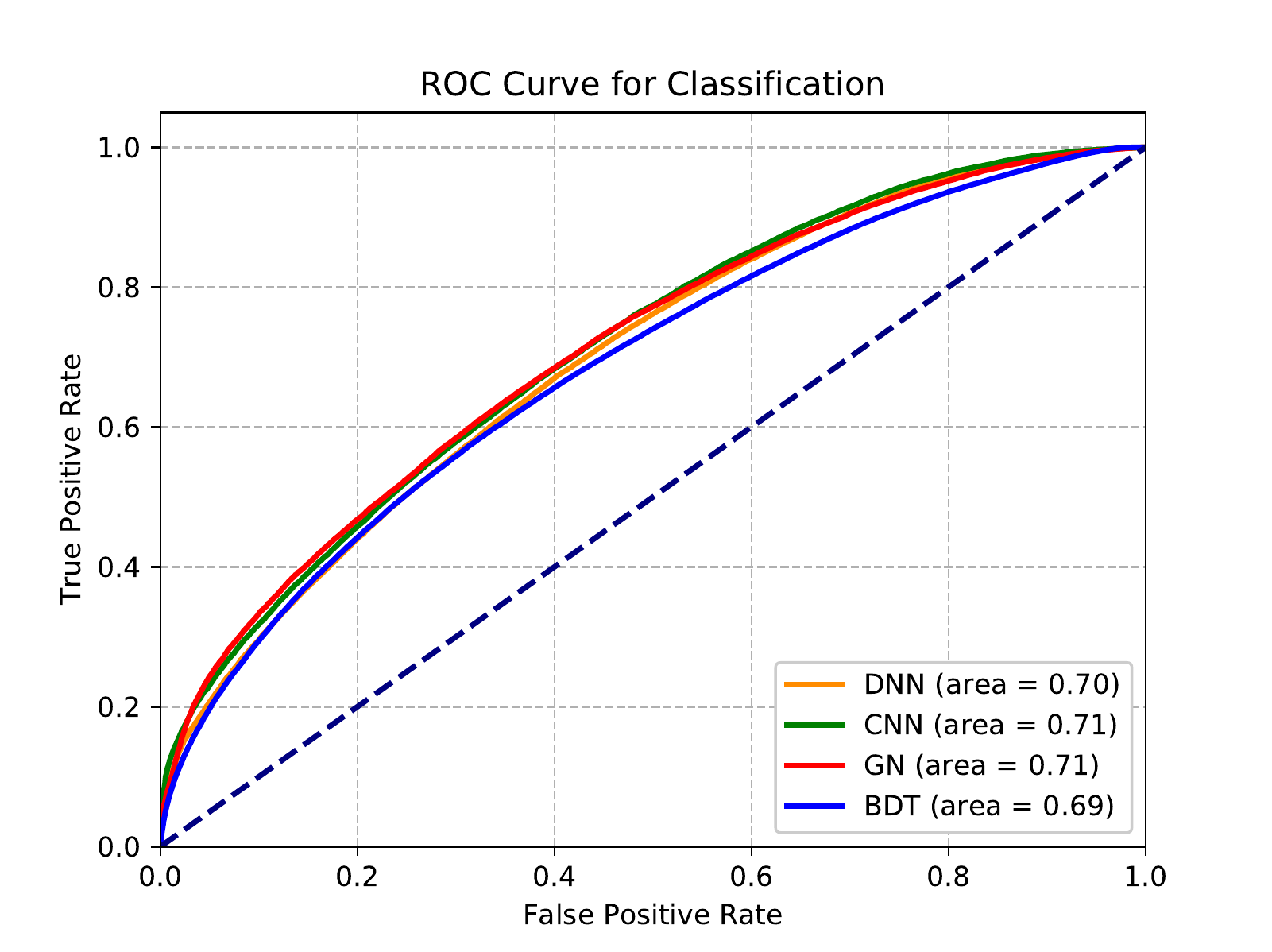}
    \caption{ROC curve comparisons for variable-angle $\gamma$/$\pi^0$ classification on data resampled to ATLAS-like (top) and CMS-like (bottom) geometries.}
    \label{fig:class_ROC_ATLAS_CMS}
\end{figure}

Comparisons of classification ROC curves between network architectures and our BDT baseline are shown in Figure~\ref{fig:class_ROC_ATLAS_CMS} for ATLAS-like and CMS-like geometries. Here we can see that the previously observed performance ranking still holds true. The GN
model performs best, followed by the CNN, then the DNN. All three networks outperform the BDT baseline. The effect is less pronounced after the CMS-like resampling, due to the low granularity and the single detector layer in the z direction.

\begin{figure*}[htbp]
\centering
\includegraphics[width=0.38\textwidth]{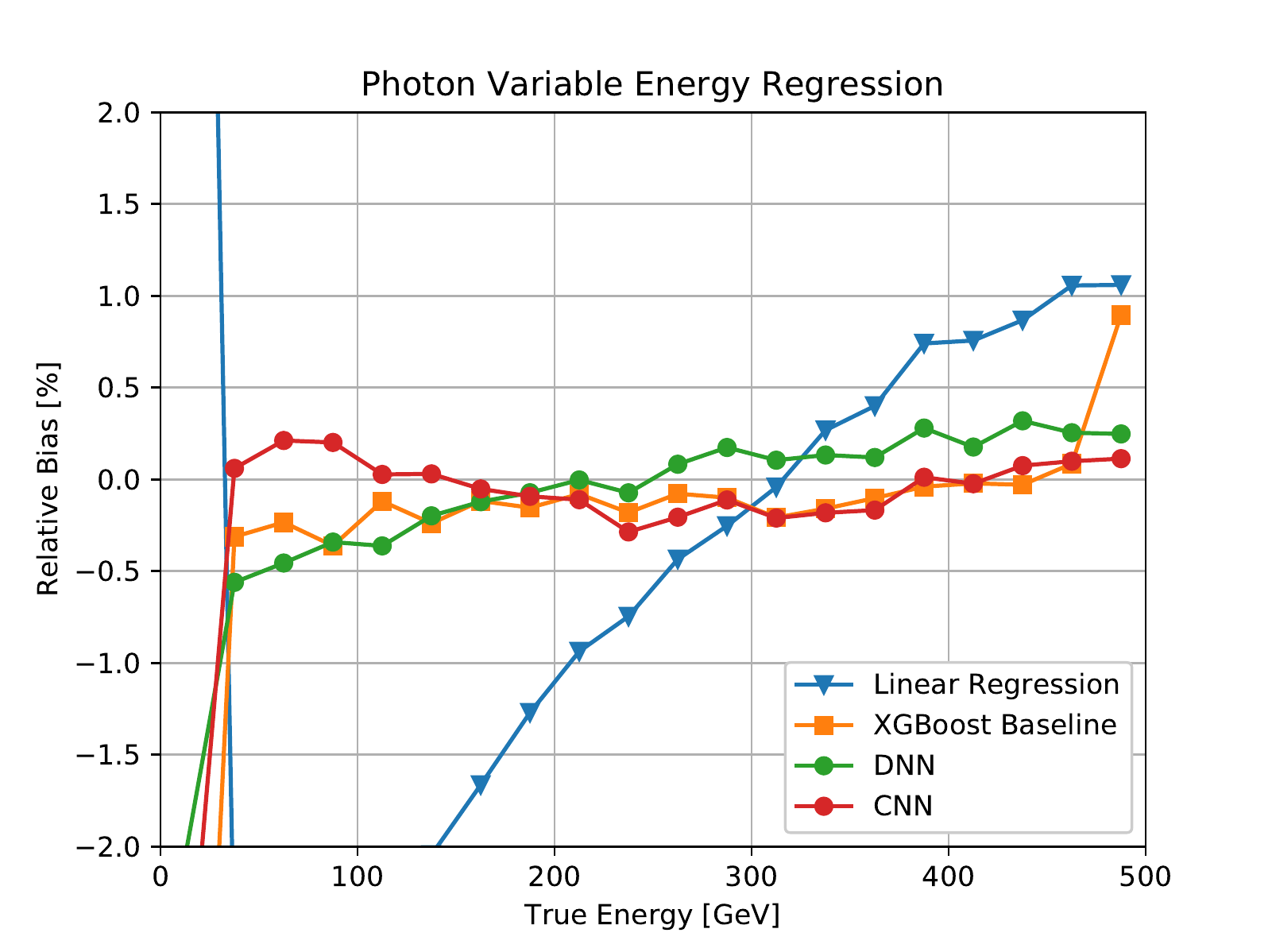}
\includegraphics[width=0.38\textwidth]{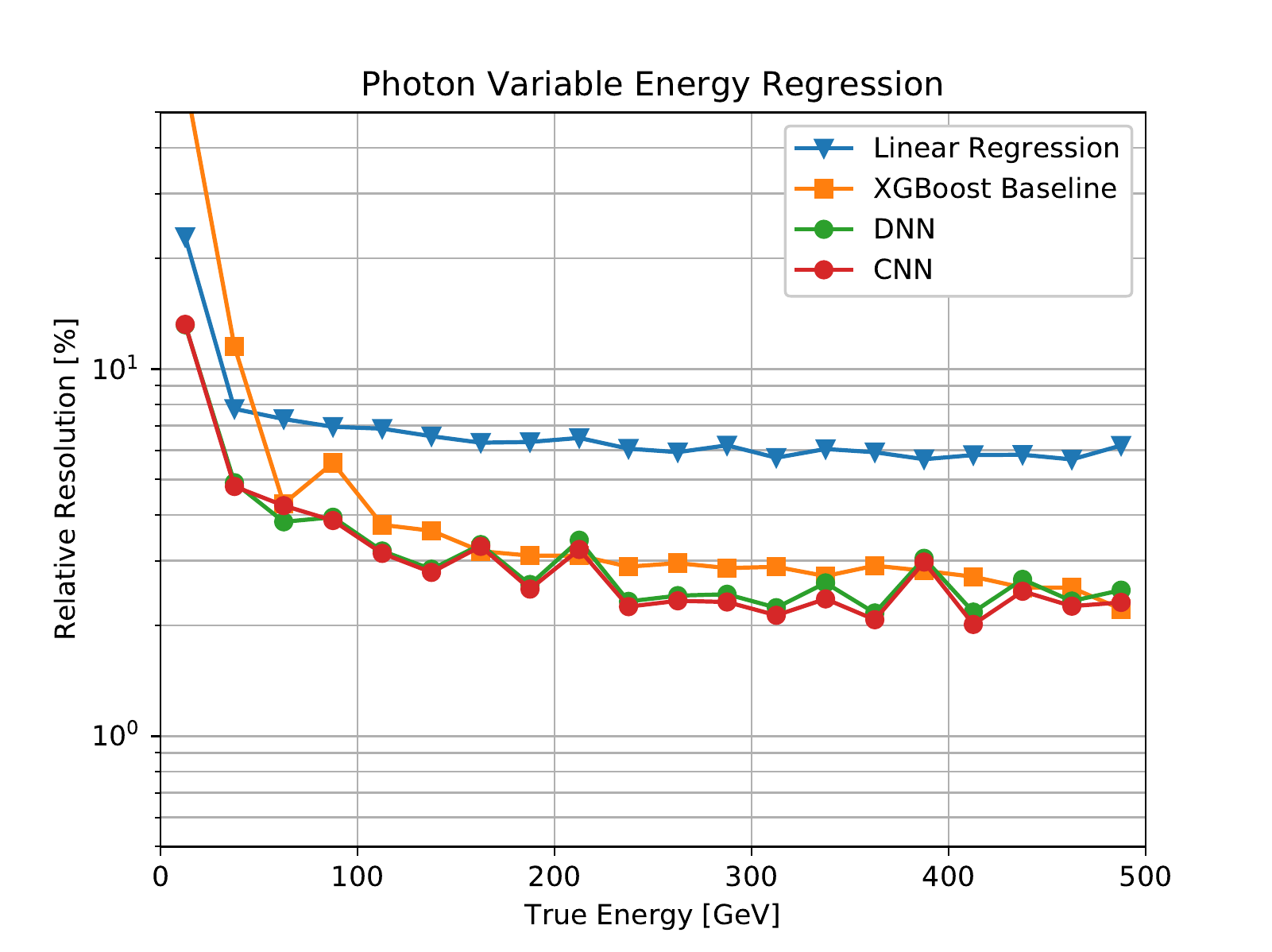} \\
\includegraphics[width=0.38\textwidth]{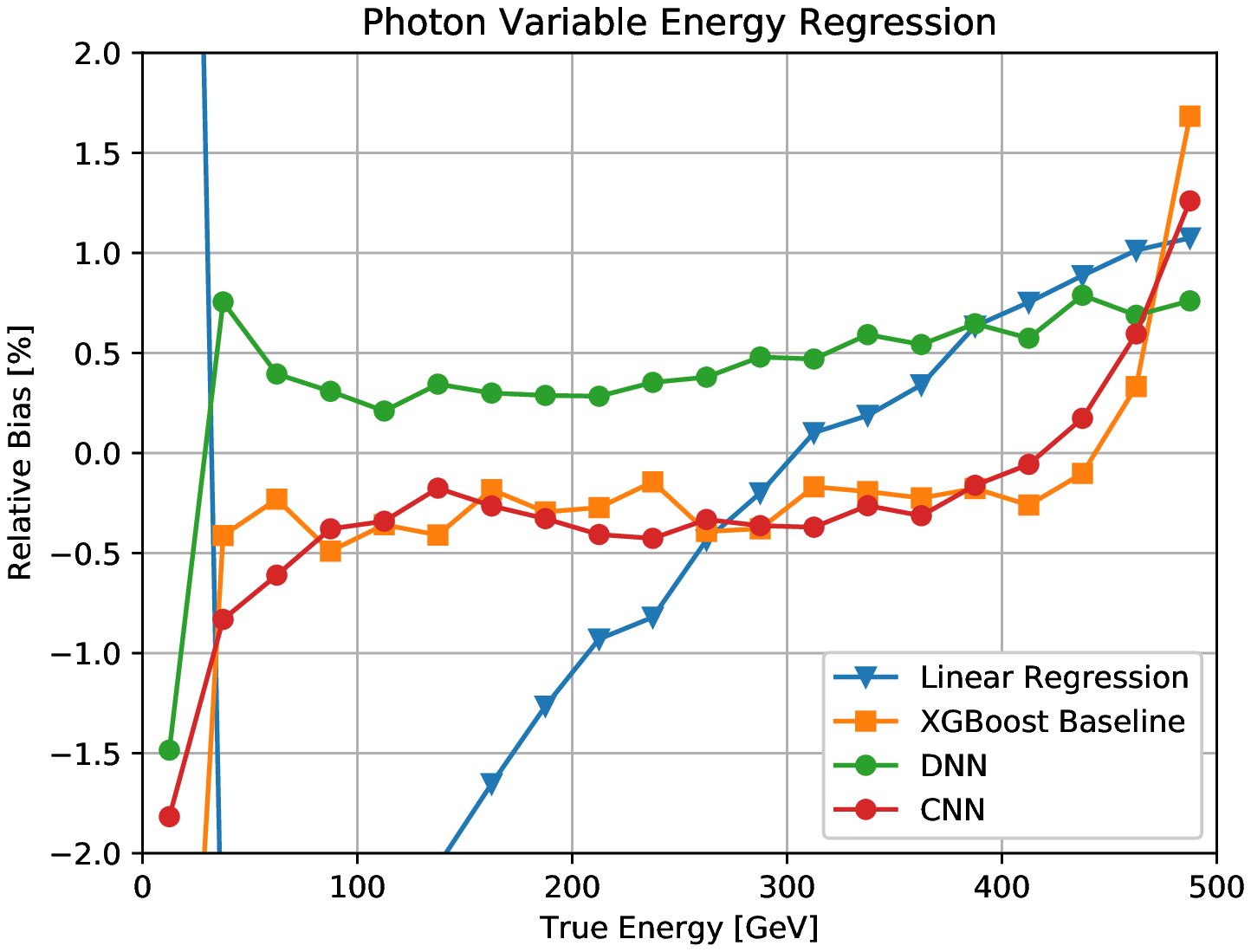}
\includegraphics[width=0.38\textwidth]{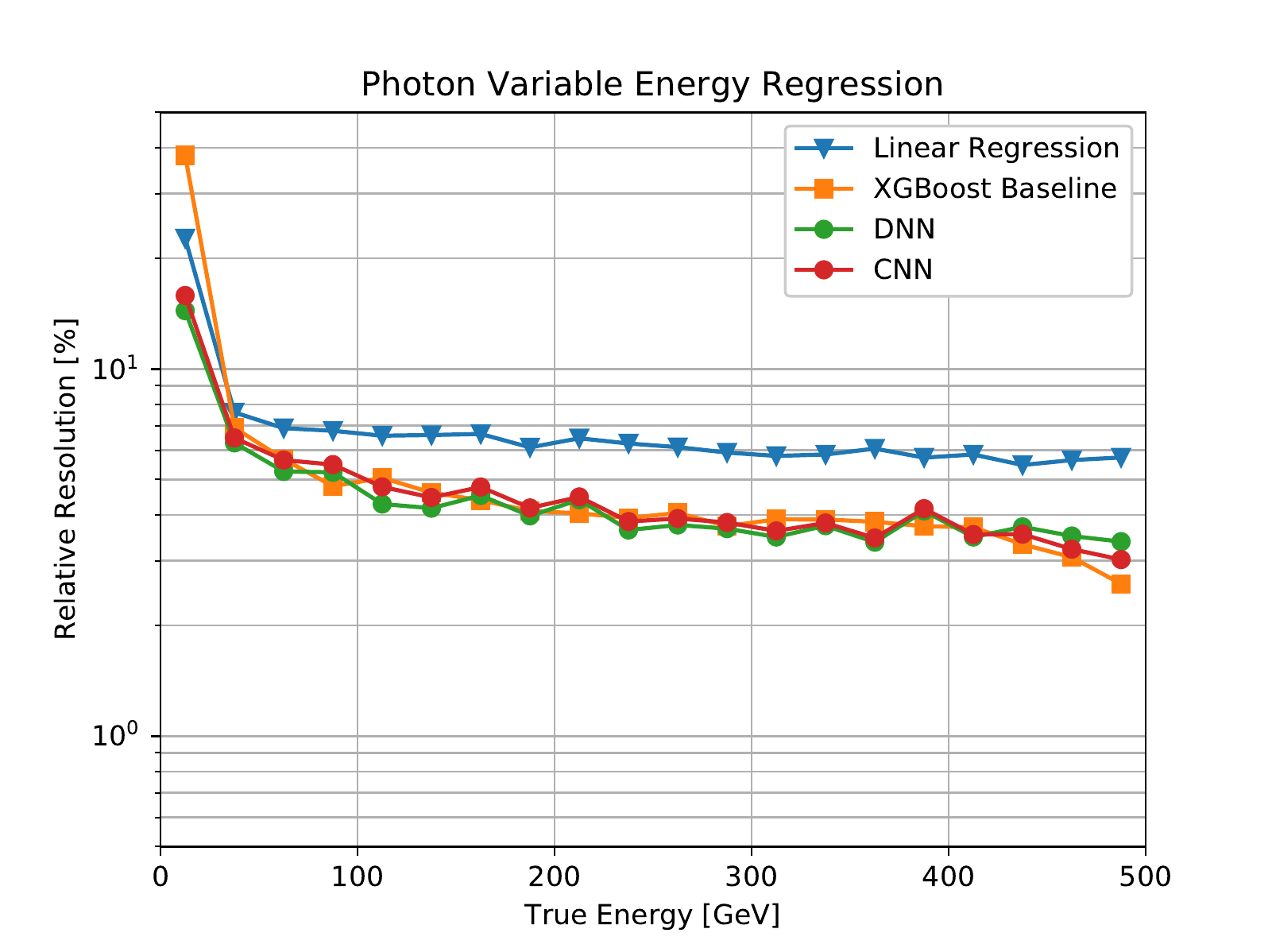}
\caption{Bias (left) and resolution (right) as a function of true energy for energy predictions for photons, on variable-angle samples resampled to ATLAS-like (top) and CMS-like (bottom) geometries.\label{fig:reg_resampled_gamma_ATLAS_CMS}}
\end{figure*}

\begin{figure*}[htbp]
\centering
\includegraphics[width=0.38\textwidth]{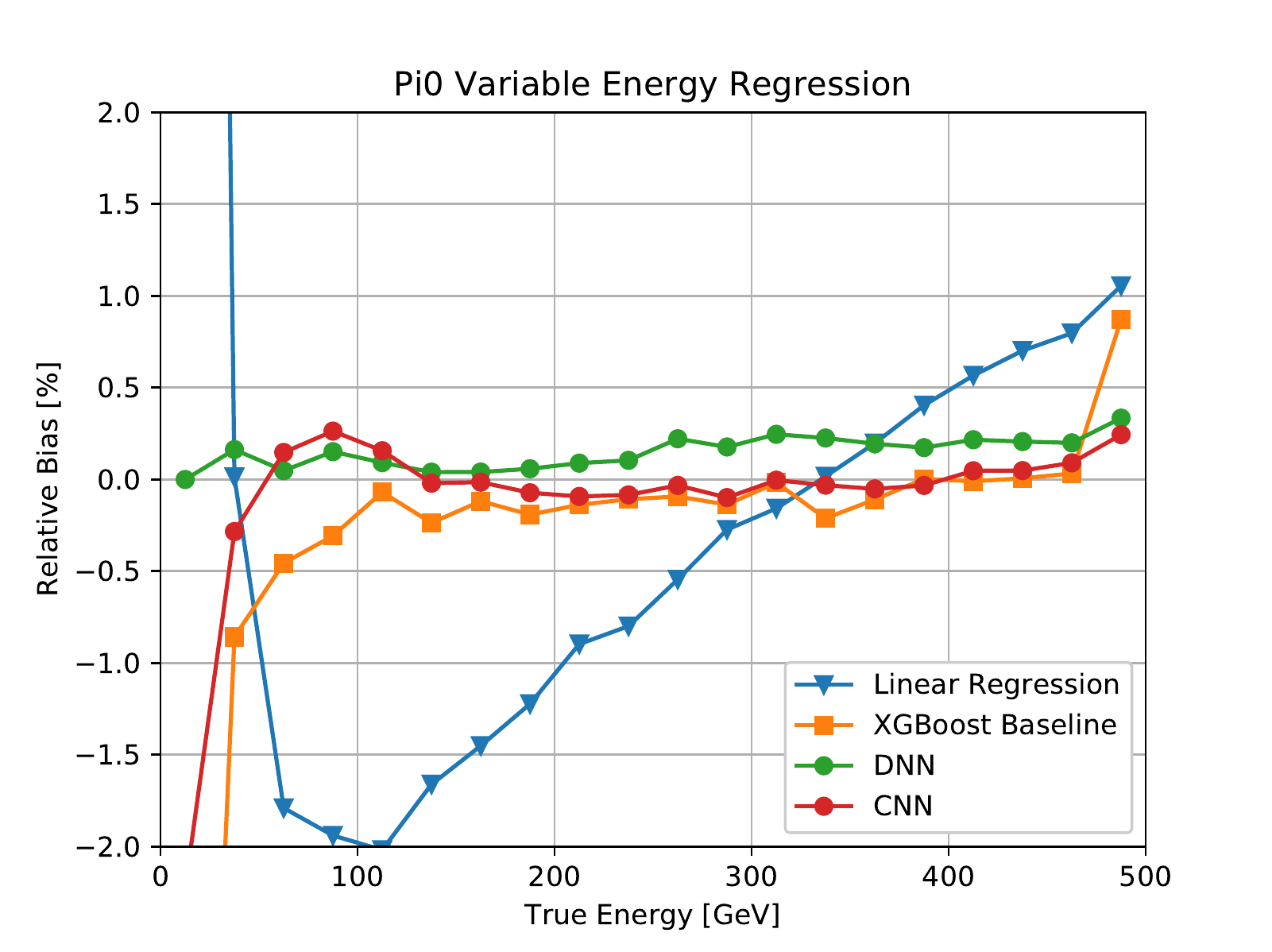}
\includegraphics[width=0.38\textwidth]{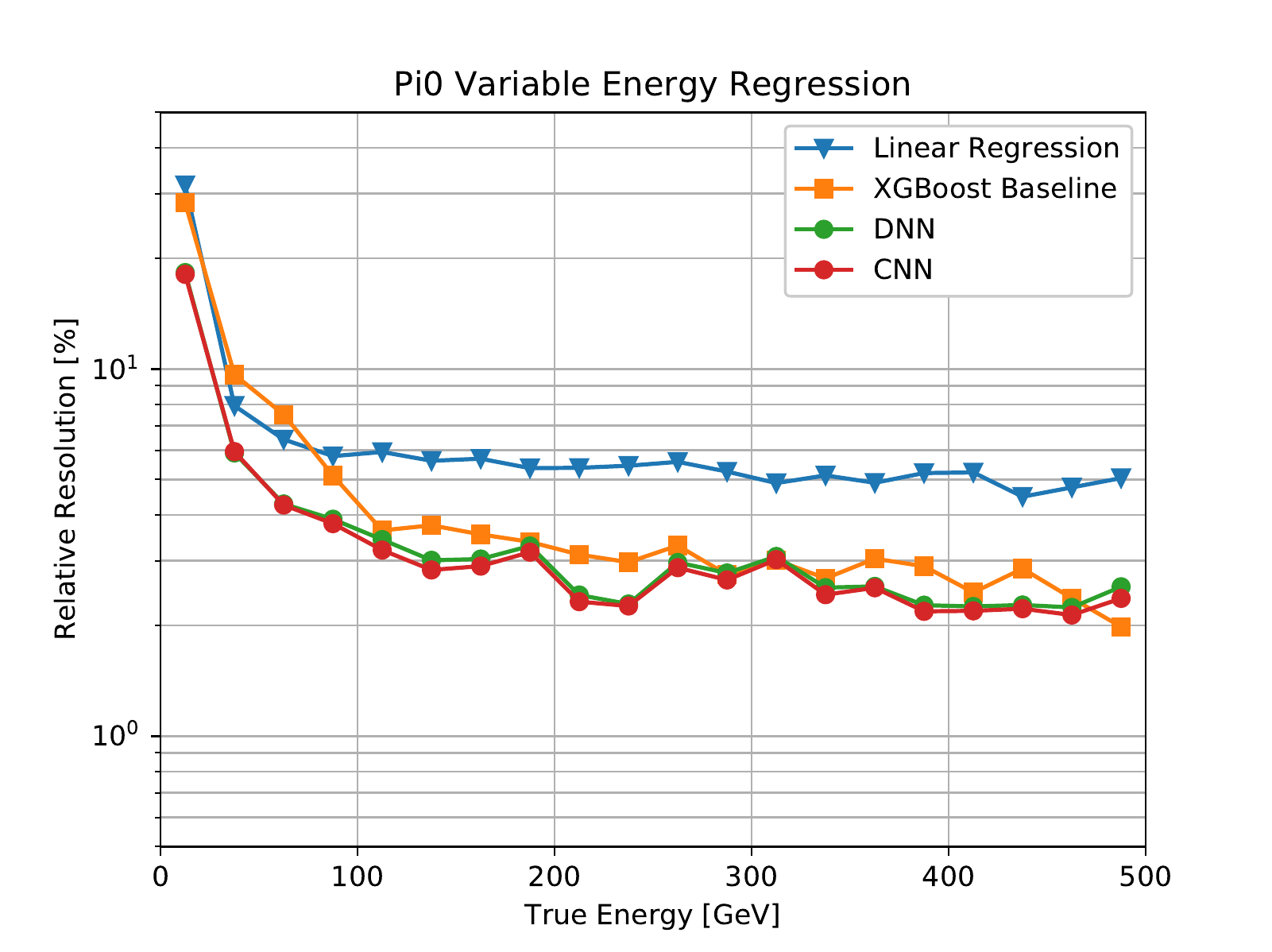} \\
\includegraphics[width=0.38\textwidth]{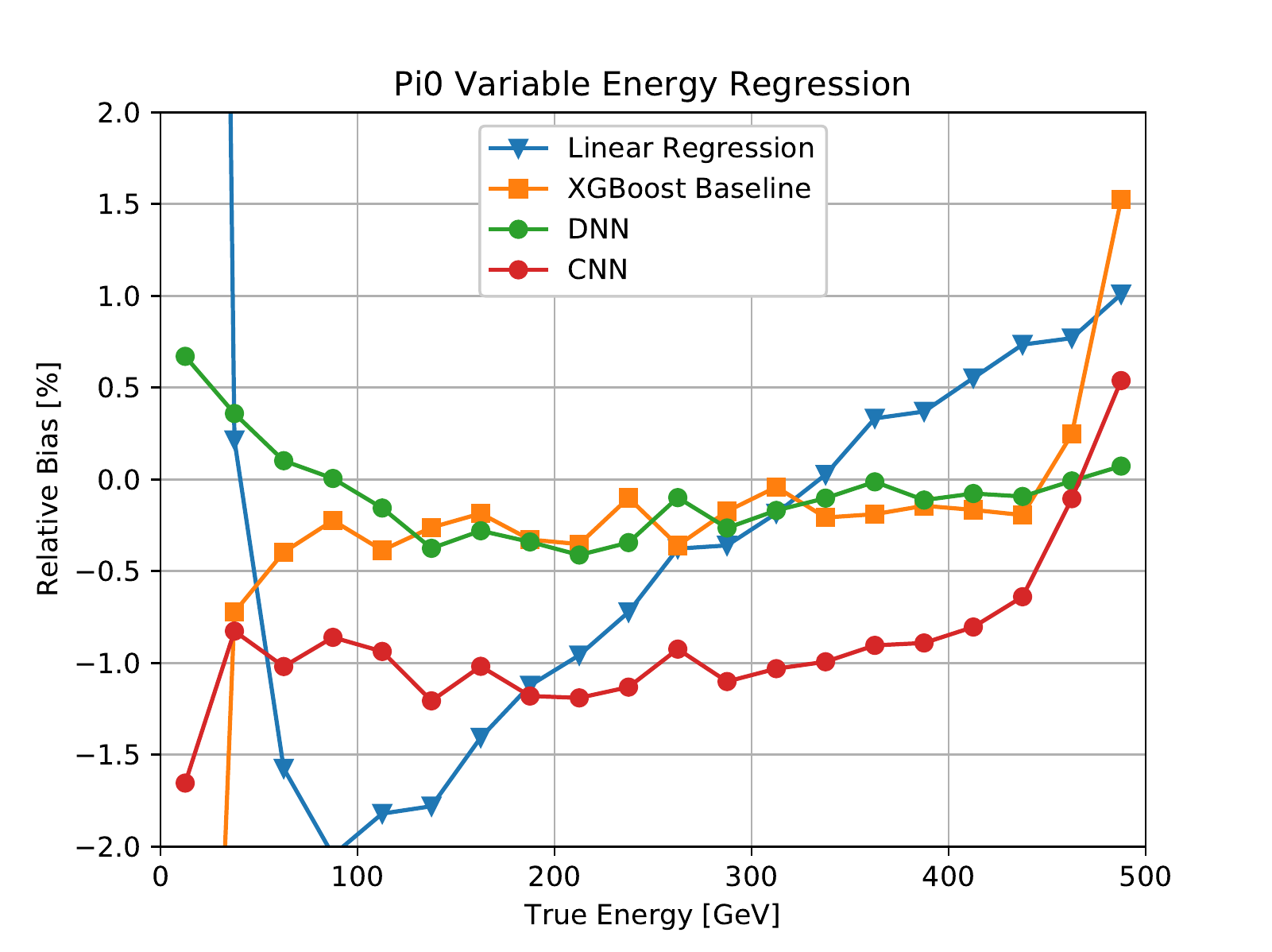}
\includegraphics[width=0.38\textwidth]{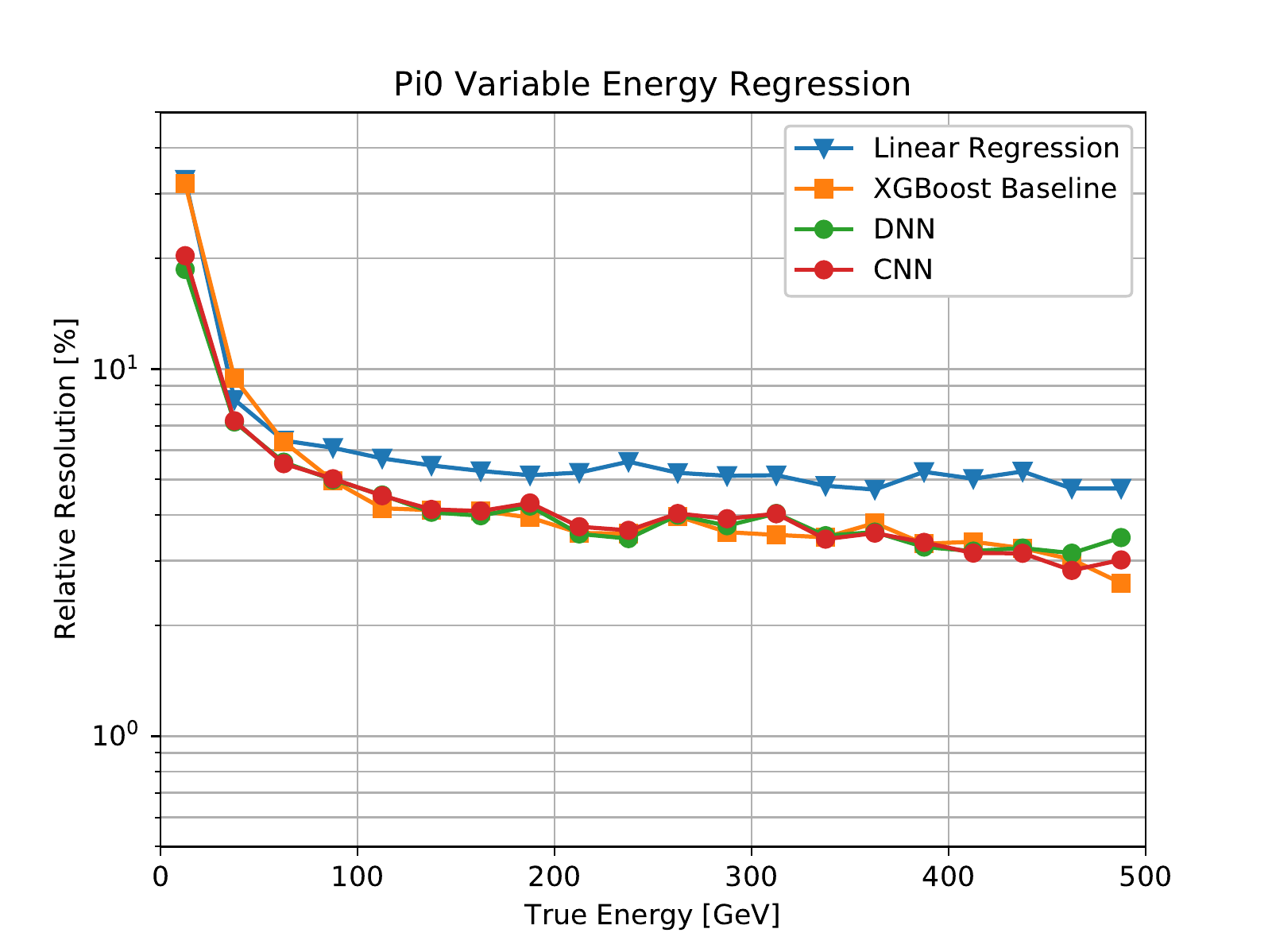}
\caption{Bias (left) and resolution (right) as a function of true energy for energy predictions for \pizero, on variable-angle samples resampled to  ATLAS-like (top) and CMS-like (bottom) geometries.\label{fig:reg_resampled_pi0_ATLAS_CMS}}
\end{figure*}

Regression results are shown in Figure~\ref{fig:reg_resampled_gamma_ATLAS_CMS} and~\ref{fig:reg_resampled_pi0_ATLAS_CMS}, for photons and neutral pions (we did not train electrons or charged pions for this comparison). Here we have included the regression baselines, DNN networks, and CNN networks, but not GN (which we did not train on resampled data). The results obtained for the ATLAS-like resampling match those on the REC dataset, with DNN and CNN matching the BDT outcome in terms of bias and surpassing it in resolution. With the CMS-like resampling the neural networks match but do not improves over the BDT energy regression resolution. Once again, this is due to the low spatial resolution in the CMS-like geometry, especially due to the lack of $z$ segmentation. We are unable to model the improved energy resolution from the actual CMS detector, so these energy regression results are based on geometry only.

\section{Conclusion and Future Work}\label{sec:conclusion}

This paper shows how deep learning techniques could outperform traditional and resource-consuming techniques in  tasks typical of physics experiments at particle colliders, such as particle shower simulation and reconstruction in a calorimeter.
We consider several model architectures, notably 3D convolutional neural networks, and we show competitive performance, matched to short execution time. In addition, this strategy comes with a GPU-friendly computing solution and would fit the current trends in particle physics towards heterogeneous computing platforms. 

We confirm findings from previous studies of this kind. On the other hand, we do so utilizing a fully accurate detector simulation, based on a complete GEANT4 simulation of a full particle detector, including several detector components, magnetic field, etc. In addition, we design the network so that different tasks are performed by a single architecture, optimized through an hyperparameter scan. 

We look forward to the development of similar solutions for current and future particle detectors, for which this kind of end-to-end solution could be extremely helpful. 
\section{Acknowledgemnts}

The authors thank Daniel Weitekamp for providing us with the event generator used in regression training. We also thank Andre Sailer from the CERN CLIC group, for guiding us on how to generate the single-particle samples.

This project is partially supported by the United States Department of Energy, Office of High Energy Physics Research under Caltech Contract No. DE-SC0011925. JR is partially supported by the Office of High Energy Physics HEP-Computation. M.~P. is supported by the European Research Council (ERC) under the European Union's Horizon 2020 research and innovation program (grant agreement n$^o$ 772369). This research is also partially supported by the Zhejiang University/University of Illinois Institute Collaborative Research Program (award 083650).

This research is part of the Blue Waters sustained-petascale computing project, which is supported by the National Science Foundation (awards OCI-0725070 and ACI-1238993) and the State of Illinois. Blue Waters is a joint effort of the University of Illinois at Urbana-Champaign and its National Center for Supercomputing Applications.

Part of this work was conducted at  "\textit{iBanks}", the AI GPU cluster at Caltech. We acknowledge NVIDIA, SuperMicro  and the Kavli Foundation for their support of "\textit{iBanks}".
The authors are grateful to  Caltech and the Kavli Foundation for their support of undergraduate student research in cross-cutting areas of machine learning and domain sciences.

\clearpage
\appendix

\section{Calorimeter Window Size}\label{app:window_size}

The optimal window size to store for ECAL and HCAL is an important issue, since this impacts not only sample storage size, but also training speed and the maximum batch sizes which we could feed to our GPUs. 

\begin{figure*}[htbp]
\centering
\includegraphics[width=0.38\textwidth]{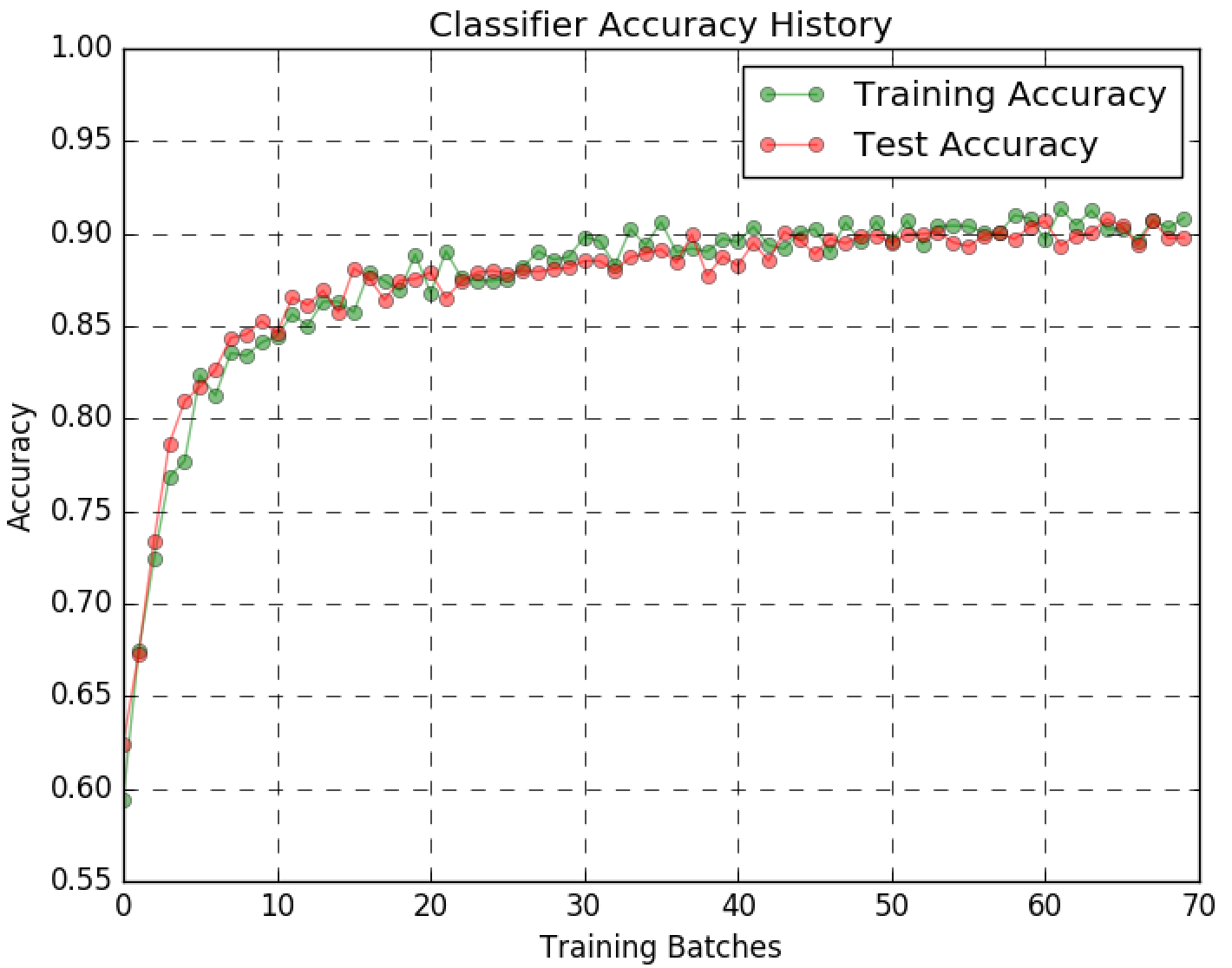}
\includegraphics[width=0.38\textwidth]{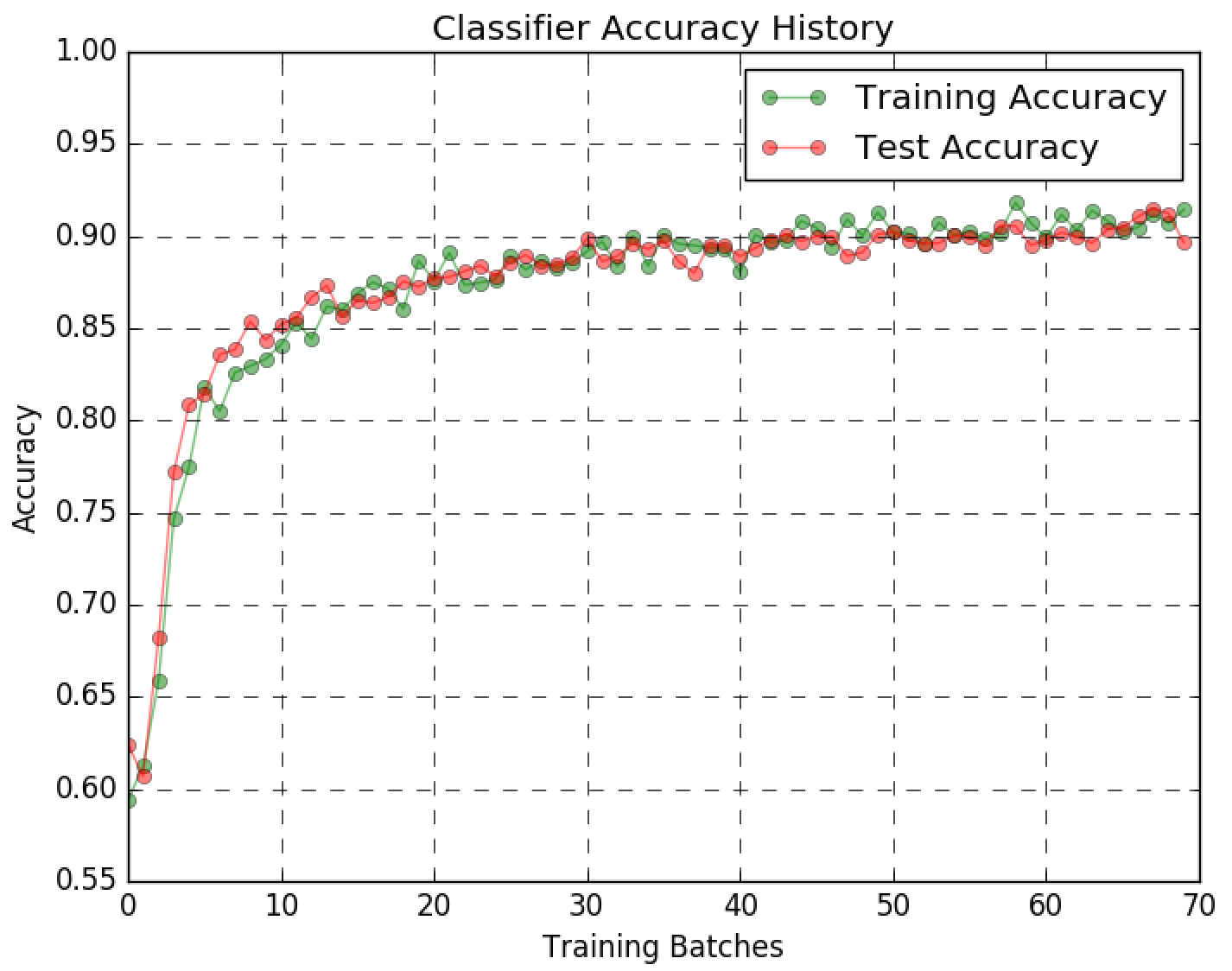}
\includegraphics[width=0.38\textwidth]{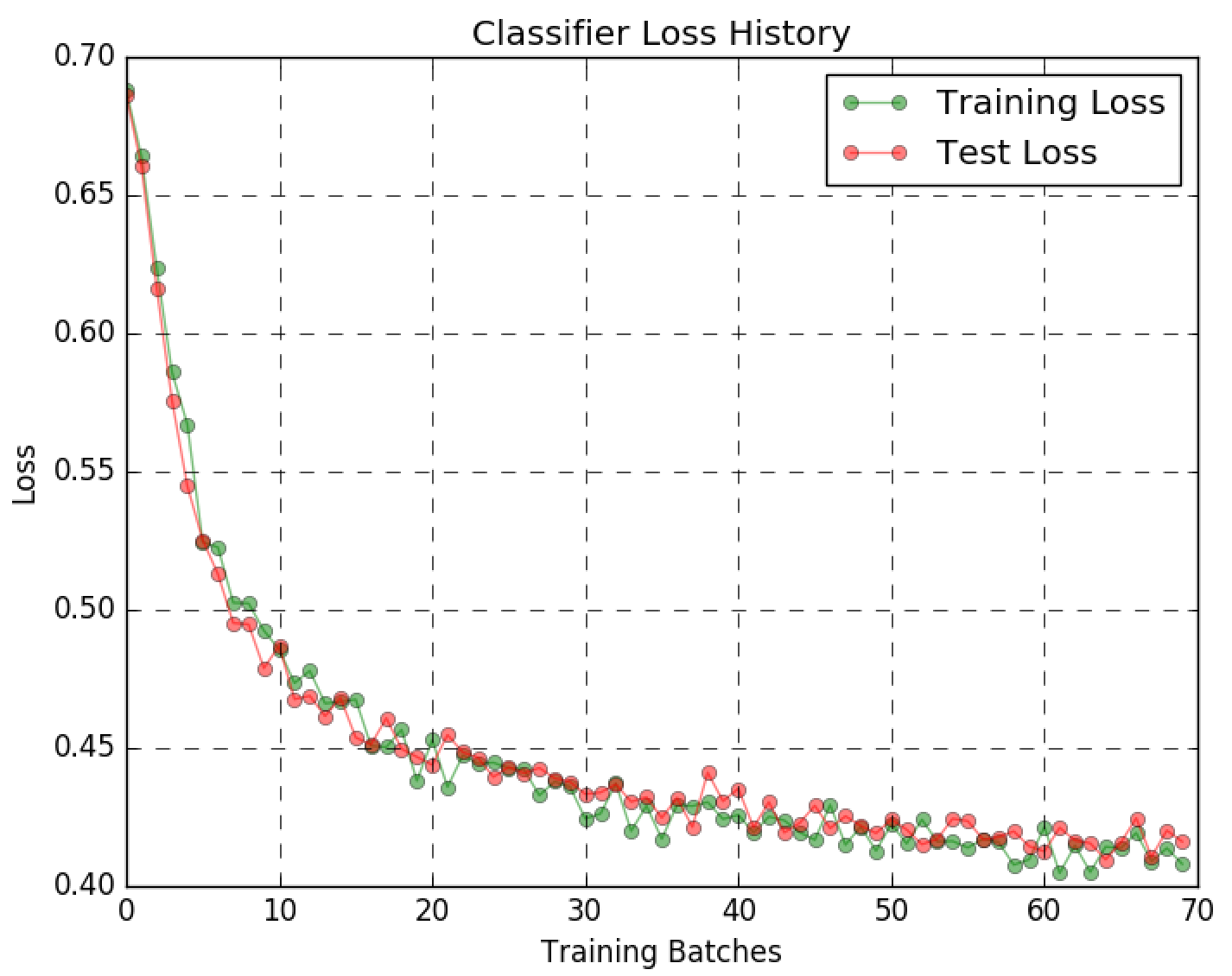}
\includegraphics[width=0.38\textwidth]{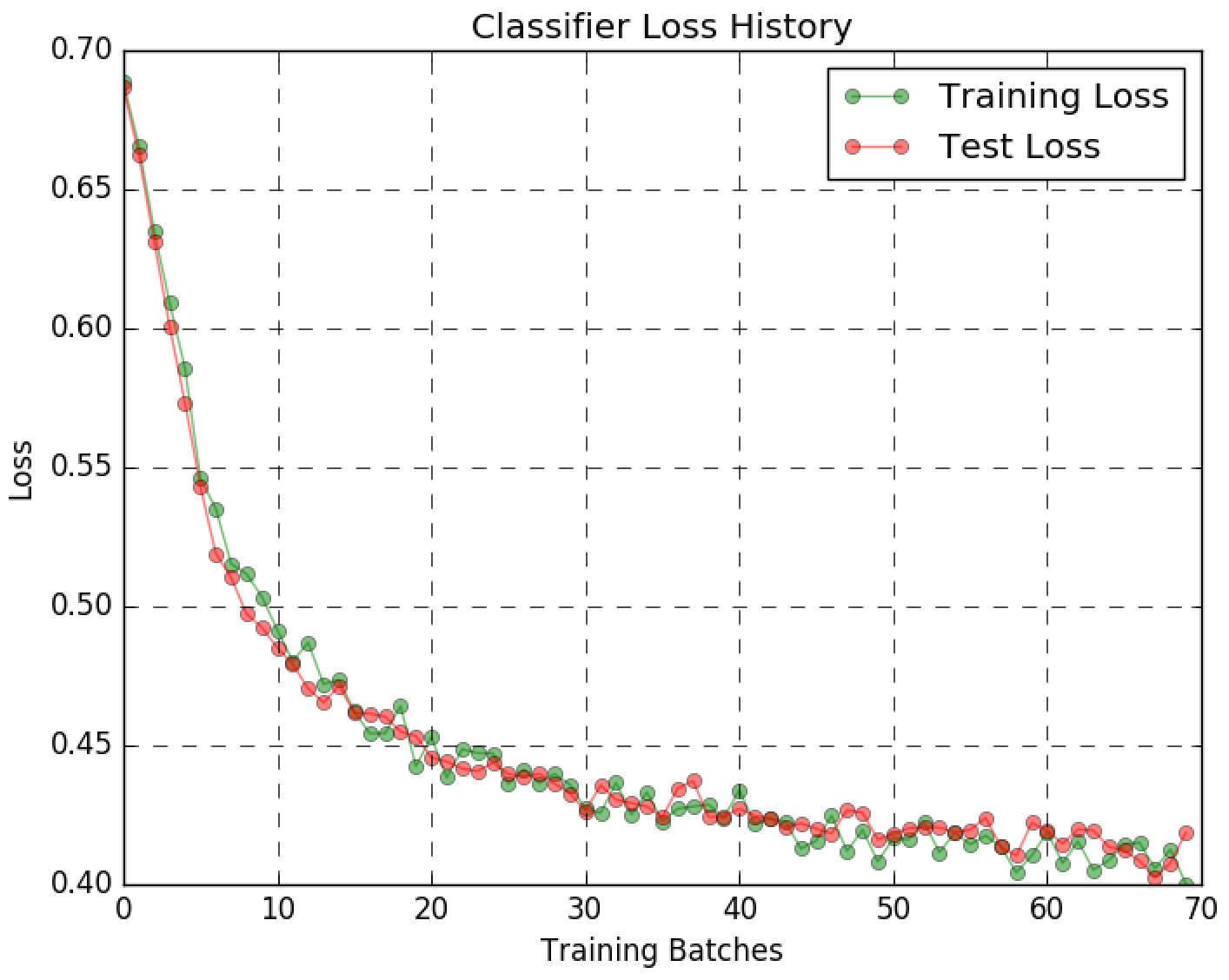}
\caption{Training history for different choices of the input 3D array zise: Accuracy (top) and loss (bottom) as a function of the training batch for photon/neutral pion classification, using a 25x25x25 (left) and 51x51x25 (right) ECAL window size.\label{fig:classification_window}}
\end{figure*}

From examinations of our generated samples, we found that an ECAL window of 25x25x51 and an HCAL window of 11x11x60 looked reasonable. To test this hypothesis, we performed training using the samples and classification architectures described in our previous studies~\cite{NIPS}, but with different-sized input samples. The architecture was altered to accommodate larger windows simply by increasing the number of neurons on the input layer. Results trained using an ECAL window of size 25x25x25 and 51x51x25 are shown in Figure~\ref{fig:classification_window}. From the similarity of these curves, we have decided that an expanded ECAL window size does not contain much additional useful information, and is thus not necessary for our problems.

\section{End-to-end reconstruction of the ECAL showers produced by the 3DGAN}
\label{appendix:RECO_on_GAN}

\begin{figure}
    \centering
    \includegraphics[scale=0.4, trim={0.5cm 0.1cm 0 1.1cm}, clip]{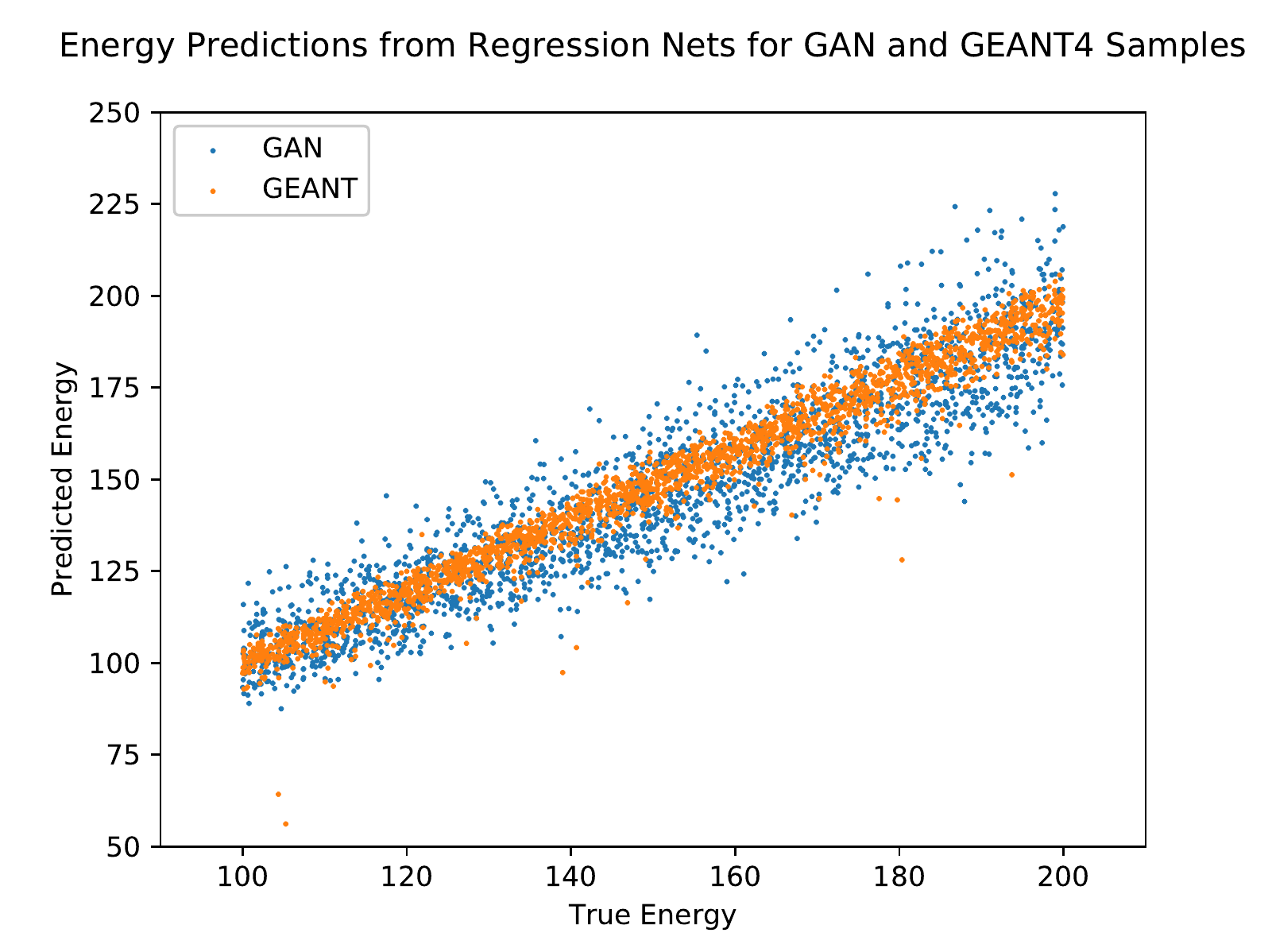}
    \caption{Predicted vs. true particle energy for GAN and GEANT
      images. Predictions were made using the reconstruction tool described in section~\ref{sec:reco}. This plot was made using 2213 electron events of each type (GAN and GEANT).\label{fig:GAN_regression}}
\end{figure}

\begin{figure}
    \centering
    \includegraphics[scale=0.4, trim={0.5cm 0.1cm 0 1.1cm}, clip]{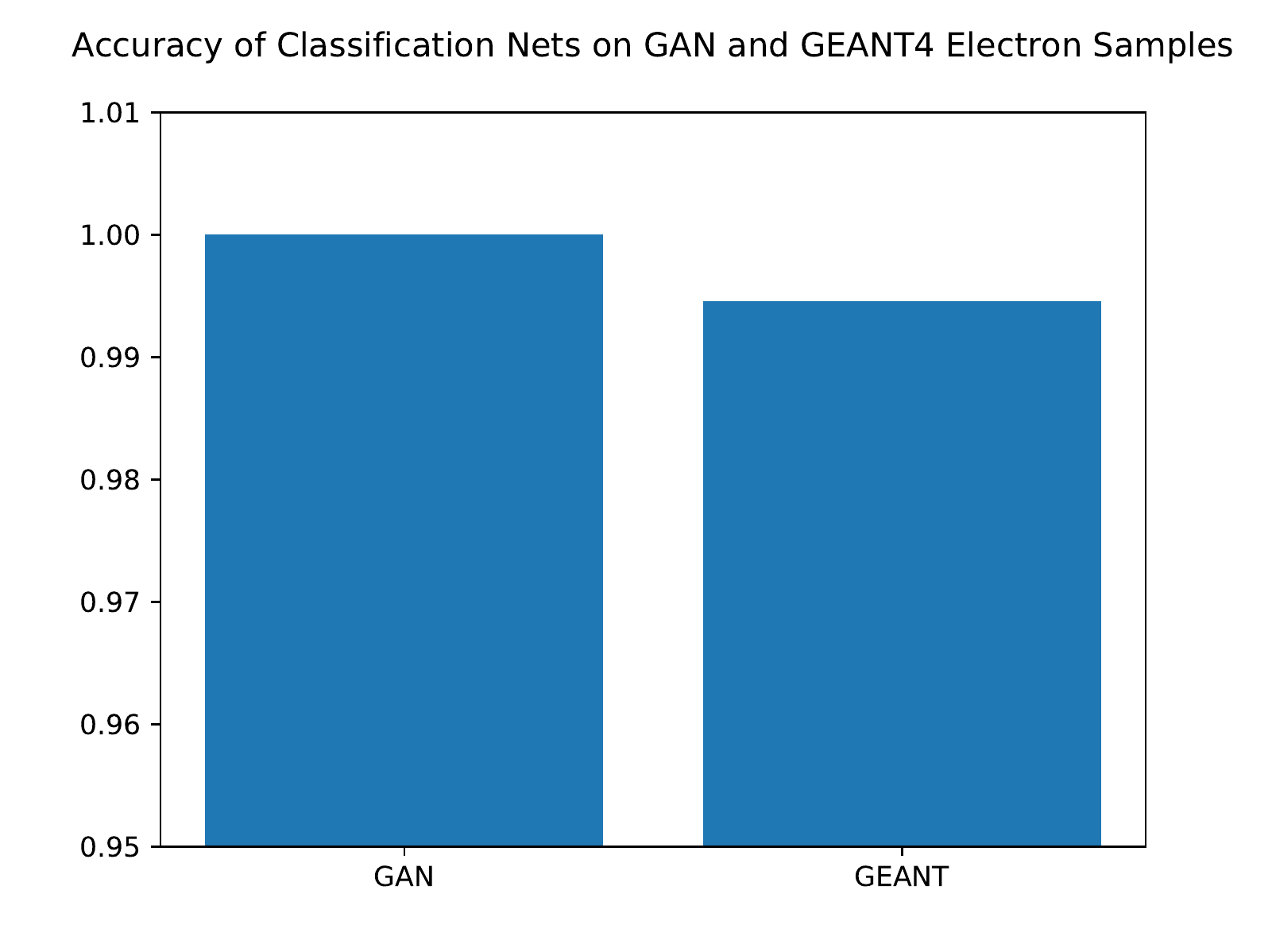}
    \caption{Predicted particle type (electron vs. charge pions) for GAN and GEANT images. There were 2213 electron events for each
      type.\label{fig:GAN_classification}}
\end{figure}

In order to further validate the GAN image quality we run the 3D CNN reconstruction network described in section~\ref{sec:reco} on the 3DGAN output and compare the response to the results obtained by running the tool on Monte Carlo data. Figure~\ref{fig:GAN_regression} shows a comparison of the energy resolution obtained on GAN and GEANT4 images. The predicted energy shows a reasonable agreement for the mean while the resolution for GAN images seems to be broader than for GEANT4 images. The classification accuracy presented in Figure~\ref{fig:GAN_classification} is very high (close to $100\%$) for both GAN and GEANT4 events. 

\section{Classification Baseline}\label{app:BDT}

Boosted Decision Trees were chosen as the baseline of comparison for our classification task, due to their popularity with HEP experiments. A BDT is effective in processing high-level features, performing complex and optimized cut-based classification in the multi-dimensional space of the input quantities. 

\begin{figure}[htbp]
\centering
\includegraphics[trim={0 0 0 1cm},clip,width=0.4\textwidth]{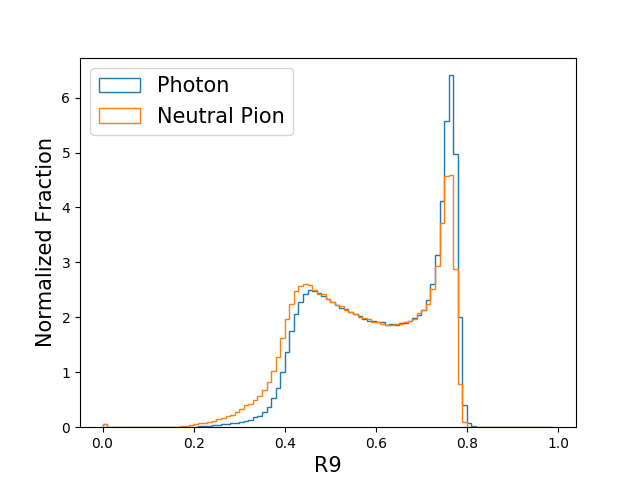}
\caption{Comparison of R9 distributions between photon and neutral pion events. In both cases, we have some events where energy is centralized, and some events where energy is "half-localized" (maybe split into two regions). Photons tend to deposit their energy more in a single location.
\label{fig:R9}}
\end{figure}
\begin{figure}[htbp]
\centering
\includegraphics[width=0.43\textwidth]{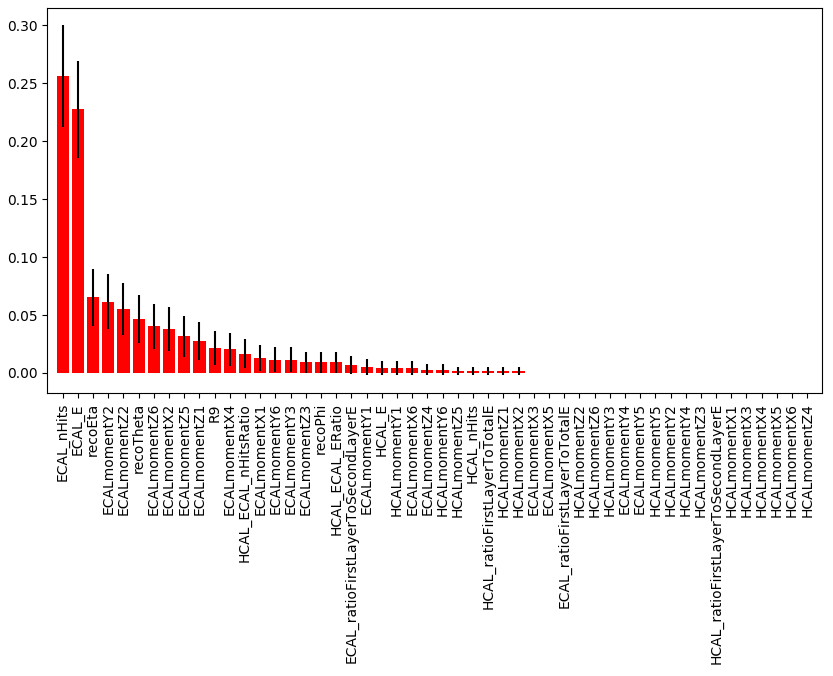}
\caption{Feature importances for inputs used in BDT training. Values shown are gini importances~\cite{Breiman}.\label{fig:BDT_ranking}}
\end{figure}

The features we use for our baseline BDT classification model, introduced in Ref.~\cite{NIPS}, are commonly used to characterize particle showers. One additional feature we added is R9, i.e., the fraction of energy contained in a 3x3 window of the $(x,y)$ projection of the shower centered around the energy barycenter. This quantity provides a measure of the "concentration" of a shower within a small region. For values near 1, the shower is highly collimated within a single region, as in electromagnetic showers. Smaller values are typical of more spread out showers, as for hadronic and multi-prong showers. A comparison of R9 values between photons and neutral pions can be seen in Figure~\ref{fig:R9}. After training, the discriminating power of various features can be seen in Figure~\ref{fig:BDT_ranking}.

\section{Energy Regression Baseline}\label{app:regression_baseline}

We use linear regression with ECAL and HCAL total energy as one of our baseline methods to compare to machine learning results (seen in Eq.~\ref{eq:linreg}).

\begin{equation}
E = a \cdot E_{ECAL} + b \cdot E_{HCAL} + c
\label{eq:linreg}
\end{equation}

\begin{figure}[htbp]
\centering
\includegraphics[width=0.38\textwidth]{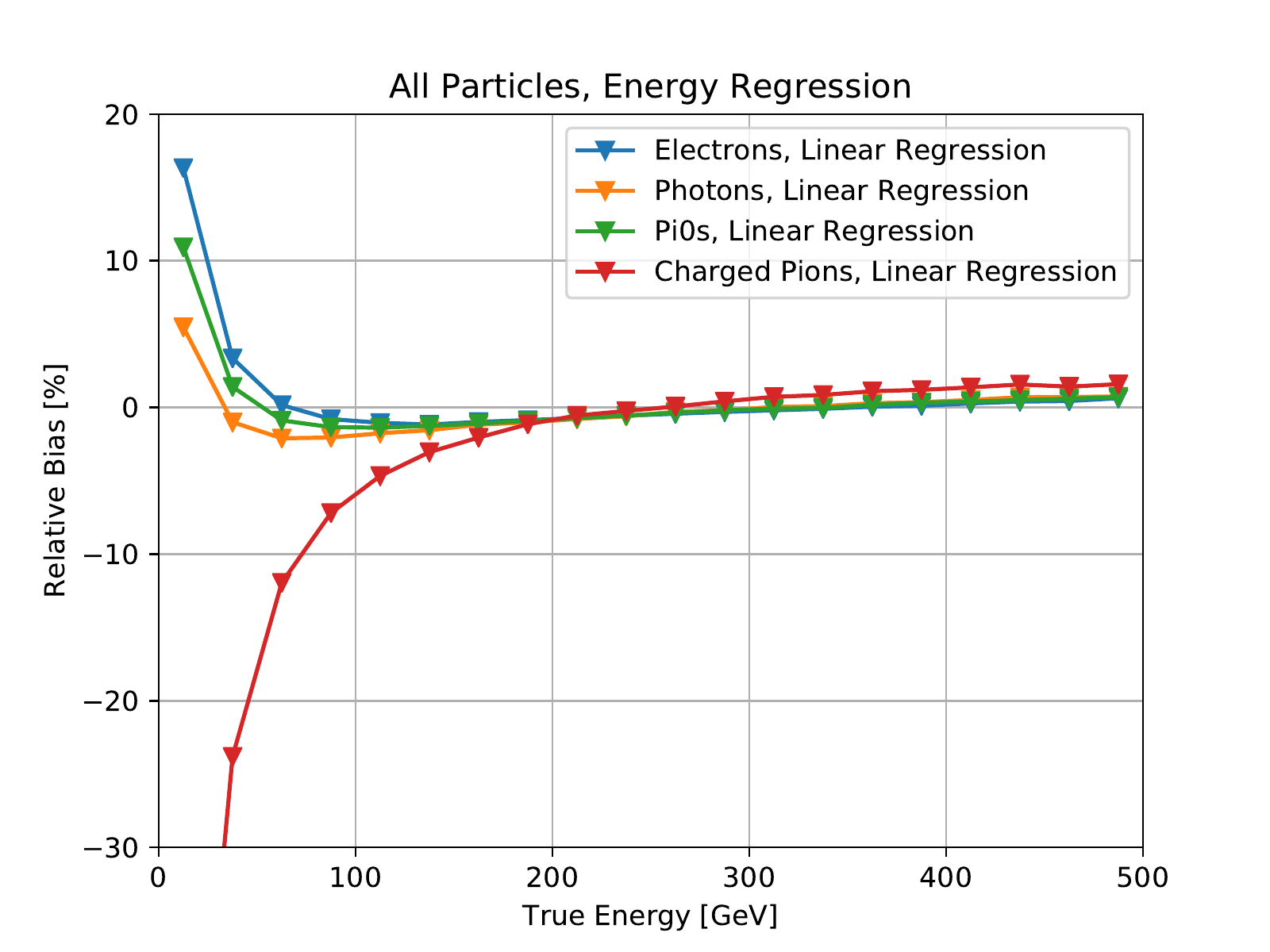}
\includegraphics[width=0.38\textwidth]{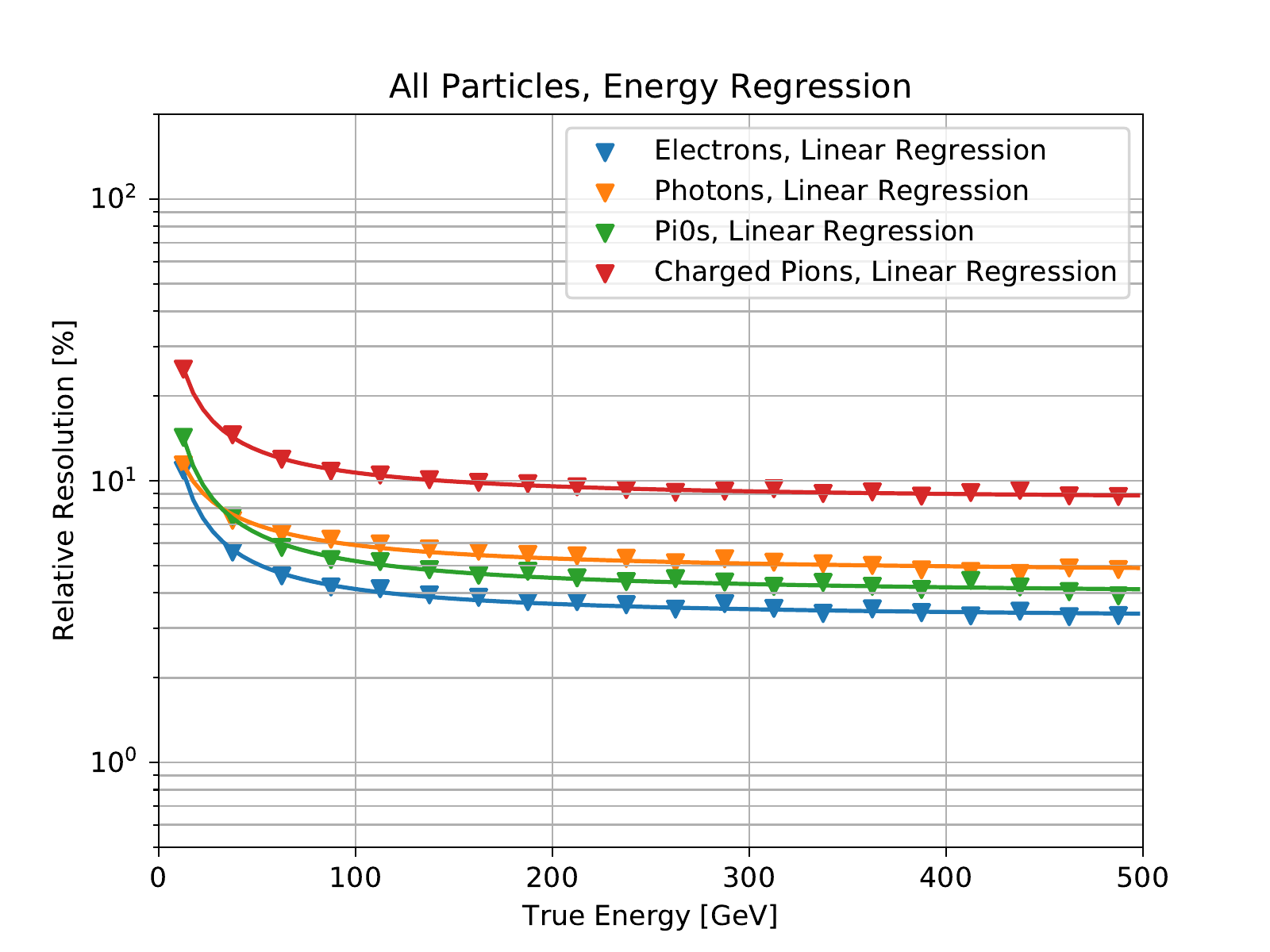}
\caption{Bias (top) and resolution (bottom) as a function of true energy for linear regression predictions of particle energy for the different particle types, trained on fixed-angle samples. \label{fig:reg_linreg}}
\end{figure}

Updated results for each of the particle types are shown in Figure~\ref{fig:reg_linreg}. In all the resolution plots shown, the points have been fitted with the expected resolution function of Eq.~\ref{eq:res}, and the fitted function is plotted as a line.

\begin{equation}
\frac{\sigma(\Delta E)}{E_{\text{true}}} = \frac{a}{\sqrt{E_{\text{true}}}} \oplus b \oplus \frac{c}{E_{\text{true}}}
\label{eq:res}
\end{equation}

It is already typical for basic ML methods like BDTs to be used for energy regression in the LHC experiments, in cases where the best resolution is critical (e.g., to study $H \rightarrow \gamma\gamma$ decays).  We tried a BDT with a few summary features as input to form an improved baseline for comparing more advanced ML techniques.  The XGBoost package was used in python, with the following hyperparameters.  
\begin{itemize}
\item maximum 1000 iterations, with early stopping if loss doesn't improve on the test set in 10 iterations
\item maximum tree depth of 3
\item minimum child weight of 1 (default)
\item learning rate $\eta = 0.3$ (default)
\end{itemize}
Varying the hyperparameters led to either worse results or negligible changes.

The following features gave good performance for electrons, photons, and \pizero:
\begin{itemize}
\item total ECAL energy
\item total HCAL energy
\item mean $z$ coordinate of the ECAL shower
\end{itemize}

\begin{figure}[htbp]
\centering
\includegraphics[width=0.38\textwidth]{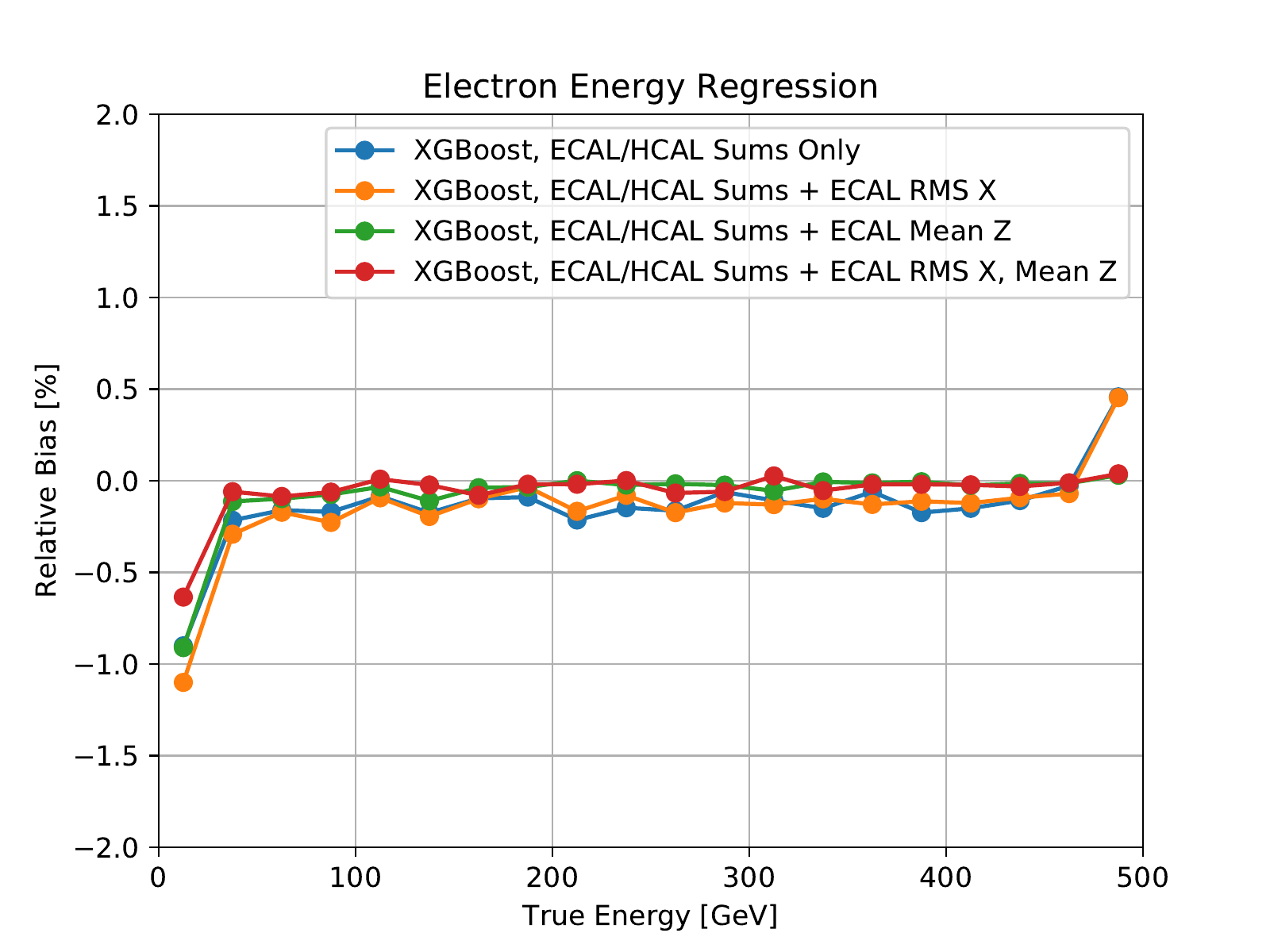}
\includegraphics[width=0.38\textwidth]{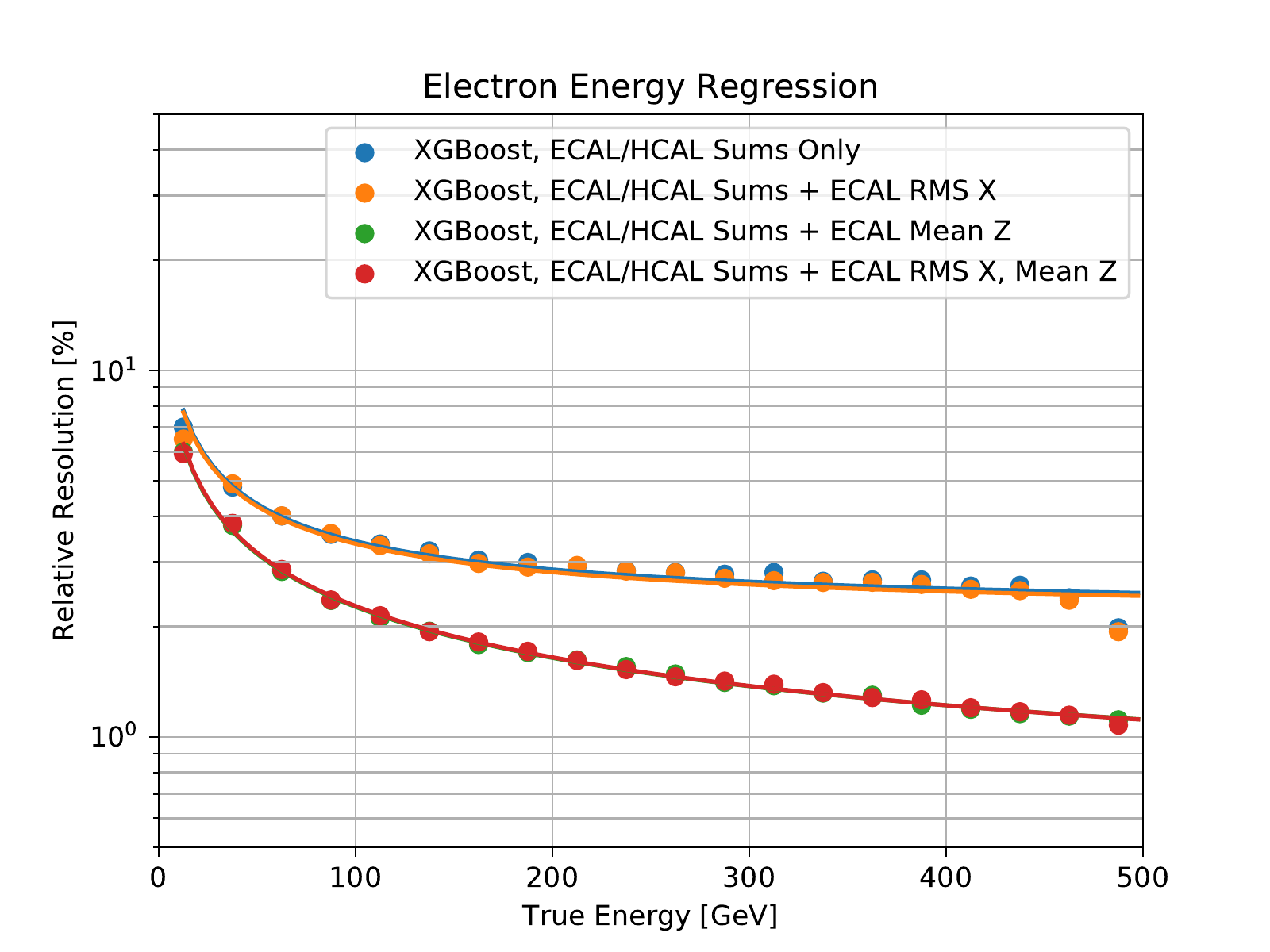}
\caption{Bias (top) and resolution (bottom) as a function of true energy for the XGBoost regression predictions of particle energy, using different input features for electrons.\label{fig:reg_xgb_ecalmoms}}
\end{figure}

Adding the mean $z$ coordinate to the ECAL and HCAL total energies improved the energy resolution for all energy values, but in particular at high energy. This is shown in Figure~\ref{fig:reg_xgb_ecalmoms} for electrons.

For \chpi, adding the following variables gave an improved result:
\begin{itemize}
\item RMS in the $x$ direction of the ECAL shower
\item RMS in the $(x,y)$ plane of the HCAL shower
\item mean $z$ coordinate of the HCAL shower
\end{itemize}

In addition, for \chpi, around 0.5\% of events were found to have almost no reconstructed energy in the selected calorimeter window.  Including these events adversely affected the algorithm training, so they were removed for all the results shown in this and the following sections. Specifically, the raw ECAL+HCAL energy is required to be at least 30\% of the true generated energy.

\begin{figure}[htbp]
\centering
\includegraphics[width=0.38\textwidth]{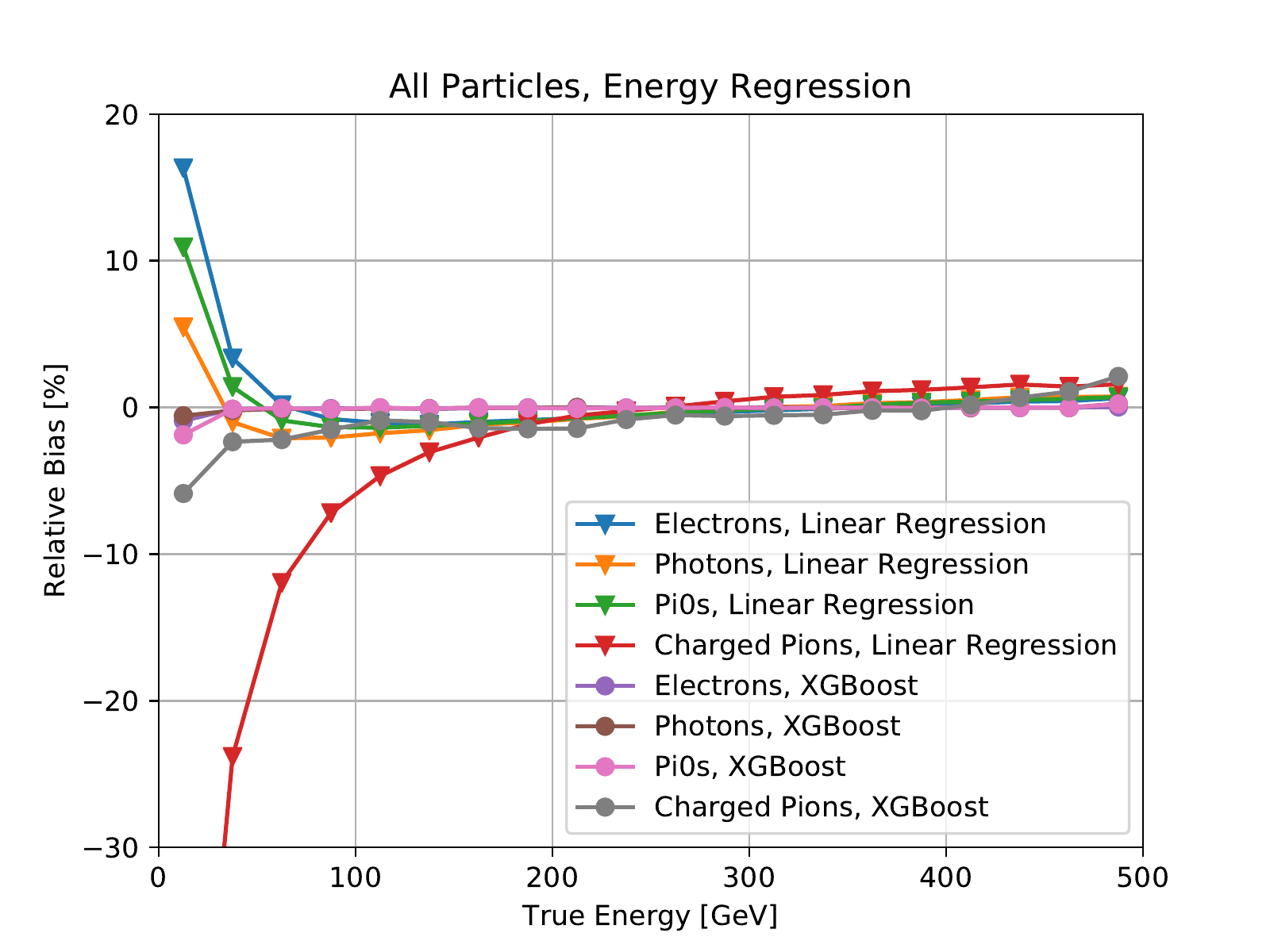}
\includegraphics[width=0.38\textwidth]{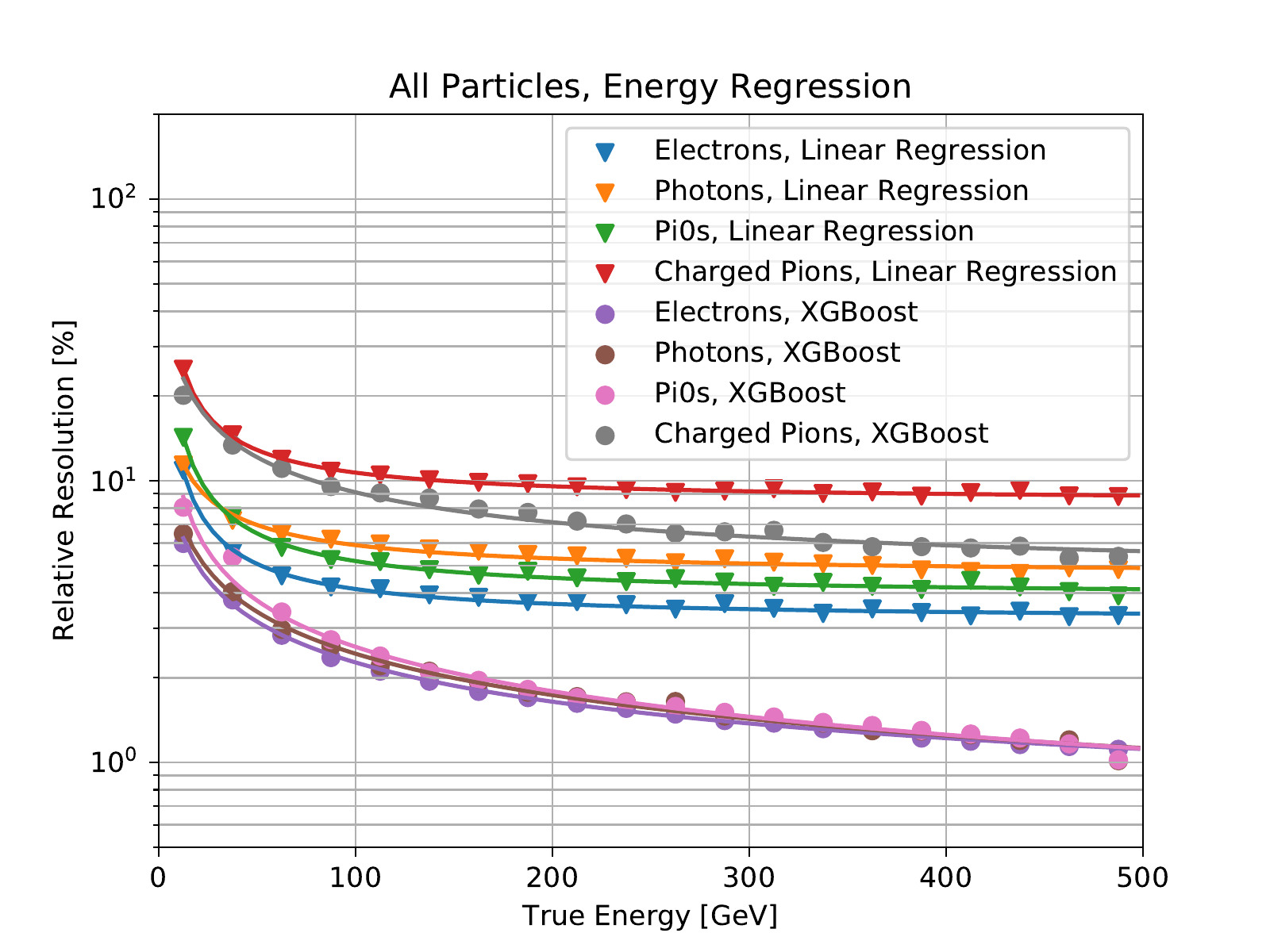}
\caption{Bias (top) and resolution (bottom) as a function of true energy for linear regression and XGBoost  predictions of particle energy for the different particle types.\label{fig:reg_xgb_linreg}}
\end{figure}

The results of the XGBoost baseline are shown in Figure~\ref{fig:reg_xgb_linreg}, where they are compared to linear regression results.
The performance of XGBoost on electrons, photons, and \pizero\ is similar, achieving relative resolutions of about 6--8\% at the lowest energies and 1.0--1.1\% at the highest energies.  Compared to the baseline linear regression, the resolution improves by a factor of about two at low energy and three to four at high energy.  For \chpi, the resolution after XGBoost regression ranges between 20 and 5.4\%, with a relative improvement over linear regression of up to 40\% at high energy.


One drawback of using a BDT algorithm in a real-world setting is that it can not be used for energy values outside the range of the training set. That is, most tree algorithms do not perform extrapolation. This can be mitigated by increasing the range of values present in the training set, or by using another algorithm like a neural network.  A small DNN was trained with the features above and was able to achieve similar performance to the BDT, with the same performance at high energy and slightly worse performance at the lowest energies.

\section{GoogLeNet Model Architecture Details}\label{app:GoogLeNet}

In our GoogLeNet architecture, we use inception modules. In these modules, inputs go through four separate branches and are then concatenated together. For an inception layer denoted as Inception(A, B, C, D, E, F, G) the branches are defined as follows:

\begin{itemize}
    \item Branch 1: A simple $1 \times 1 \times 1$ convolution, taking A input channels to B output channels. This is followed by a batch normalization and a ReLU activation function.
    \item Branch 2: A $1 \times 1 \times 1$ convolution followed by a $3 \times 3 \times 3$ convolution. The first convolution takes A input channels to C output channels, followed by batch normalization and ReLU. This then goes to the next convolution layer, which outputs D channels using a kernel of size $3 \times 3 \times 3$. This is again followed by batch normalization and ReLU.
    \item Branch 3: A $1 \times 1 \times 1$ convolution followed by a $5 \times 5 \times 5$ convolution. The details are the same as for the other branches, but the first convolution takes A input channels to E output channels, and the next convolution outputs F channels.
    \item Branch 4: A max pool of kernel size $3 \times 3 \times 3$ is followed by a convolution of kernel size $1 \times 1 \times 1$ that takes A input channels to G output channels. This is followed once again by batch normalization and ReLU.
\end{itemize}

Here are full details for each layer of the GoogLeNet-based architecture:

\begin{itemize}
    \item Apply instance normalization to ECAL input.
    \item Convolution with 3D kernel of size 3, going from 1 input channel to 192 channels, with a padding of 1. This is followed by batch normalization and ReLU.
    \item Inception(192,  64,  96, 128, 16, 32, 32)
    \item Inception(256, 128, 128, 192, 32, 96, 64)
    \item Max pooling with a 3D kernel of size 3, a stride of 2, and padding of 1.
    \item Inception(480, 192,  96, 208, 16,  48,  64)
    \item Inception(512, 160, 112, 224, 24,  64,  64)
    \item Inception(512, 128, 128, 256, 24,  64,  64)
    \item Inception(512, 112, 144, 288, 32,  64,  64)
    \item Inception(528, 256, 160, 320, 32, 128, 128)
    \item Max pooling with a 3D kernel of size 3, a stride of 2, and padding of 1.
    \item Inception(832, 256, 160, 320, 32, 128, 128)
    \item Inception(832, 384, 192, 384, 48, 128, 128)
    \item Average pooling with a 3D kernel of size 7 and a stride of 1.
    \item The output array is flattened and concatenated with input $\phi$, $\eta$, total ECAL energy, and total HCAL energy.
    \item A densely connected layer with 1024 outputs, followed by ReLU.
    \item The output array is once again concatenated with the same input values.
    \item A final densely connected layer outputs 5 values, as in the architectures of the other two models.
\end{itemize}

\section{Use of HCAL in Classification}\label{app:classification_HCAL}

\begin{figure}[htbp]
\centering
\includegraphics[width=0.38\textwidth]{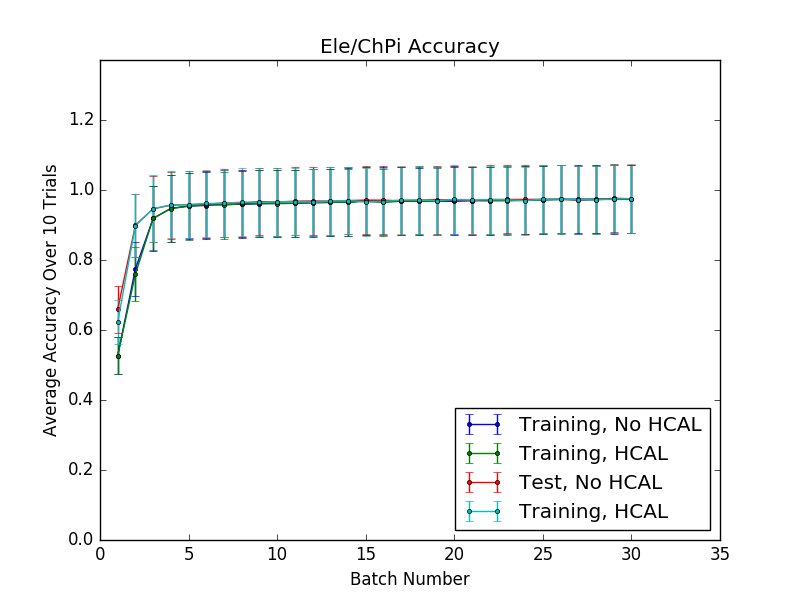}
\includegraphics[width=0.38\textwidth]{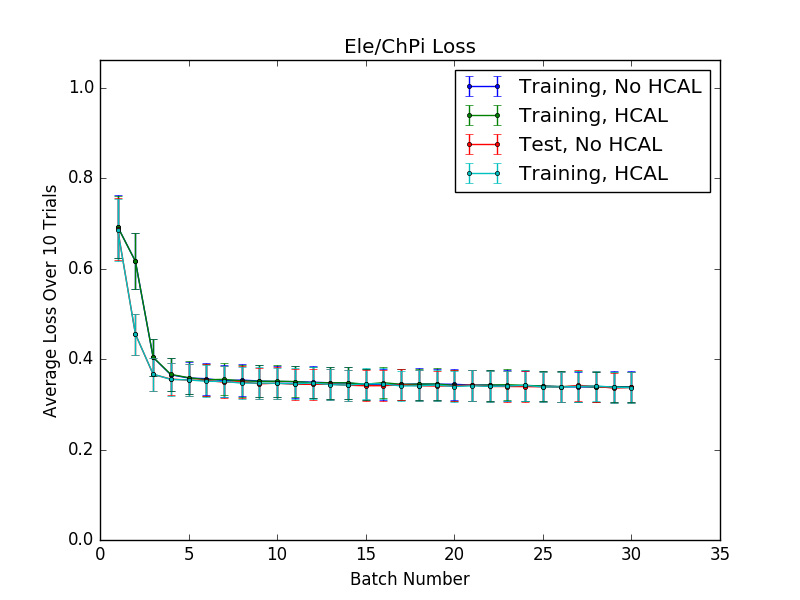}
\caption{Accuracy and loss curves for electron/charged pion classification, with and without HCAL cells, using best DNN architecture.}
\label{fig:HCAL_study_elechpi}
\end{figure}

\begin{figure}[htbp]
\centering
\includegraphics[width=0.38\textwidth]{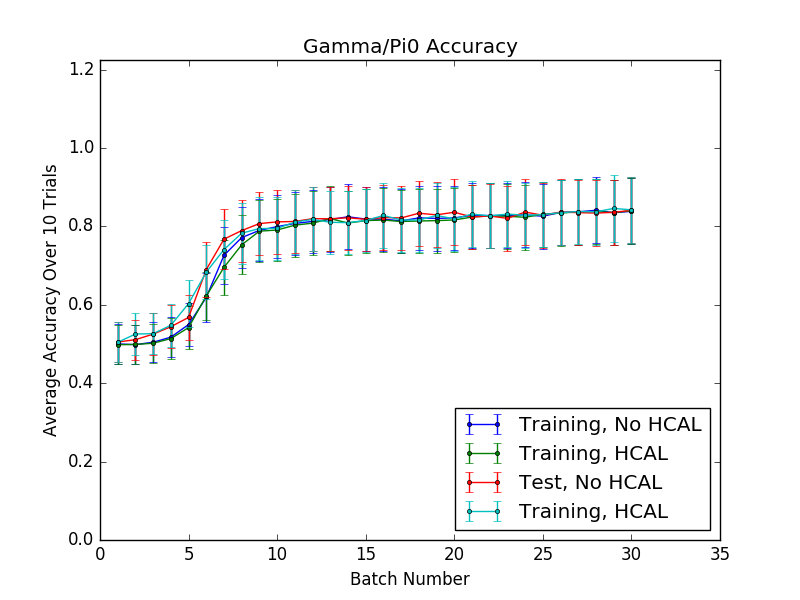}
\includegraphics[width=0.38\textwidth]{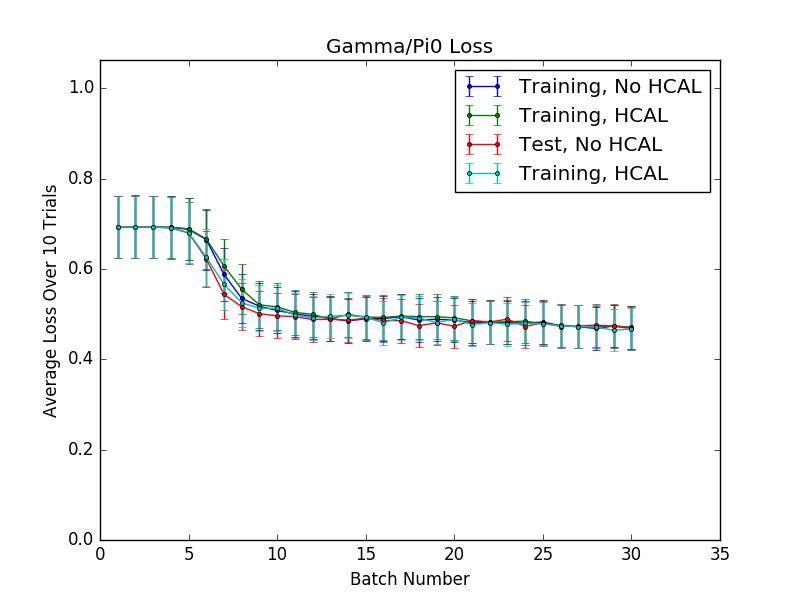}
\caption{Accuracy and loss curves for photon/neutral pion classification, with and without HCAL cells, using best DNN architecture.}
\label{fig:HCAL_study_gammapi0}
\end{figure}

Since the GoogLeNet architecture was quite large and required significant memory usage and computational power, we decided to investigate the possibility of leaving out HCAL cell-level information, since most of the particle shower occurs in the ECAL. Using our best-performing DNN architecture, we ran ten training sessions with HCAL information, and ten training sessions without HCAL. Averaged training curves from these runs are shown in Figures~\ref{fig:HCAL_study_elechpi} and~\ref{fig:HCAL_study_gammapi0}. These studies demonstrated that including the HCAL caused little to no improvement in classification accuracy. For memory purposes, we thus kept HCAL cell-level information out of our GN architecture. Summed HCAL energy was still fed as an input to the combined classification-regression net, for use in energy regression.

\section{Skip Connections for Regression}\label{app:skip_connections}

\begin{figure}[htbp]
\centering
\includegraphics[width=0.38\textwidth]{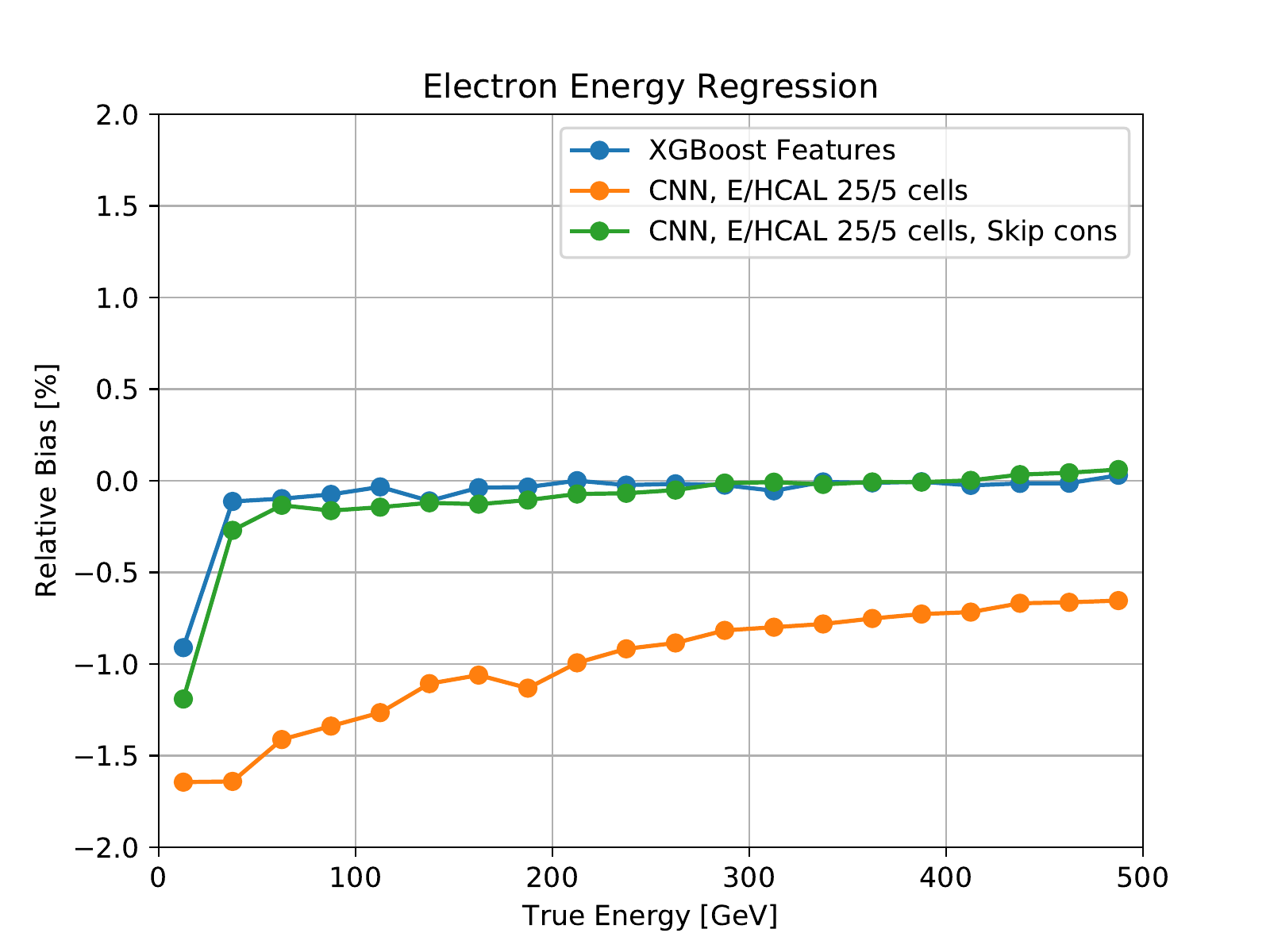}
\includegraphics[width=0.38\textwidth]{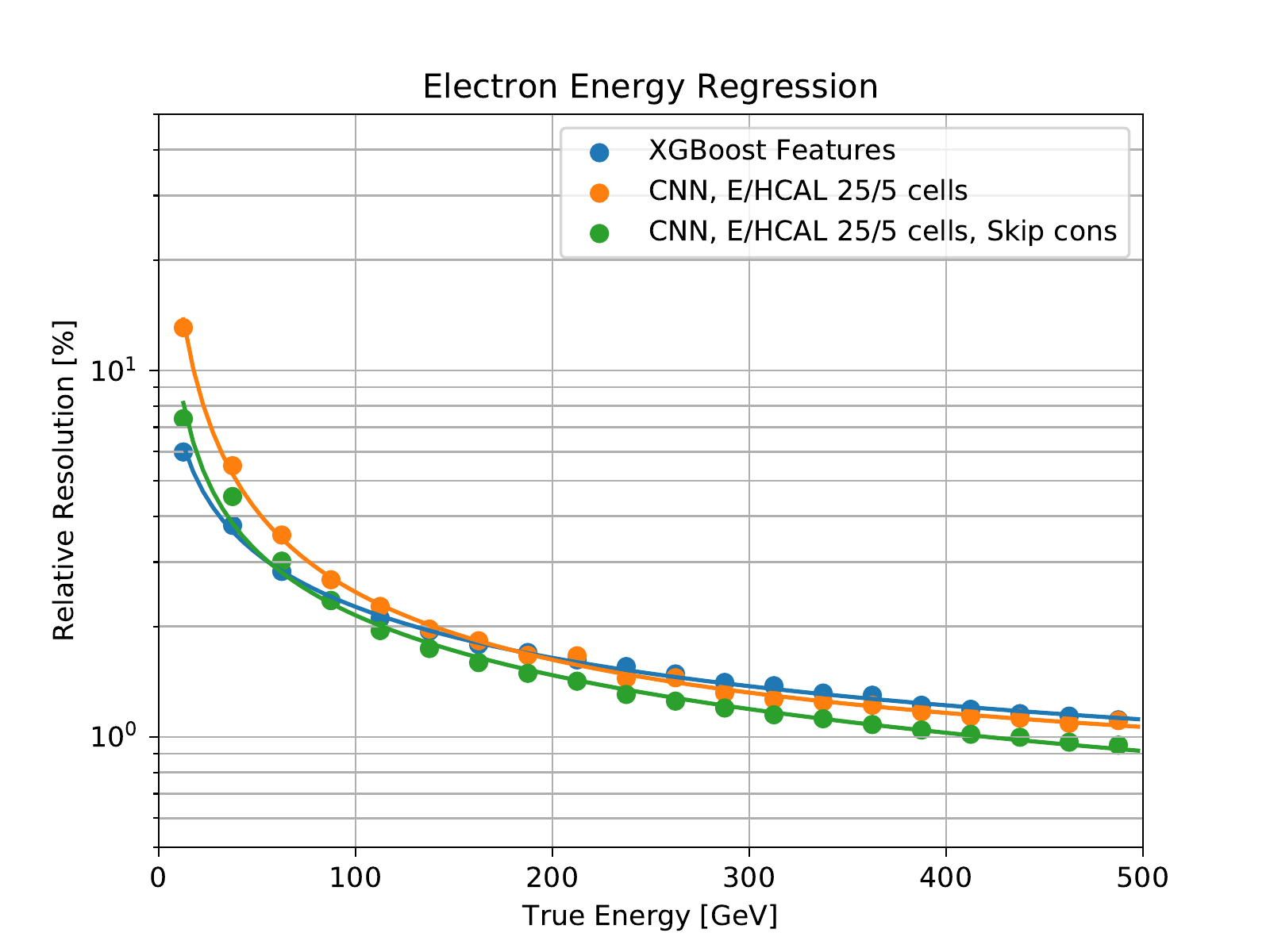}
\caption{Bias (top) and resolution (bottom) as a function of true energy for CNN energy predictions for electrons, with or without skip connections in the architecture.
}
\label{fig:reg_cnn_skip}
\end{figure}

A design choice that improved convergence time, and improved performance for the CNN, is including ``skip connections'' for the total ECAL and HCAL energies in the network.  In addition to the individual cell energy values, the total ECAL and HCAL energy values are given as inputs to both the first dense layer and to the last output layer.  The weights for these energy values are initialized to 1, as linear regression with coefficients near 1 is observed to reasonably reproduce the true energy values.  The impact of adding skip connections on performance using a CNN architecture for a fixed number of 5 training epochs is shown in Figure~\ref{fig:reg_cnn_skip}.

\section{Training for Regression Using Energy Summed in $z$}\label{app:z_sum_regression}

\begin{figure}[htbp]
\centering
\includegraphics[width=0.38\textwidth]{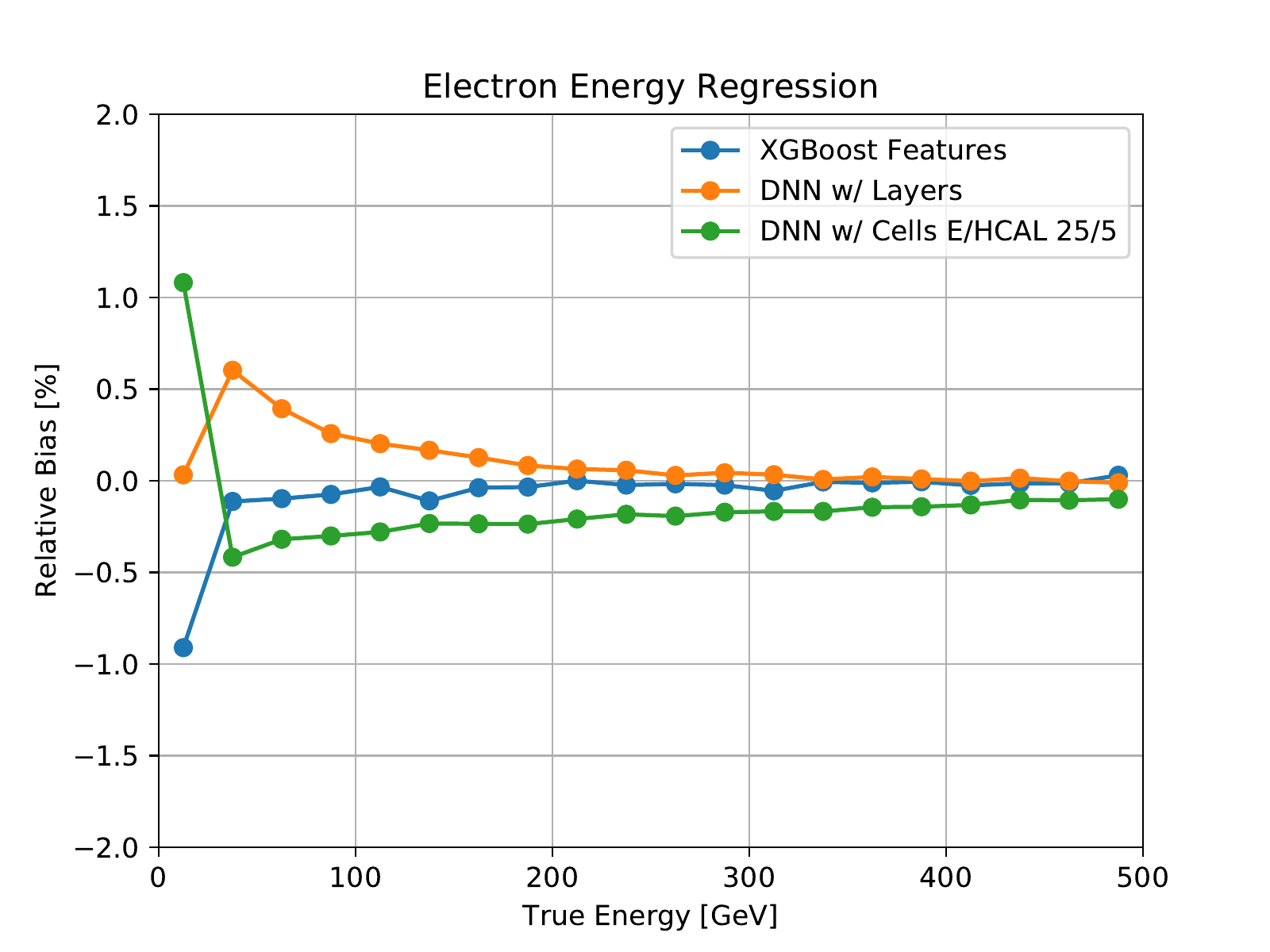}
\includegraphics[width=0.38\textwidth]{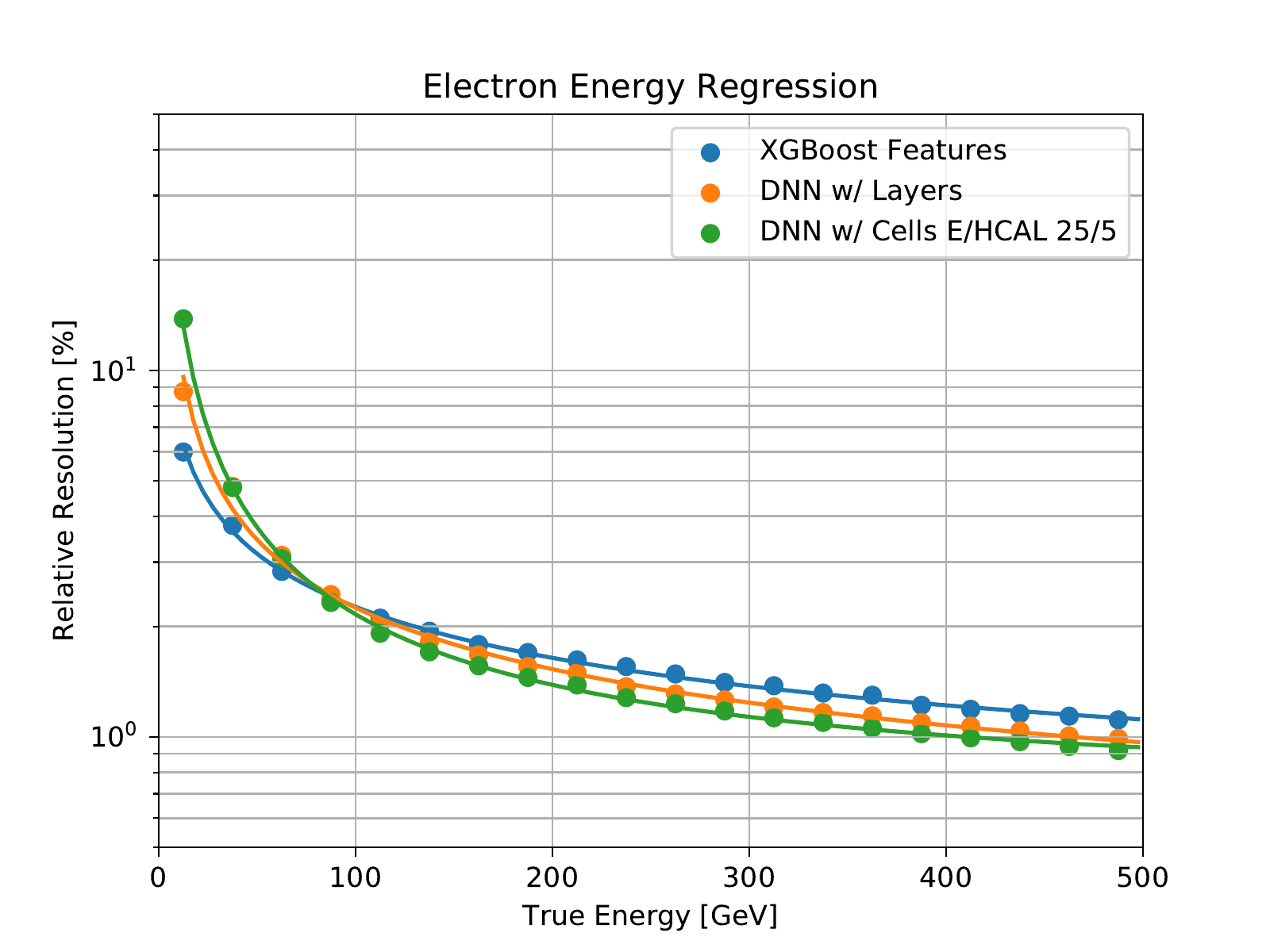}
\caption{Bias (top) and resolution (bottom) as a function of true energy for DNN energy predictions for electrons, using as input either the energy summed in layers of $z$, or the full cell information.\label{fig:reg_dnn_inputs}}
\end{figure}

For regression, we tried using only the energy summed in layers in the $z$ direction, instead of the full array of cell energies, as the mean $z$ coordinate was seen to be the most important additional feature in the XGBoost baseline.  The performance is better than the XGBoost baseline at high energies but worse than using the full cell-level information, as shown in Figure~\ref{fig:reg_dnn_inputs}.

\section{Energy Regression at Fixed Angles}\label{app:regression_fixed_angle}

\begin{figure*}[htbp]
\centering
\includegraphics[width=0.38\textwidth]{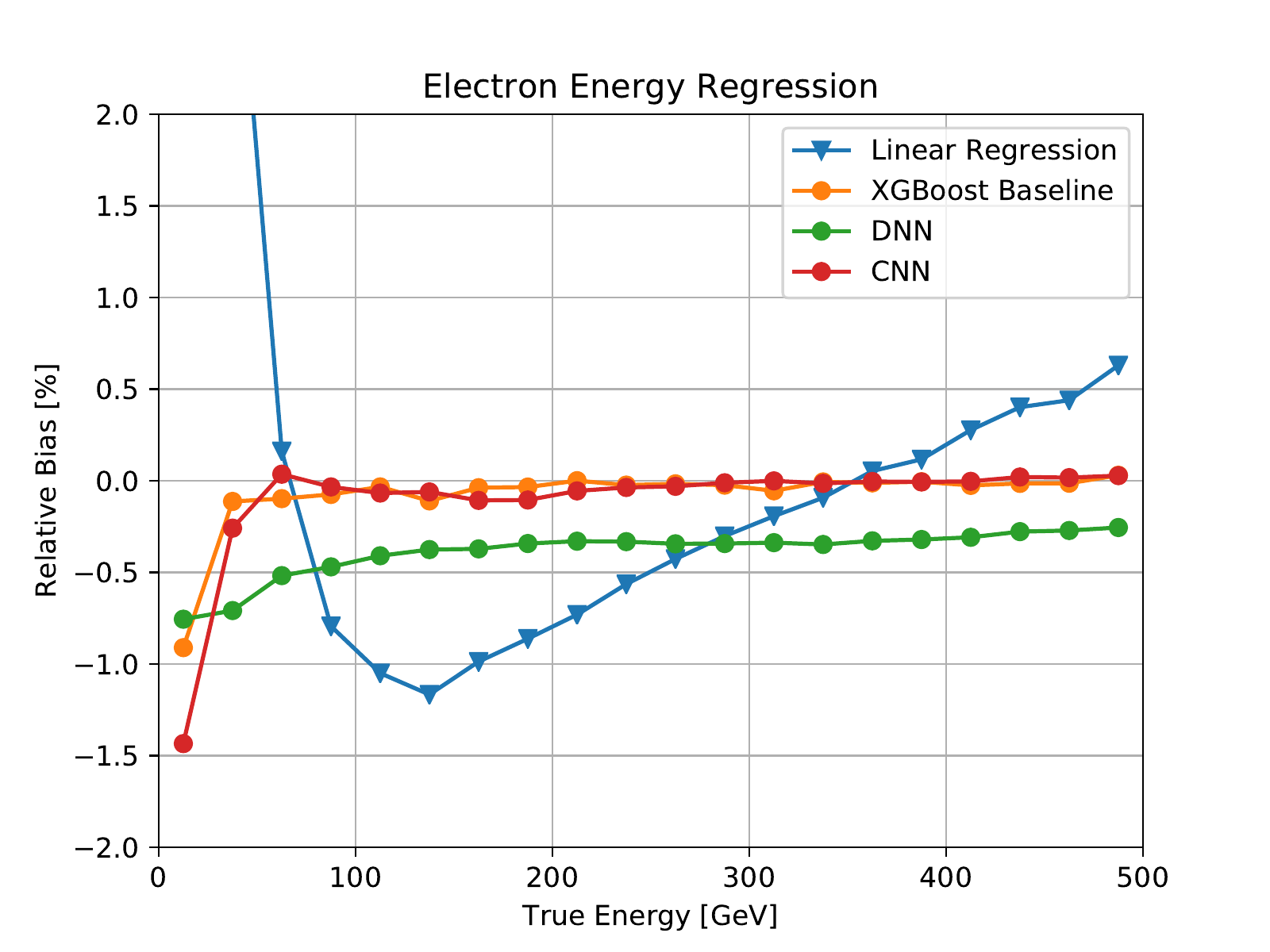}
\includegraphics[width=0.38\textwidth]{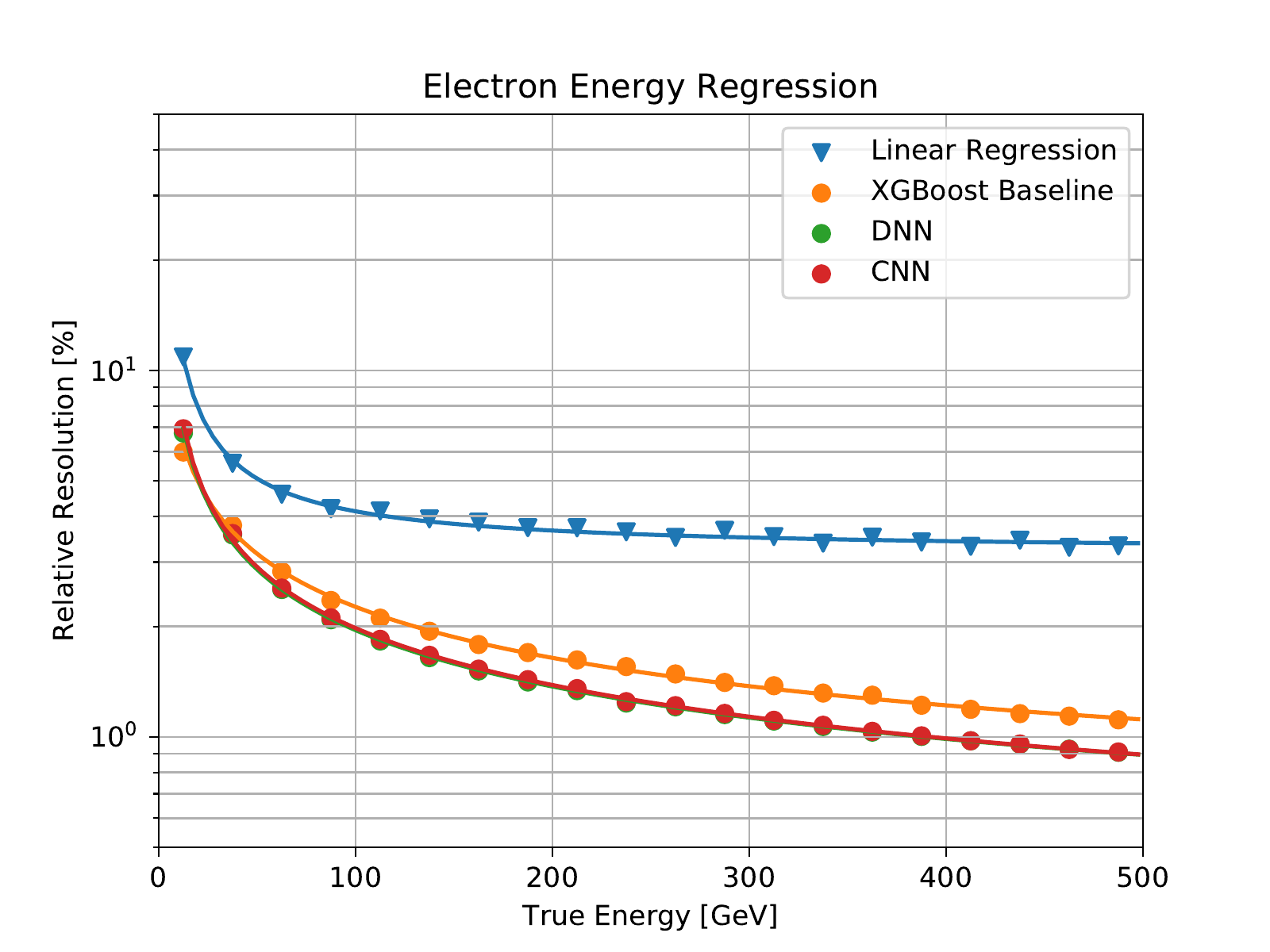} \\
\includegraphics[width=0.38\textwidth]{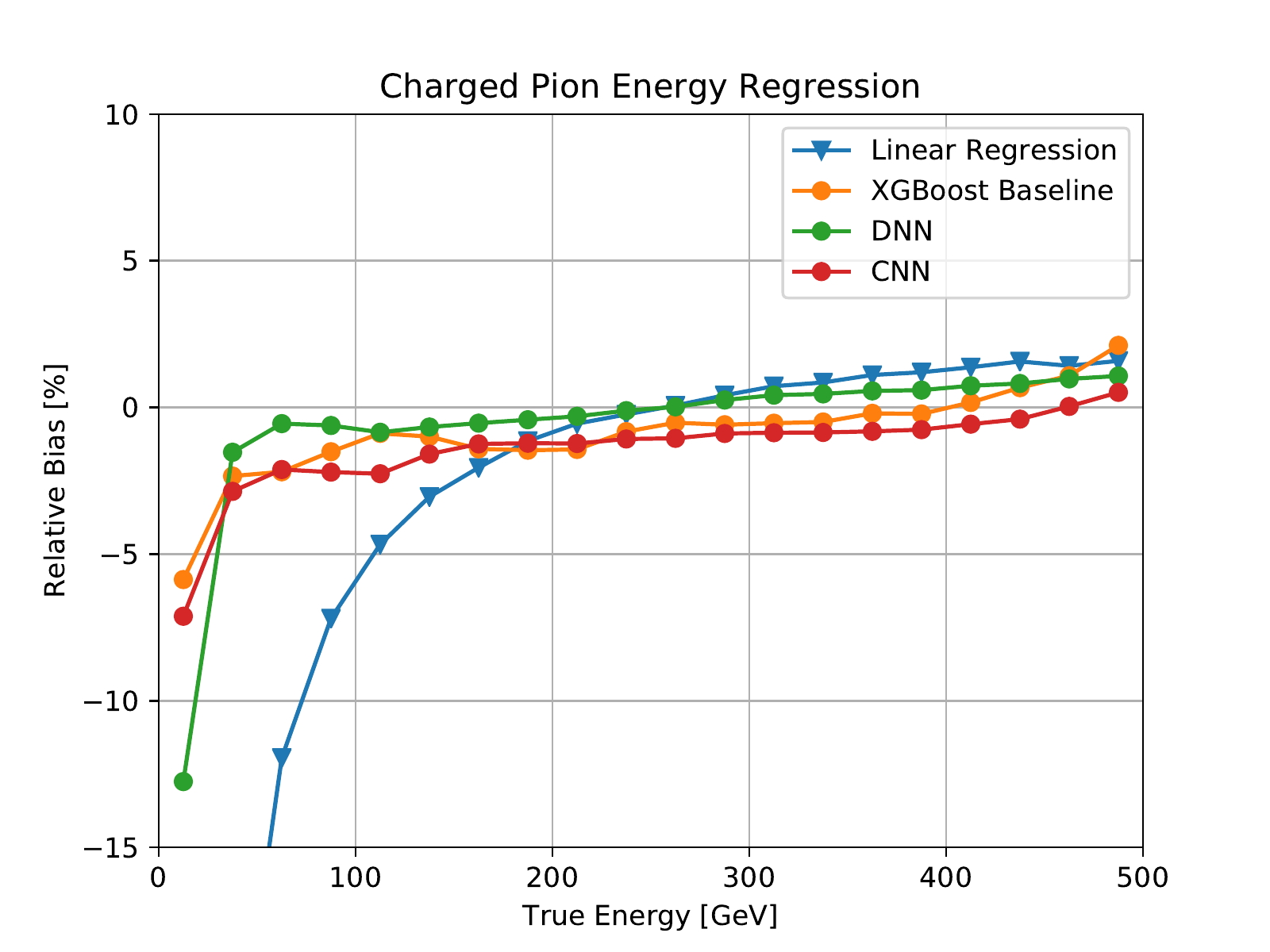}
\includegraphics[width=0.38\textwidth]{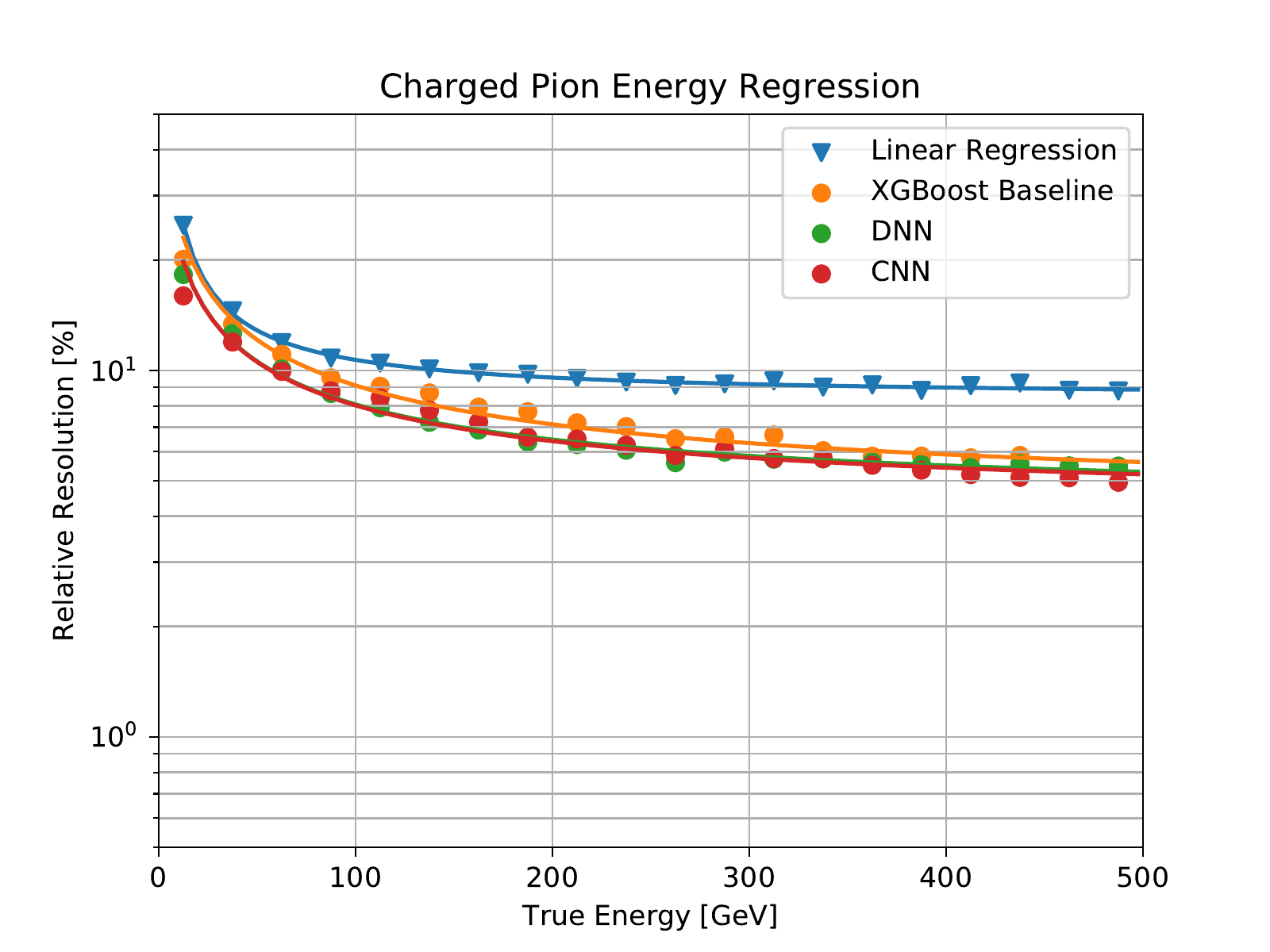}\\
\includegraphics[width=0.38\textwidth]{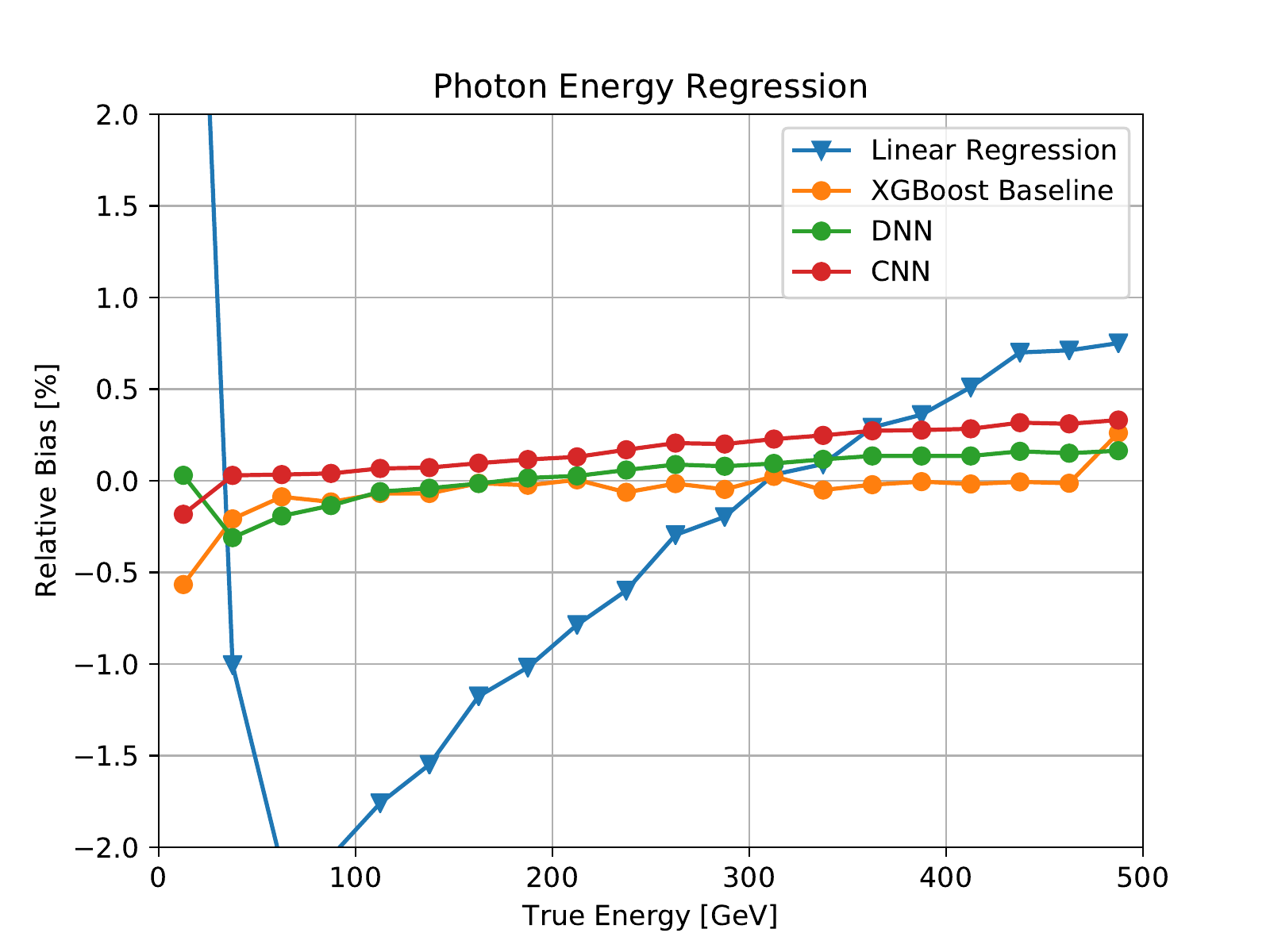}
\includegraphics[width=0.38\textwidth]{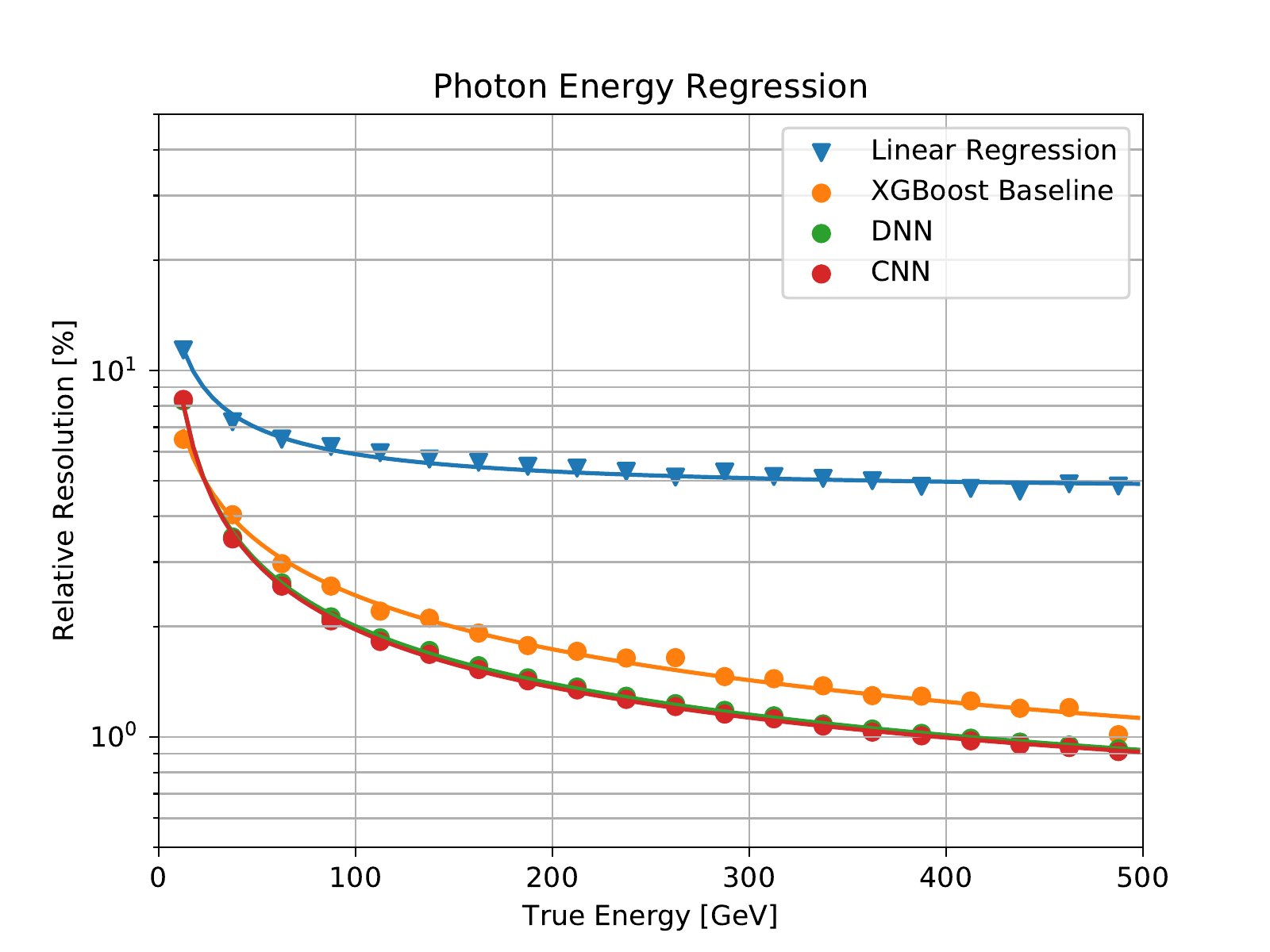} \\
\includegraphics[width=0.38\textwidth]{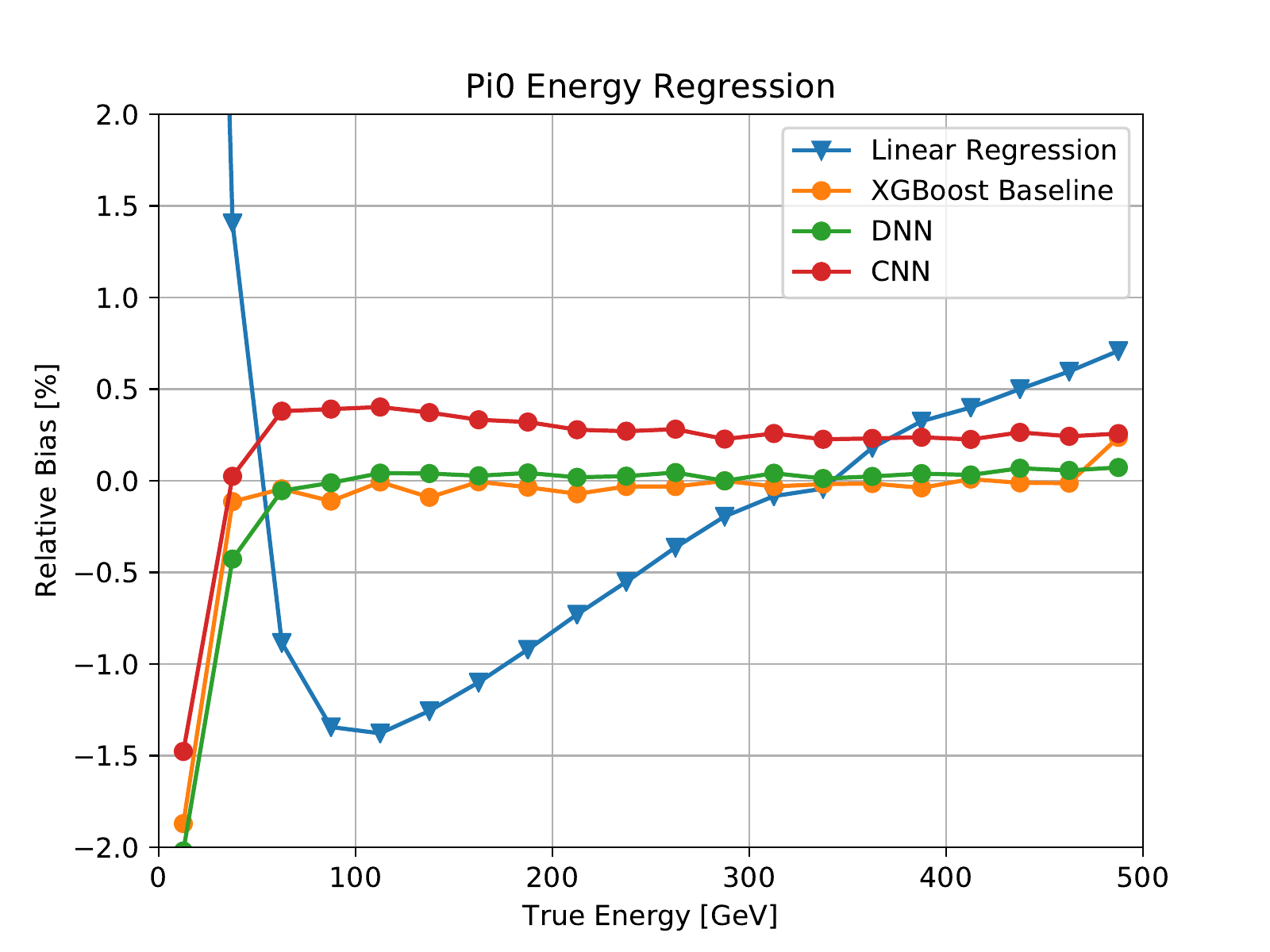}
\includegraphics[width=0.38\textwidth]{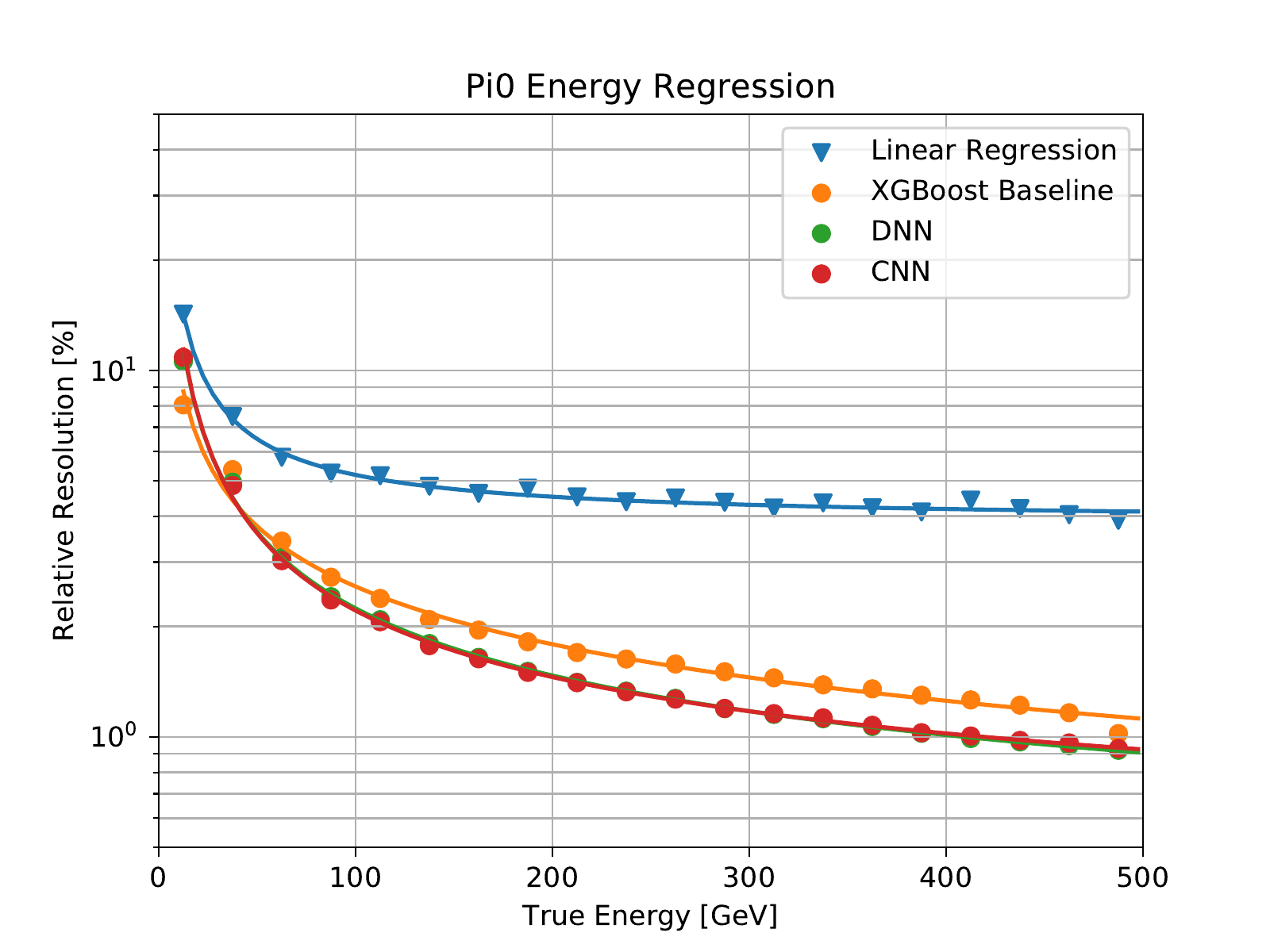}
\caption{Regression bias (top) and resolution (bottom) as a function
  of true energy for energy predictions on the REC dataset with fixed
  incident angle ($90^\mathrm{o}$). From top to bottom: electrons,
  charged pions, photons, and neutral
  pions.\label{fig:reg_dnn_vs_cnn_fixed}}
\end{figure*}

In Figure~\ref{fig:reg_dnn_vs_cnn_fixed} we show energy regression results when particles impact the calorimeter inner surface at a fixed angle of $90^{\mathrm o}$. All neural architectures and baseline algorithms are able to perform with great accuracy in this regime.

\begin{figure}[htbp]
\centering
\includegraphics[width=0.38\textwidth]{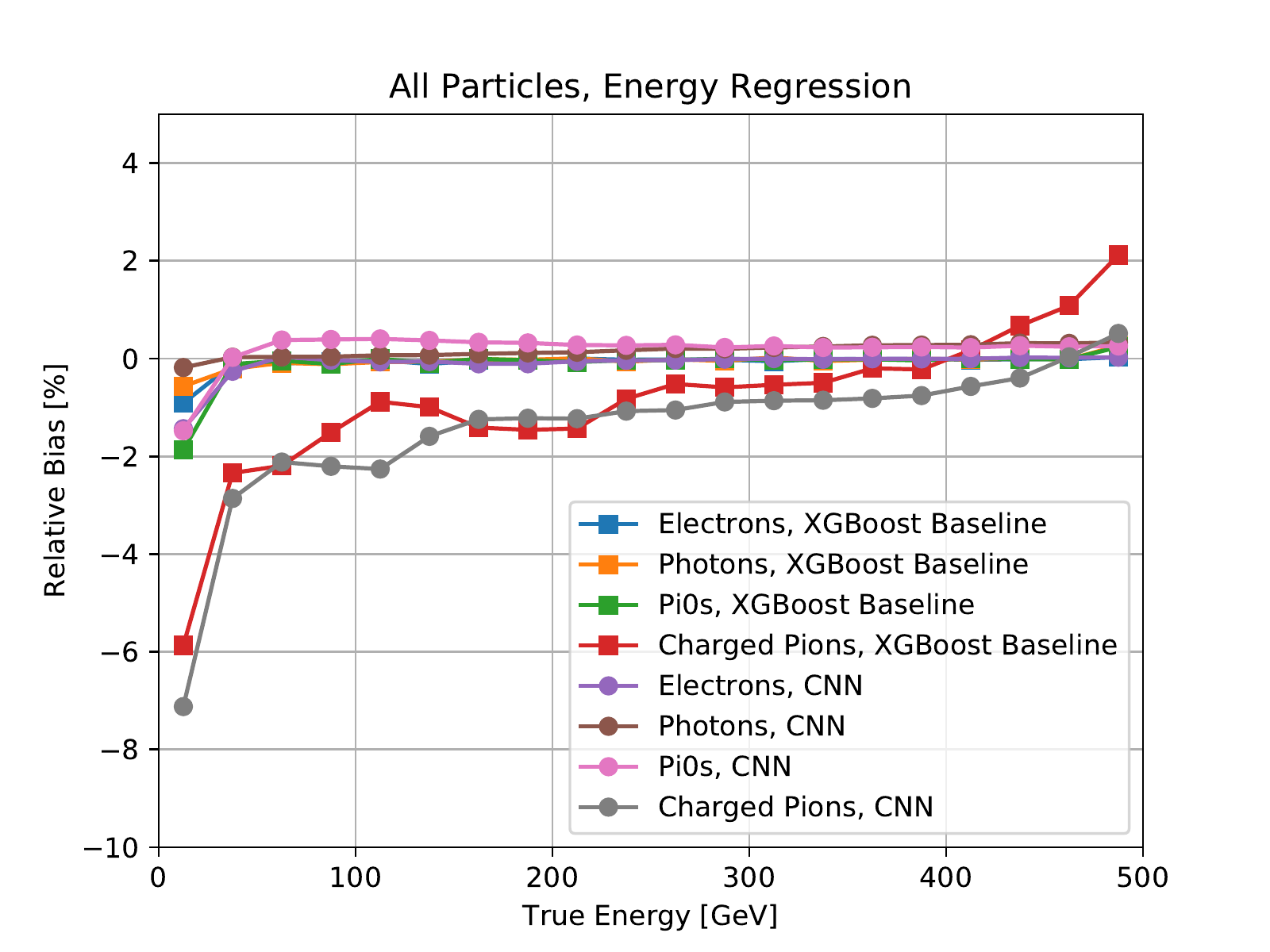}
\includegraphics[width=0.38\textwidth]{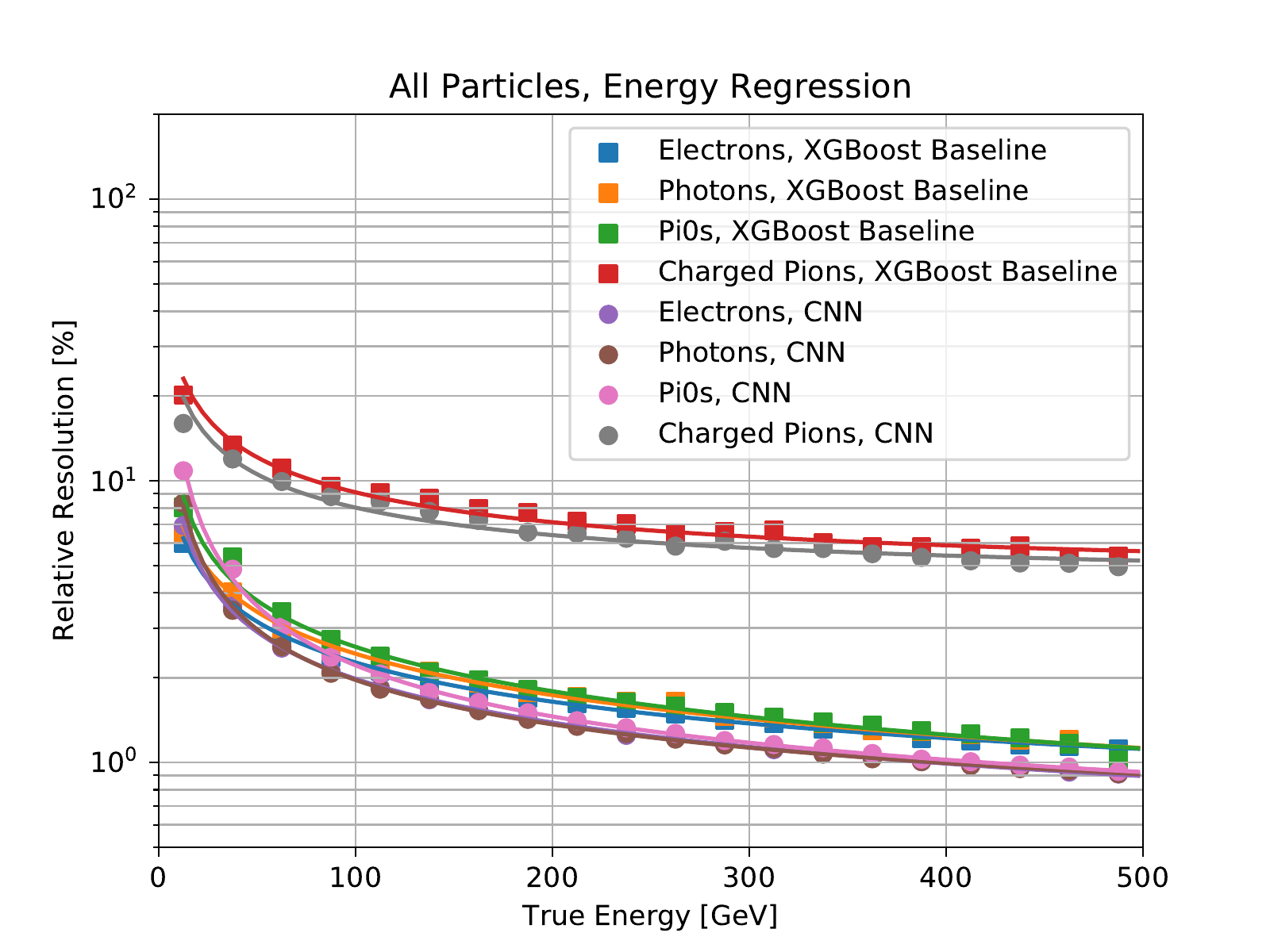}
\caption{Regression bias (top) and resolution (bottom) as a function of true energy for all particles, comparing the XGBoost baseline with the best CNN model on fixed-angle samples.\label{fig:reg_xgb_cnn_allparts_fixed}}
\end{figure}

Furthermore, in Figure~\ref{fig:reg_xgb_cnn_allparts_fixed} we summarize performance results on fixed-angle samples for all particle types with the XGBoost baseline and the CNN model.

\section{Regression performance training on a different particle type}\label{app:xtrain_regression}

All the tests so far have assumed that we know exactly what type of particle led to the reconstructed energy deposits.  In a real world situation, the particle identities are not known with complete confidence.  To see how the algorithms above would cope with that situation, we tried training each algorithm on an input sample of electron events, and then we used the trained algorithm to predict the energies for other particle types.

\begin{figure}[hbp]
\centering
\includegraphics[width=0.38\textwidth]{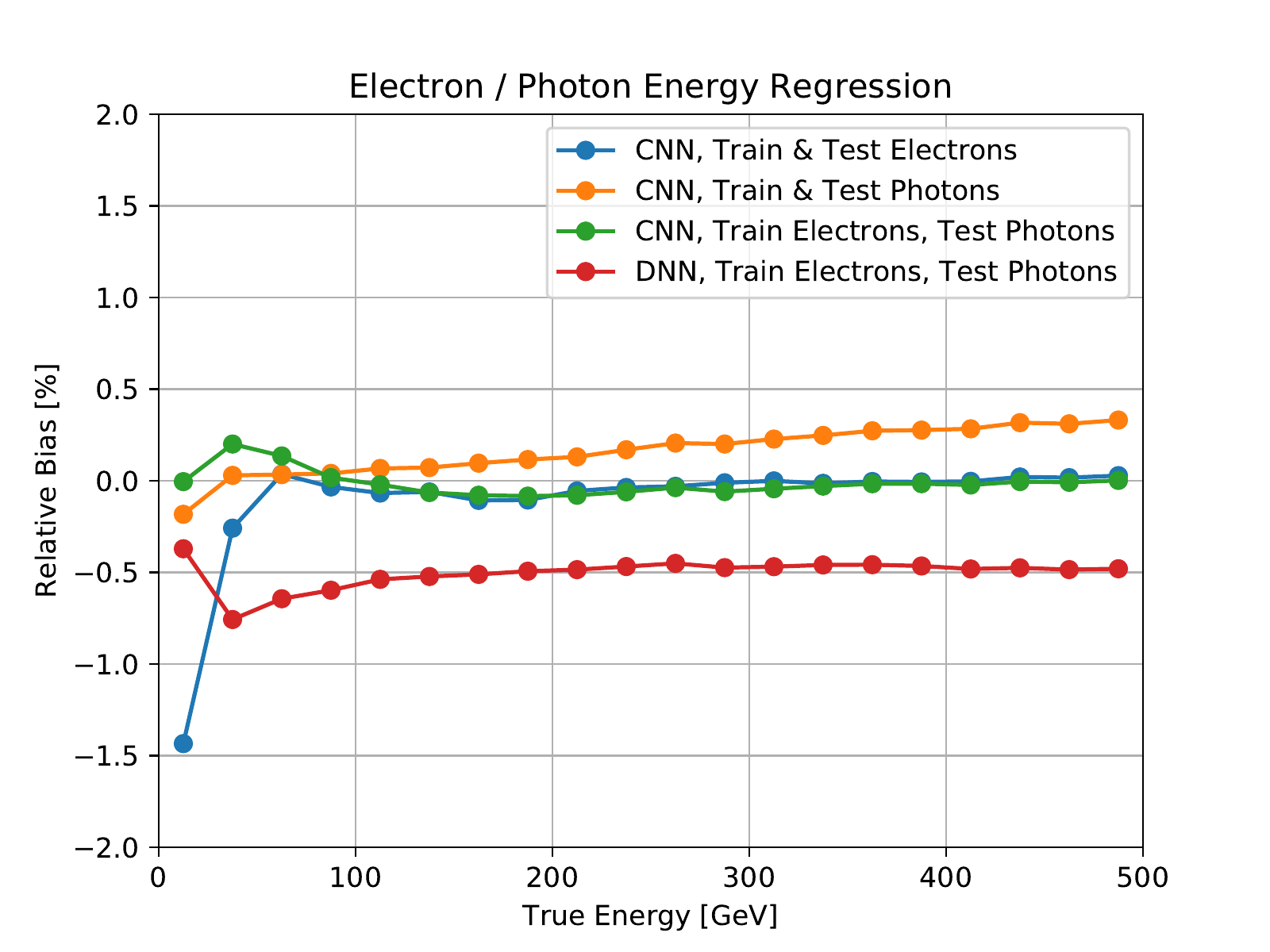}
\includegraphics[width=0.38\textwidth]{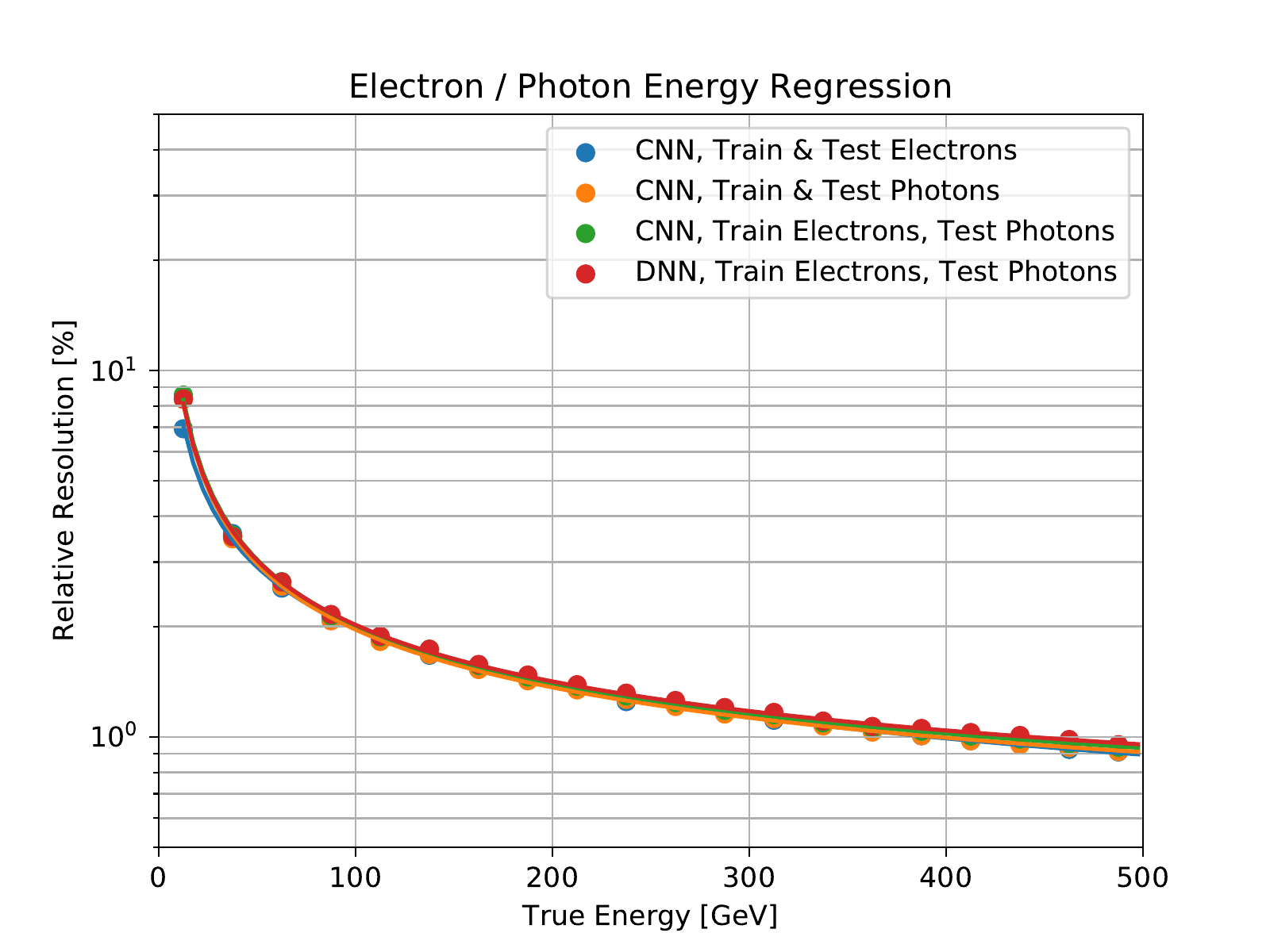}
\caption{Bias (top) and resolution (bottom) as a function of true energy, for electrons and photons.  The particles used to train and test each algorithm are given in the legend.
}
\label{fig:reg_nn_cross_gamma}
\end{figure}

\begin{figure}[htbp]
\centering
\includegraphics[width=0.38\textwidth]{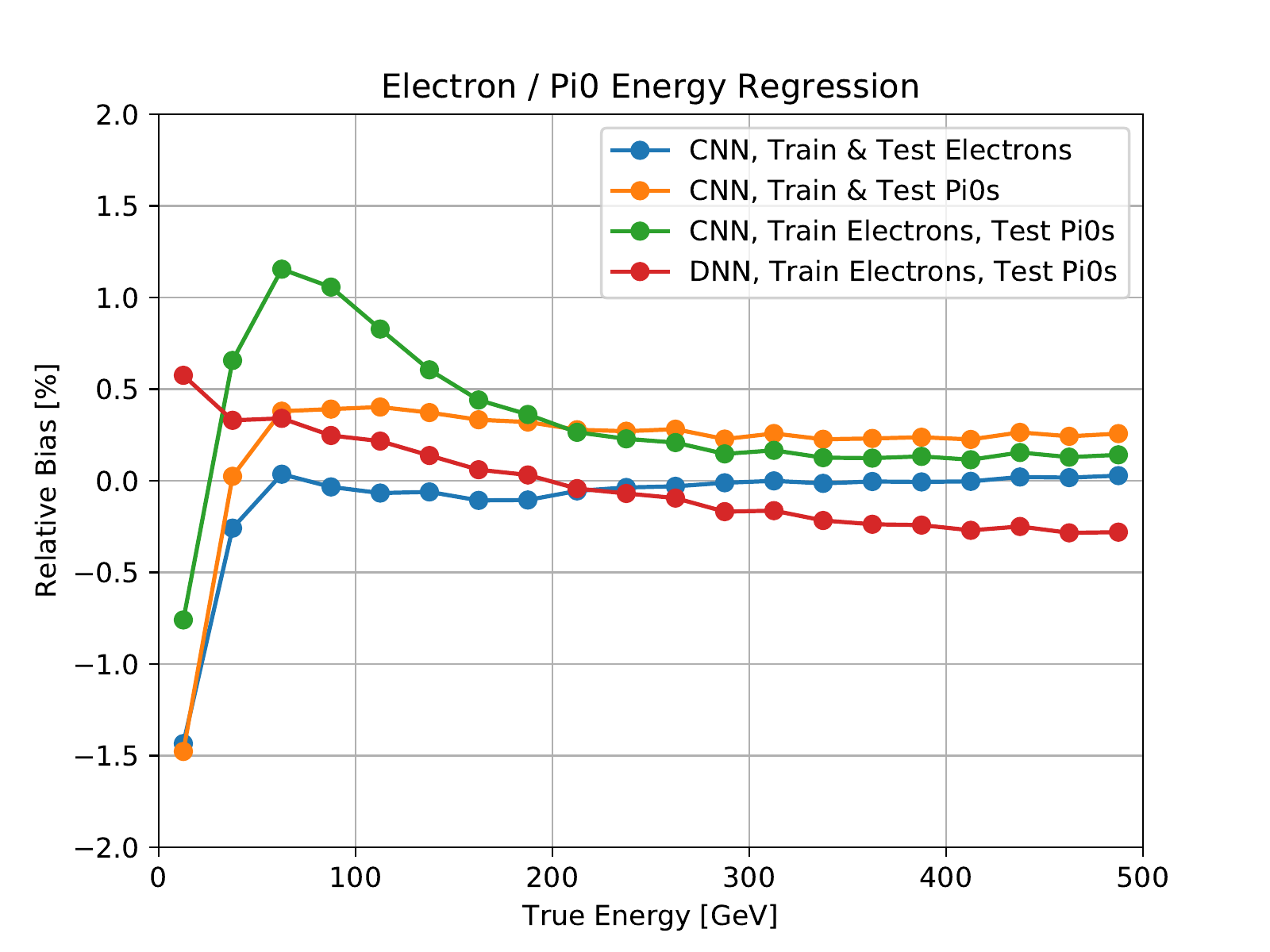}
\includegraphics[width=0.38\textwidth]{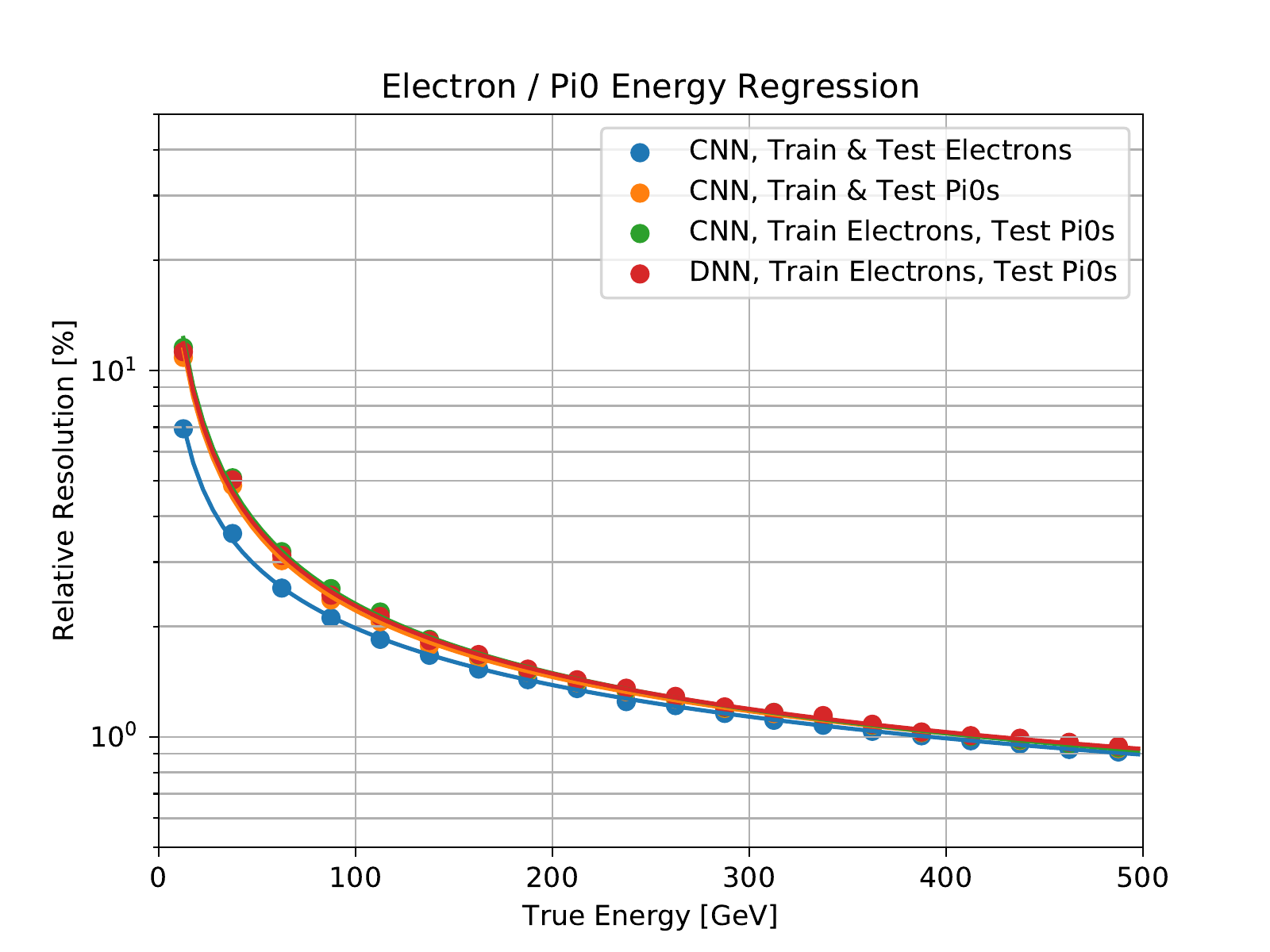}
\caption{Bias (top) and resolution (bottom) as a function of true energy, for electrons and \pizero.  The particles used to train and test each algorithm are given in the legend.
}
\label{fig:reg_nn_cross_pi0}
\end{figure}

The results are shown in Figure~\ref{fig:reg_nn_cross_gamma} for predicting photon energies and Figure~\ref{fig:reg_nn_cross_pi0} for predicting \pizero\ energies, and are compared to algorithms that are both trained and tested on the same particle type.  In each case, a DNN or CNN trained on electrons is able to achieve the same resolution as a CNN trained on photons or \pizero.  The bias is slightly larger in some cases.

Models trained on electrons, photons, or \pizero\ were found to not describe \chpi\ well at all.  This is not surprising given that \chpi\ have a hadronic shower, with a large fraction of energy deposited in the HCAL, compared to the other particles depositing almost all of their energy in the ECAL.

\begin{figure}[htbp]
\centering
\includegraphics[width=0.38\textwidth]{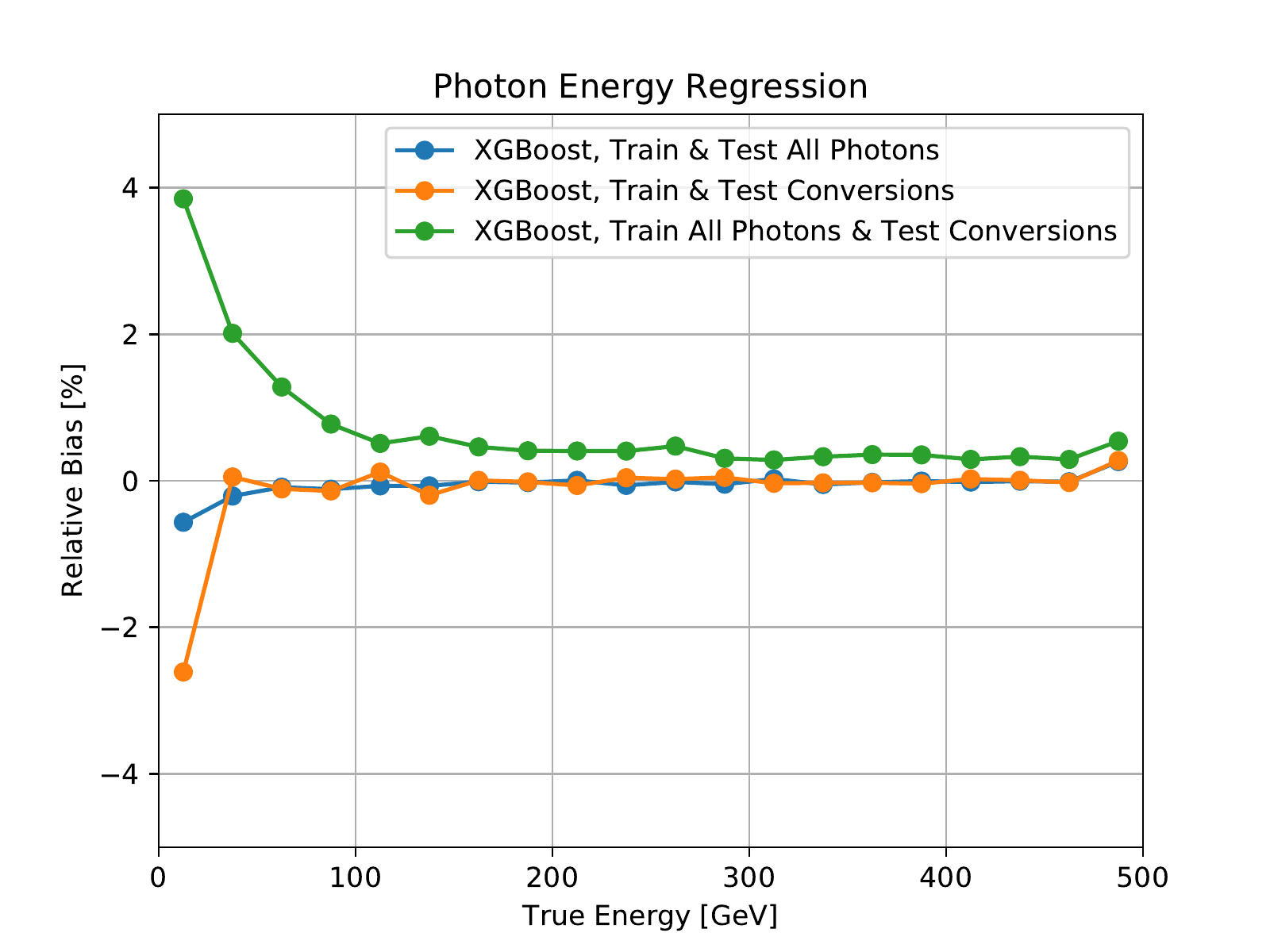}
\includegraphics[width=0.38\textwidth]{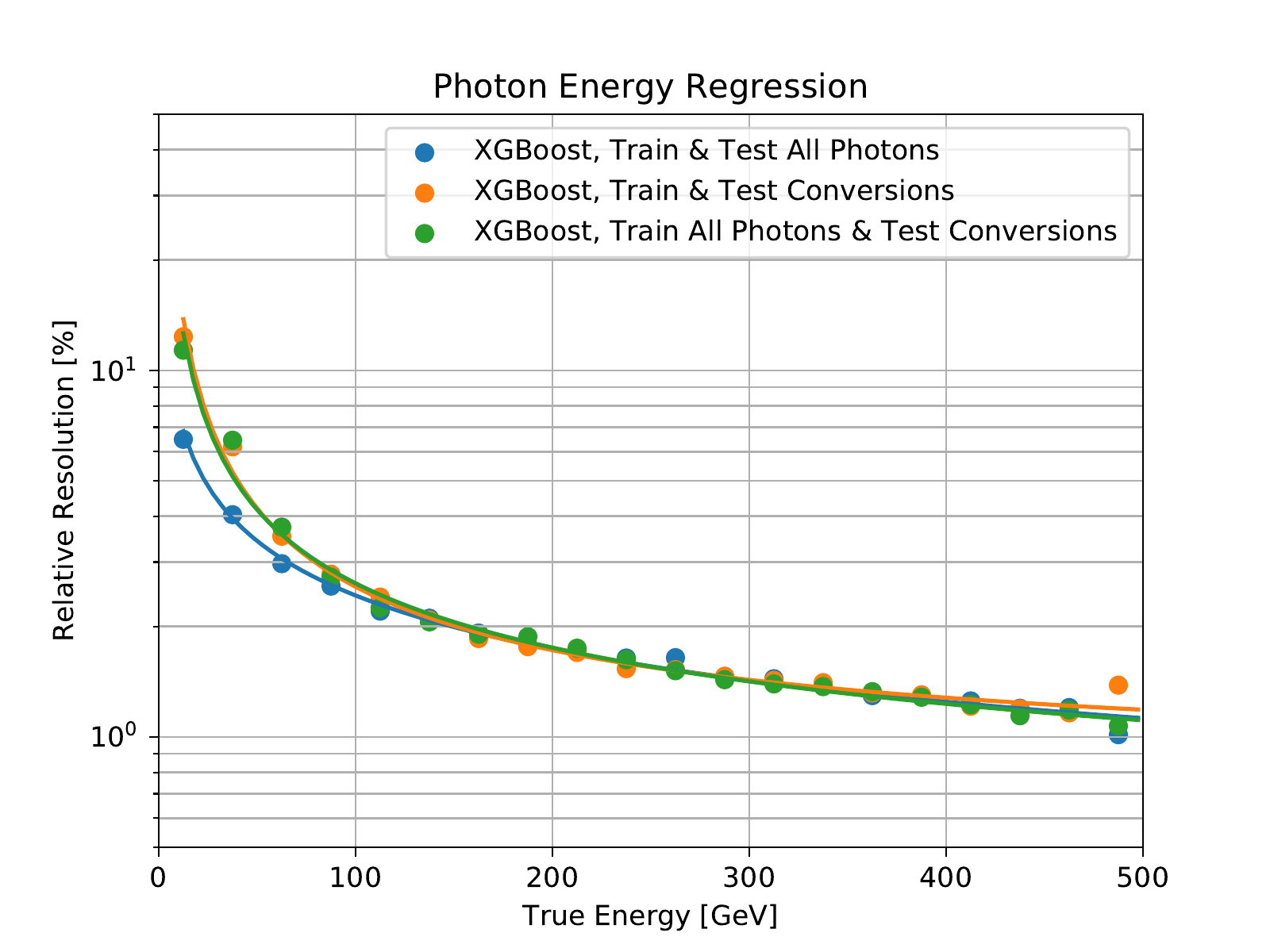}
\caption{Bias (top) and resolution (bottom) as a function of true energy, for photons using XGBoost regression.  We look at the photon sample when split up by conversions.
}
\label{fig:reg_xgb_conv_gamma}
\end{figure}

\begin{figure}[htbp]
\centering
\includegraphics[width=0.38\textwidth]{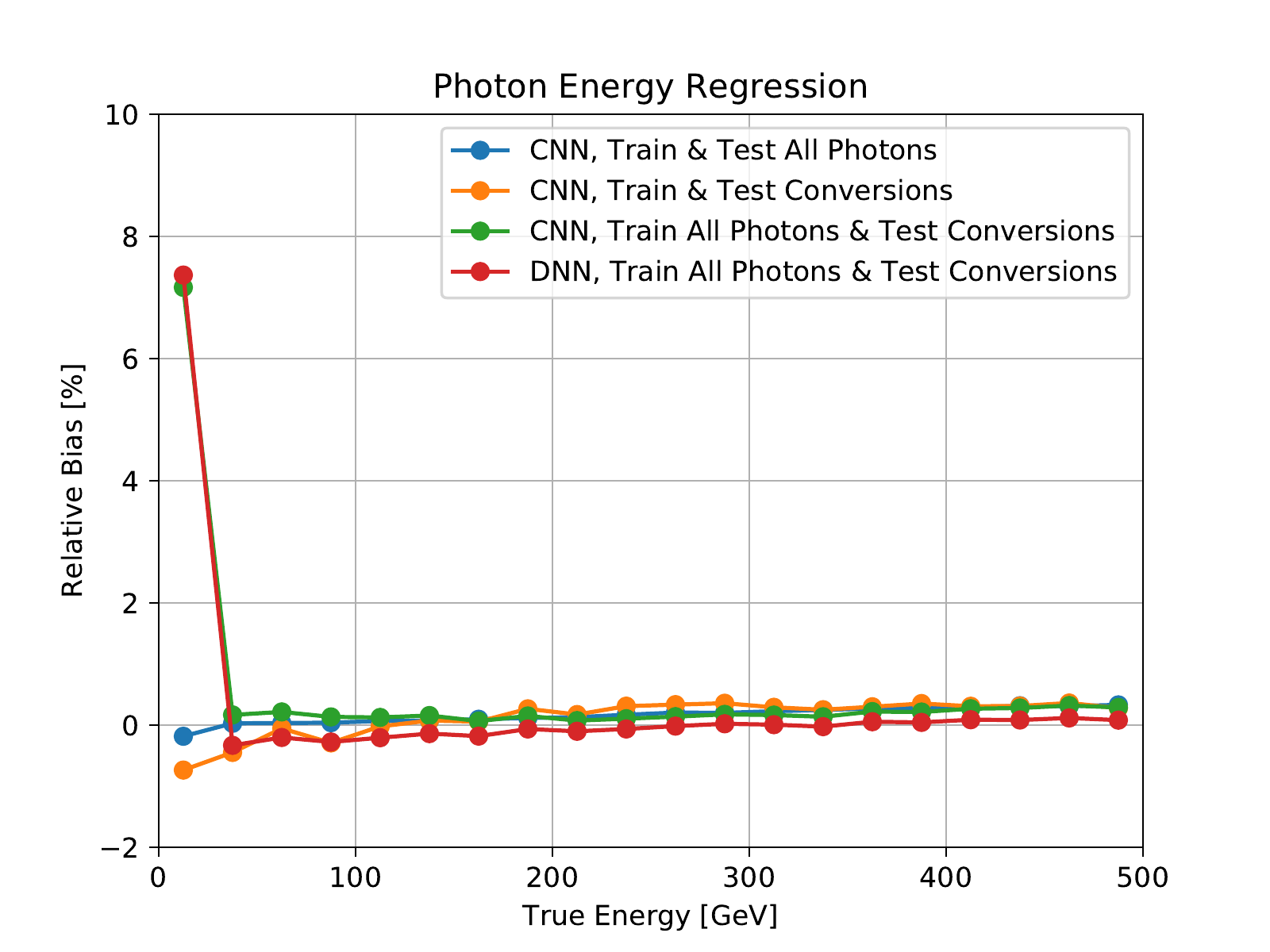}
\includegraphics[width=0.38\textwidth]{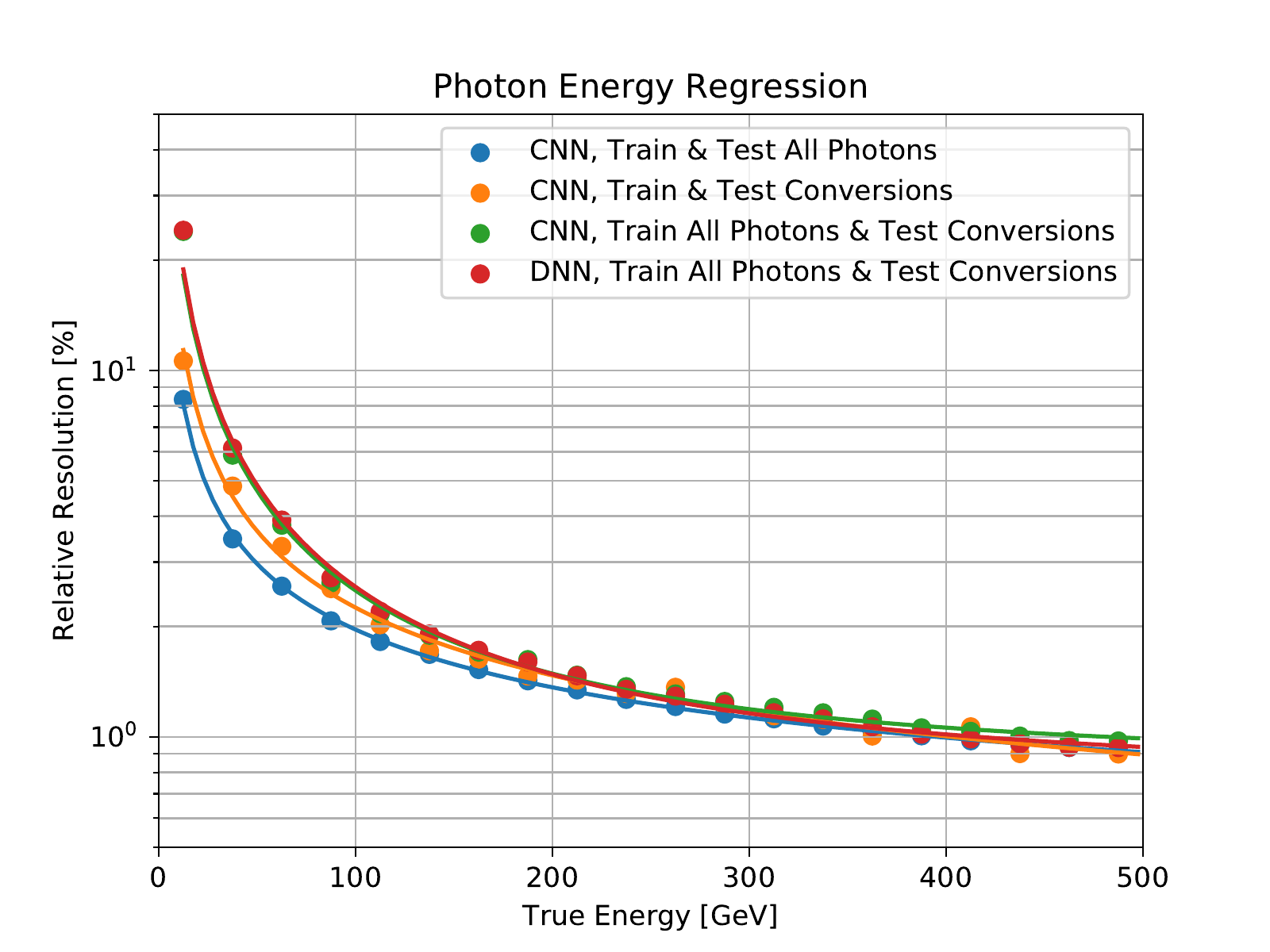}
\caption{Bias (top) and resolution (bottom) as a function of true energy, for photons using CNN or DNN regression.  We look at the photon sample when split up by conversions.
}
\label{fig:reg_nn_conv_gamma}
\end{figure}

We also checked whether the energy regression was different for photons that have converted into an $e^{+}e^{-}$ pair through interaction with the detector material.  These conversion photons comprise about 9\% of the photon sample.  We tried training and/or evaluating regression models separately on converted photons compared to all photons (which are dominated by unconverted).  The results are shown for XGBoost in Figure~\ref{fig:reg_xgb_conv_gamma} and for CNN/DNN models in Figure~\ref{fig:reg_nn_conv_gamma}.  Worse resolution is seen in each case for converted photons below around 100~GeV, which can be attributed to the subsequent electrons forming two showers instead of one in the calorimeter.  With XGBoost, the resolution remains the same for converted photons when training on the full sample, while for CNN or DNN, the resolution is worse below around 100~GeV.  The bias is also worse for converted photons at lower energy when training on all photons.

\section{Regression Studies with Large Sample Windows}\label{app:large_window_regression}

The studies in this section were performed using the full large window samples, of size 51x51x25 in ECAL and 11x11x60 in HCAL.
The samples consist of approximately 800,000 events for each particle type.  2/3 of the events were used for training and 1/3 of the events were used for testing.

\begin{figure}[htbp]
\centering
\includegraphics[width=0.38\textwidth]{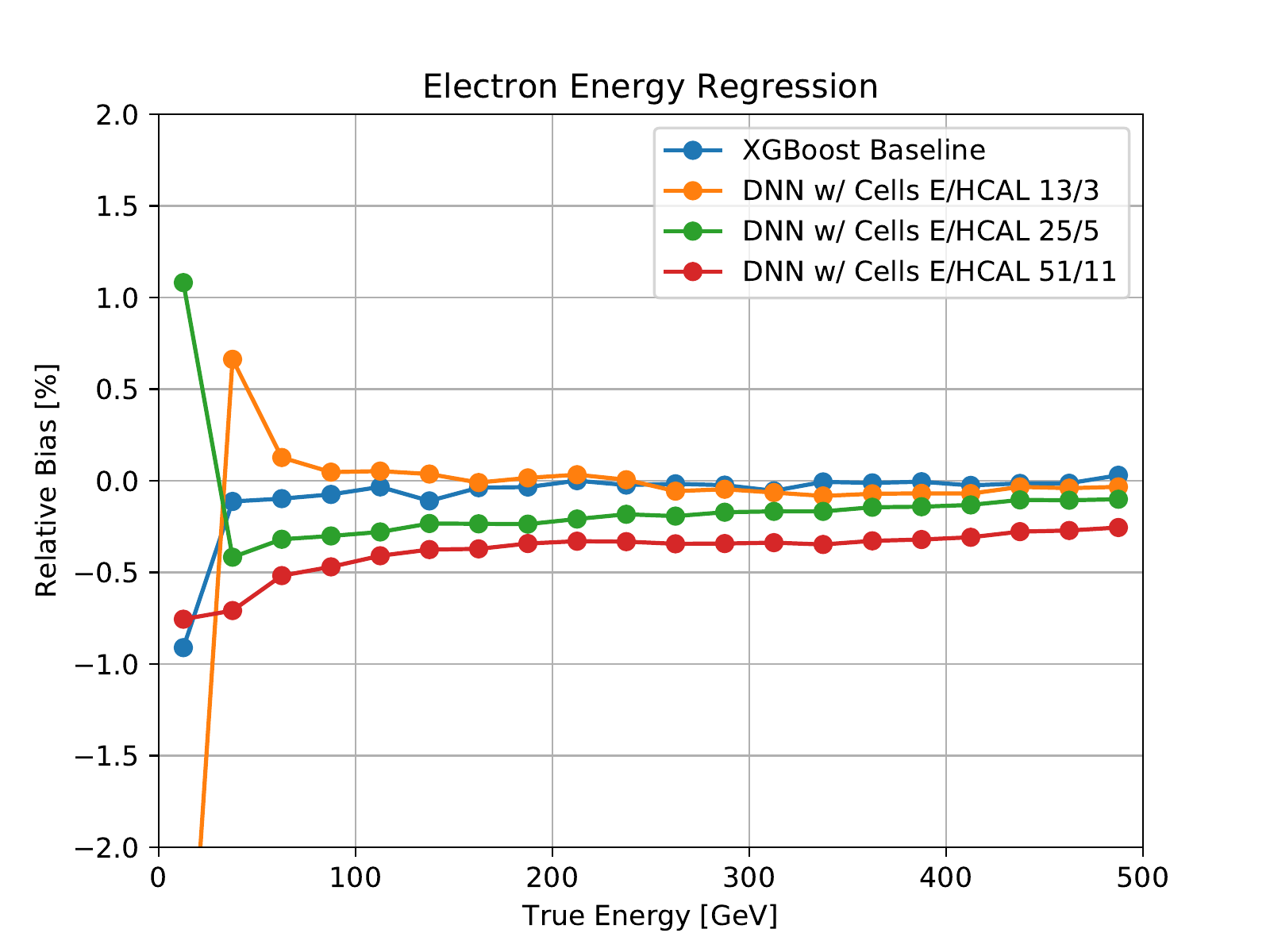}
\includegraphics[width=0.38\textwidth]{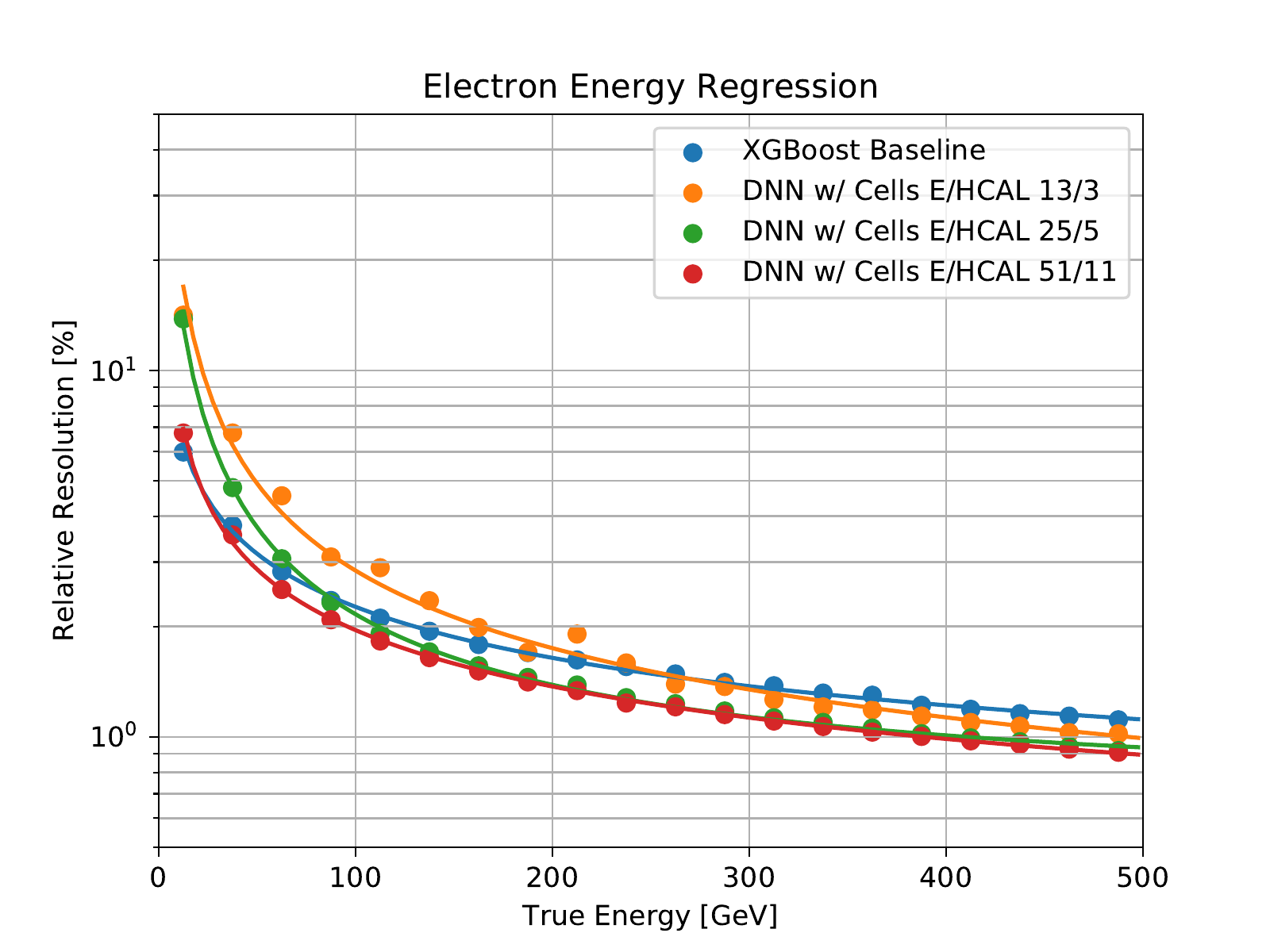}
\caption{Bias (top) and resolution (bottom) as a function of true energy for DNN energy predictions for electrons, with varying input window sizes.
}
\label{fig:reg_dnn_numcells}
\end{figure}

The most important design choice found for the DNN/CNN networks is the size of the window used as input.  For both DNN and CNN, to achieve the best performance for energies above 150~GeV, a minimum $(x,y)$ size of 25x25 in the ECAL and 5x5 in the HCAL is needed.  For energies below 150~GeV, the optimal performance is observed for a window size of 51x51 in the ECAL and 11x11 in the HCAL.  This is presumably due to wider showers at low energy.  The impact of the choice of window size is shown for DNN in Figure~\ref{fig:reg_dnn_numcells}, with the results for CNN being similar.  Drawbacks to the larger window size, however, include larger files, more memory usage, and that training takes about 5 times longer per epoch.

\begin{figure}[htbp]
\centering
\includegraphics[width=0.38\textwidth]{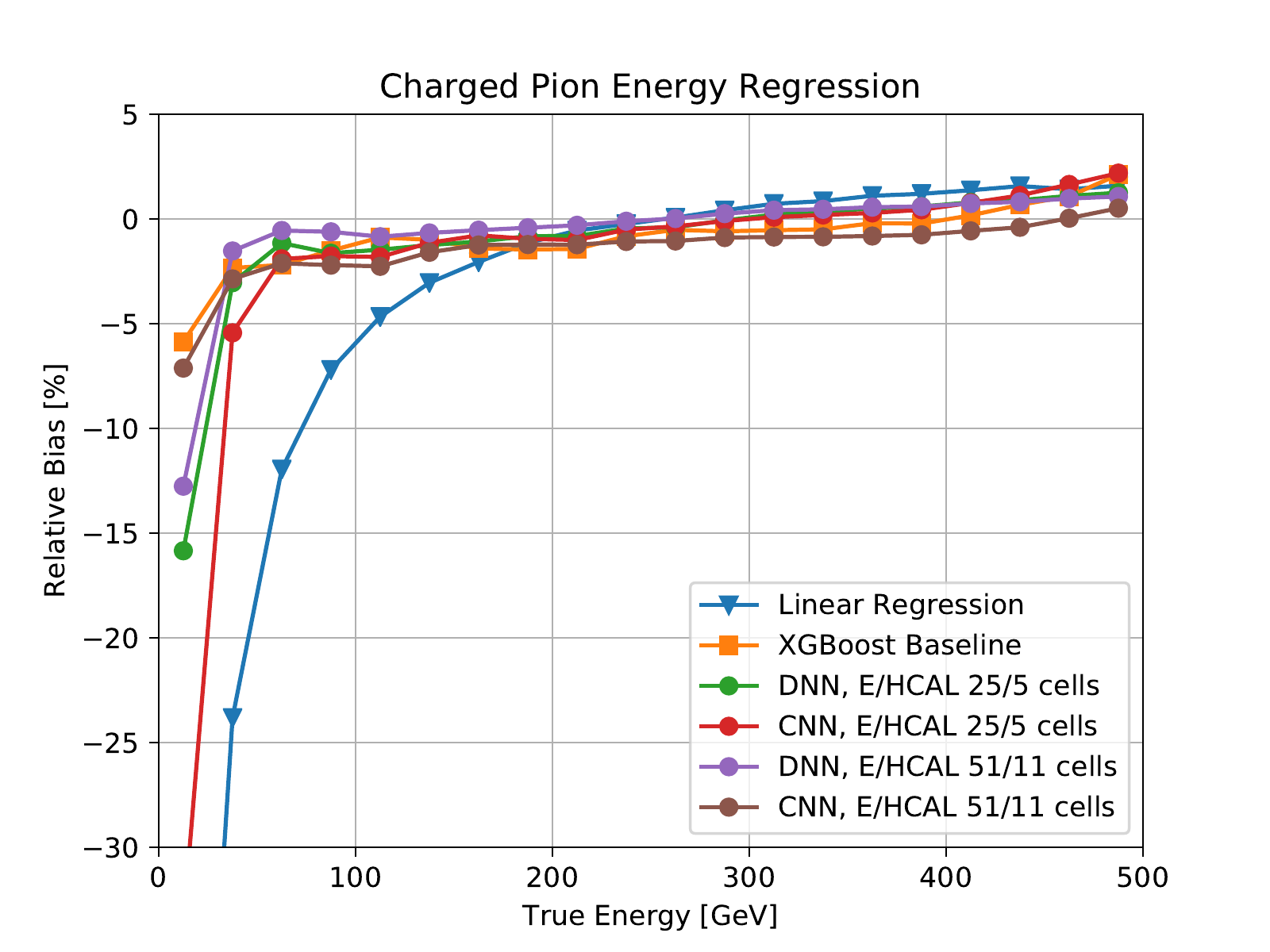}
\includegraphics[width=0.38\textwidth]{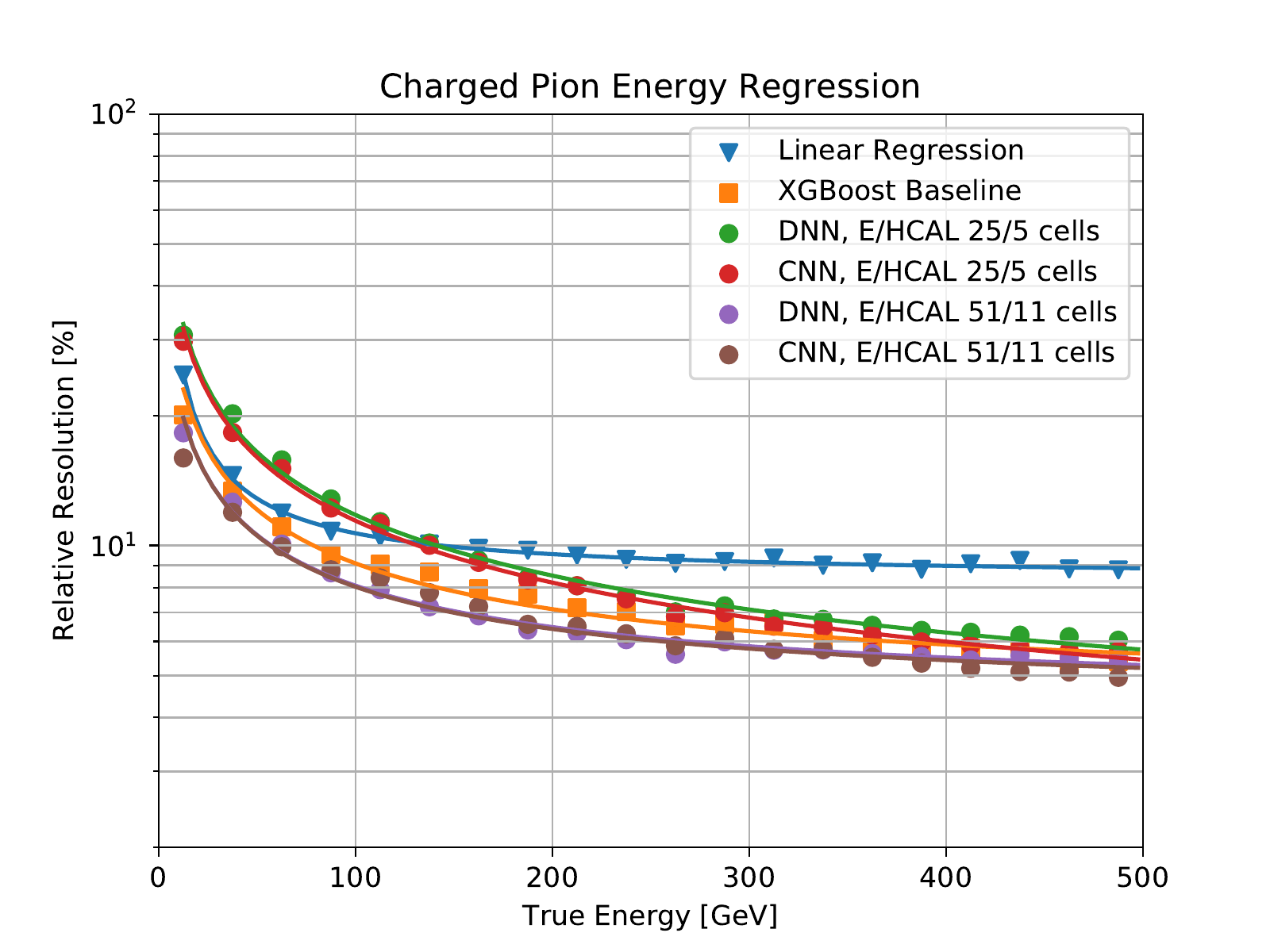}
\caption{Bias (top) and resolution (bottom) as a function of true energy for energy predictions for \chpi, comparing calorimeter window sizes for the CNN and DNN models.
}
\label{fig:reg_nn_numcells_chpi_large_window}
\end{figure}

Showers for \chpi\ were observed to be wider than the other particle types, especially at low energies, and so we compare the effect of the calorimeter window size choice for \chpi\ in Figure~\ref{fig:reg_nn_numcells_chpi_large_window}.  The wider window of 51x51 in $(x,y)$ in the ECAL and 11x11 in the HCAL gives better performance, especially at the lowest energies where the resolution is improved by a factor of about 2 over the smaller window size (25x25 ECAL, 5x5 HCAL).

\clearpage
\bibliographystyle{unsrt} 
\bibliography{biblio}

\begin{thebibliography}{10}

\bibitem{Denby:1987rk}
Bruce~H. Denby.
\newblock {Neural Networks and Cellular Automata in Experimental High-energy
  Physics}.
\newblock {\em Comput. Phys. Commun.}, 49:429--448, 1988.

\bibitem{Peterson:1988gs}
Carsten Peterson.
\newblock {Track Finding With Neural Networks}.
\newblock {\em Nucl. Instrum. Meth.}, A279:537, 1989.

\bibitem{Abreu:1992jp}
P.~Abreu et~al.
\newblock {Classification of the hadronic decays of the Z0 into b and c quark
  pairs using a neural network}.
\newblock {\em Phys. Lett.}, B295:383--395, 1992.

\bibitem{HiggsATLAS}
Georges Aad et~al.
\newblock {Observation of a new particle in the search for the Standard Model
  Higgs boson with the ATLAS detector at the LHC}.
\newblock {\em Phys. Lett.}, B716:1--29, 2012.

\bibitem{HiggsCMS}
Serguei Chatrchyan et~al.
\newblock {Observation of a new boson at a mass of 125 GeV with the CMS
  experiment at the LHC}.
\newblock {\em Phys. Lett.}, B716:30--61, 2012.

\bibitem{Aad:2008zzm}
G.~Aad et~al.
\newblock {The ATLAS Experiment at the CERN Large Hadron Collider}.
\newblock {\em JINST}, 3:S08003, 2008.

\bibitem{Chatrchyan:2008aa}
S.~Chatrchyan et~al.
\newblock {The CMS Experiment at the CERN LHC}.
\newblock {\em JINST}, 3:S08004, 2008.

\bibitem{Apollinari:2284929}
Apollinari G., Béjar~Alonso I., Brüning O., Fessia P., Lamont M., Rossi L.,
  and Tavian L.
\newblock {\em {High-Luminosity Large Hadron Collider (HL-LHC): Technical
  Design Report V. 0.1}}.
\newblock CERN Yellow Reports: Monographs. CERN, Geneva, 2017.

\bibitem{ILC}
Ties Behnke, James~E. Brau, Brian Foster, Juan Fuster, Mike Harrison,
  James~McEwan Paterson, Michael Peskin, Marcel Stanitzki, Nicholas Walker, and
  Hitoshi Yamamoto.
\newblock {The International Linear Collider Technical Design Report - Volume
  1: Executive Summary}.
\newblock 2013.

\bibitem{CLIC}
L.~{Linssen}, A.~{Miyamoto}, M.~{Stanitzki}, and H.~{Weerts}.
\newblock {Physics and Detectors at CLIC: CLIC Conceptual Design Report}.
\newblock {\em ArXiv e-prints}, February 2012.

\bibitem{CMSCollaboration:2015zni}
V.~Khachatryan et~al.
\newblock {Technical Proposal for the Phase-II Upgrade of the CMS Detector}.
\newblock 2015.

\bibitem{GEANT4}
S.~Agostinelli et~al.
\newblock {GEANT4: A Simulation toolkit}.
\newblock {\em Nucl. Instrum. Meth.}, A506:250--303, 2003.

\bibitem{GEANT_usage}
Roland Jansky.
\newblock {The ATLAS Fast Monte Carlo Production Chain Project}.
\newblock {\em J. Phys.: Conf. Ser. 664 072024}, 2015.

\bibitem{ML1}
Luke de~Oliveira, Michael Kagan, Lester Mackey, Benjamin Nachman, and Ariel
  Schwartzman.
\newblock {Jet-images — deep learning edition}.
\newblock {\em JHEP}, 07:069, 2016.

\bibitem{ML2}
Luke de~Oliveira, Michela Paganini, and Benjamin Nachman.
\newblock {Learning Particle Physics by Example: Location-Aware Generative
  Adversarial Networks for Physics Synthesis}.
\newblock {\em Comput. Softw. Big Sci.}, 1(1):4, 2017.

\bibitem{ML3}
Michela Paganini, Luke de~Oliveira, and Benjamin Nachman.
\newblock {CaloGAN: Simulating 3D High Energy Particle Showers in Multi-Layer
  Electromagnetic Calorimeters with Generative Adversarial Networks}.
\newblock 2017.

\bibitem{ML4}
Josh Cogan, Michael Kagan, Emanuel Strauss, and Ariel Schwarztman.
\newblock {Jet-Images: Computer Vision Inspired Techniques for Jet Tagging}.
\newblock {\em JHEP}, 02:118, 2015.

\bibitem{Goodfellow}
I.~J. {Goodfellow}, J.~{Pouget-Abadie}, M.~{Mirza}, B.~{Xu}, D.~{Warde-Farley},
  S.~{Ozair}, A.~{Courville}, and Y.~{Bengio}.
\newblock {Generative Adversarial Networks}.
\newblock {\em ArXiv e-prints}, June 2014.

\bibitem{Mau2017}
Federico Carminati, Gulrukh Khattak, Maurizio Pierini, Sofia Vallecorsafa, Amir
  Farbin, Benjamin Hooberman, Wei Wei, Matt Zhang, Barin Pacela, Vitorial,
  Maria Spiropulu, and Jean-roch Vlimant.
\newblock {Calorimetry with Deep Learning : Particle Classification , Energy
  Regression , and Simulation for High-Energy Physics}.
\newblock In {\em NIPS}, 2017.

\bibitem{keras}
Fran\c{c}ois Chollet et~al.
\newblock Keras.
\newblock \url{https://github.com/fchollet/keras}, 2015.

\bibitem{tensorflow2015-whitepaper}
Mart\'{\i}n~Abadi et~al.
\newblock {TensorFlow}: Large-scale machine learning on heterogeneous systems,
  2015.
\newblock Software available from tensorflow.org.

\bibitem{PyTorch}
Adam Paszke, Sam Gross, Soumith Chintala, Gregory Chanan, Edward Yang, Zachary
  DeVito, Zeming Lin, Alban Desmaison, Luca Antiga, and Adam Lerer.
\newblock Automatic differentiation in pytorch.
\newblock In {\em NIPS-W}, 2017.

\bibitem{Lebrun:2012hj}
P.~Lebrun, L.~Linssen, A.~Lucaci-Timoce, D.~Schulte, F.~Simon, S.~Stapnes,
  N.~Toge, H.~Weerts, and J.~Wells.
\newblock {The CLIC Programme: Towards a Staged e+e- Linear Collider Exploring
  the Terascale : CLIC Conceptual Design Report}.
\newblock 2012.

\bibitem{NIPS}
Federico Carminati, Amir Farbin, Benjamin Hooberman, Gulrukh Khattak, Vitória
  Barin~Pacela, Maurizio Pierini, Maria Spiropulu, Sofia Vallecorsafac,
  Jean-Roch Vlimant, Wei Wei, and Matt Zhang.
\newblock Calorimetry with deep learning : Particle classification , energy
  regression , and simulation for high-energy physics.
\newblock 2017.

\bibitem{Nachman_DNN}
Michela~Paganini Luke~de Oliveira, Benjamin~Nachman.
\newblock Electromagnetic showers beyond shower shapes, 2018.

\bibitem{Nachman_GAN}
{Deep generative models for fast shower simulation in ATLAS}.
\newblock Technical Report ATL-SOFT-PUB-2018-001, CERN, Geneva, Jul 2018.

\bibitem{Nachman_GAN2}
Benjamin~Nachman Luke~de Oliveira, Michela~Paganini.
\newblock Controlling physical attributes in gan-accelerated simulation of
  electromagnetic calorimeters, 2017.

\bibitem{BDT_thesis}
Nicolas~Palm Perez.
\newblock Electron identification using machine learning in the atlas
  experiment with 2016 data, 2017.

\bibitem{acgan}
A.~{Odena}, C.~{Olah}, and J.~{Shlens}.
\newblock {Conditional Image Synthesis With Auxiliary Classifier GANs}.
\newblock {\em ArXiv e-prints}, October 2016.

\bibitem{rmsProp}
Geoffrey Hinton, Nitish Srivastava, and Kevin Swersky.
\newblock {Lecture 6a overview of mini–batch gradi-ent descent.}
\newblock 2012.

\bibitem{GoogLeNet}
Yangqing Jia Pierre Sermanet Scott Reed Dragomir Anguelov Dumitru Erhan Vincent
  Vanhoucke Andrew~Rabinovich Christian~Szegedy, Wei~Liu.
\newblock Going deeper with convolutions, 2014.

\bibitem{skopt}
Tim Head et~al.
\newblock scikit-optimize/scikit-optimize: v0.5.2, March 2018.

\bibitem{Breiman}
Friedman Breiman.
\newblock {\em Classification and regression trees}.
\newblock 1984.

\end{thebibliography}

\end{document}